\documentclass[12pt]{iopart}  

\usepackage{color}
\usepackage{graphics}
\usepackage{psfig,subfigure}
\usepackage{upgreek}
\usepackage{graphicx}
\usepackage{dcolumn}
\usepackage{bm}
\usepackage{epsfig}
\usepackage{iopams}

\eqnobysec

\begin{document}

\title{The magnetic fields of forming solar-like stars}

\author{S G Gregory$^1$, M Jardine$^2$, C G Gray$^3$ and J-F Donati$^4$}

\address{$^1$ School of Physics, University of Exeter, Stocker Road, Exeter, EX4~4QL, UK}
\address{$^2$ School of Physics and Astronomy, University of St Andrews, North 
Haugh, St Andrews, Fife, KY16~9SS, UK}
\address{$^3$ Department of Physics, University of Guelph, Guelph, Ontario, N1G 2W1, Canada}
\address{$^4$ LATT - CNRS/Universit\'{e} de Toulouse, 14 Av. E. Belin, F-31400 Toulouse, France}
 \ead{scott@astro.ex.ac.uk; mmj@st-andrews.ac.uk; cgg@physics.uoguelph.ca; donati@ast.obs-mip.fr}

\begin{abstract}
Magnetic fields play a crucial role at all stages of the formation of low mass stars and planetary systems.  In the final stages, in particular, they control the kinematics of in-falling
gas from circumstellar discs, and the launching and collimation of spectacular outflows.  The magnetic coupling with the disc is thought to influence
the rotational evolution of the star, while magnetised stellar winds control the braking of more evolved stars and may influence the migration of planets. 
Magnetic reconnection events trigger energetic flares which irradiate circumstellar discs with high energy particles that influence the disc chemistry and set the initial 
conditions for planet formation.  
However, it is only in the past few years that the current generation of optical spectropolarimeters have allowed 
the magnetic fields of forming solar-like stars to be probed in unprecedented detail.  
In order to do justice to the recent extensive observational programs new theoretical models are being developed
that incorporate magnetic fields with an observed degree of complexity.  In this review we draw together disparate results from the classical electromagnetism, 
molecular physics/chemistry, and the geophysics literature, and demonstrate how they can be adapted to construct models of the large scale magnetospheres of stars and planets.  We 
conclude by examining how the incorporation of multipolar magnetic fields into new theoretical models will drive future progress in the field through the 
elucidation of several observational conundrums.
\end{abstract}
\pacs{97.21.$+$a, 97.10.Bt, 97.10.Ex, 97.10.Ld}
\submitto{\RPP}

\maketitle

\section{Introduction}\label{intro}
The current generation of spectropolarimeters, ESPaDOnS at the Canada-France-Hawai'i telescope, and its twin instrument NARVAL 
at T{\'e}lescope Bernard Lyot in the French Pyr{\'e}n{\'e}es, are revolutionising our understanding of stellar magnetism as a function
of stellar age and spectral type.  Results include (but are not limited to) the possible detection of a remnant fossil field on a hot massive star \cite{don06}; 
the first ever magnetic surface maps of pre-main sequence stars in the classical T Tauri phase of their evolution \cite{don07,don08c,don10,hus09};
the discovery of successive global magnetic polarity switches on a star other than the Sun, whose short cycle may be caused by the known presence of an 
orbiting close-in giant planet \cite{don08b}; the rapid increase in field 
complexity at the transition from completely convective low-mass stars to those with radiative cores \cite{mor08,don08a};  and the discovery of globally 
structured magnetic fields on the intermediate mass Herbig Ae-Be stars \cite{cat07}.   
        
Knowledge of the medium and large-scale topology of stellar magnetospheres provided by the spectropolarimetric observations 
is crucial to our understanding of many important processes.  For low-mass pre-main sequence stars the magnetic star-disc interaction is believed to control the 
spin evolution of the central star, and may also be responsible for the collimation and launching of outflows 
\cite{mat05b,mat08b,fer06,kwa07}.
As both low and high mass stars evolve on the main sequence, the angular momentum that can be extracted by winds depends on the amount and 
distribution of open field \cite{hol07,udd09}.  Orbiting close-in giant planets may also interact magnetically with stellar magnetospheres, 
which in principle provides a mechanism for characterising planetary magnetic fields and therefore their internal structure 
\cite{mou07,shk08,jar08a}.       

In order to model such physical processes, new theories and simulations that incorporate magnetic fields with an observed degree of complexity 
are required.  Over the past few years a series of models which move beyond the assumption that stellar magnetic fields
are dipolar have been developed.  In this paper we review such models, and provide a thorough derivation of the magnetic potential in the region exterior to a star, deriving general
expressions for a large-scale multipolar stellar magnetosphere.  
In this review we concentrate on the theoretical study of stellar magnetospheres, briefly discussing observational results where appropriate.  An authoritative review
of the observational study of stellar magnetic fields is provided in \cite{don09}.  We focus on the magnetic fields of forming solar-like stars, although the analytic expressions derived 
herein are applicable to models of stellar and planetary magnetospheres generally.  In the remainder of \S\ref{intro} we 
review the basic techniques that allow stellar magnetic fields to be detected and mapped, before focussing specifically on the magnetic fields of accreting T Tauri stars -
low mass stars still surrounded by planet forming discs.  In \S\ref{dipole_accn} we discuss the observational support for the magnetospheric accretion scenario and 
briefly review previous models with dipolar stellar magnetic fields.  Following this, in \S\ref{derivations}, we draw together results from molecular physics and 
classical electromagnetism to derive self-consistent analytic expressions for multipolar stellar magnetic fields.  In \S\ref{mag_models} we discuss the first models of the 
magnetospheres of forming solar-like stars that take account of non-dipolar magnetic fields.  We conclude in \S\ref{summary} by highlighting several open problems where 
consideration of the true complexity of stellar magnetic fields may be crucial for future progress.  


\subsection{Detecting and mapping stellar magnetic fields}
Stellar magnetic fields can be probed using two complementary techniques.  Measuring the Zeeman broadening of unpolarised spectral lines
has proved to be a successful method of determining average stellar field strengths.  References \cite{rob80a} and \cite{rob80b} demonstrated that by measuring
changes in the shapes of magnetically sensitive lines in intensity spectra, estimates of total field strength, and the fraction
of the stellar surface covered in fields (the magnetic filling factor) could be estimated.  Zeeman broadening measurements were carried out on a 
number of stars (for example \cite{saa88,joh96}), however, for young stars this proved problematic due to rotational broadening 
dominating the line profiles \cite{bas92}.  Broadening measurements are easier to carry out at infrared (IR) wavelengths, as the Zeeman splitting 
increases more rapidly at longer wavelengths compared to Doppler broadening \cite{joh99b}.  The use of IR line profiles to measure
stellar magnetic fields was pioneered by the authors of \cite{saa85} and \cite{saa94}.  Subsequently the analysis of various features in IR spectra 
has proved to be an extremely successful method of detecting stellar magnetic fields (for example \cite{rei06,rei07,joh07}).
Zeeman broadening measurements, however, give no information on the magnetic field topology.  In contrast, measuring the circular polarisation 
signature in spectral lines gives access to the field topology (see \cite{don01} and \cite{don09} for reviews of the basic methodology). However, like all 
polarisation techniques, this suffers from flux cancellation effects and yields limited information regarding the field strength.

If a stellar atmosphere is permeated by a magnetic field, spectral lines forming in that region will be polarised, with the sense of 
polarisation depending on the field polarity.  In practise the polarisation signals detected in photospheric absorption lines
are small, and cross-correlation techniques (such as Least-Squares Deconvolution; \cite{don97c}) are employed
in order to extract information from as many spectral lines as possible.  The signal-to-noise ratio of 
the resulting average Zeeman signature is several tens of times larger than that of a single spectral line \cite{don97c}.   
Magnetic surface features produce distortions in
the Zeeman signature that depend on the latitude and longitude of the magnetic region, and on the orientation of
the field lines.  By monitoring how such distortions move through the Zeeman signature as the star rotates, a method 
referred to as Zeeman-Doppler imaging (ZDI; \cite{sem89}), the 2D distribution of magnetic polarities across the surface of stars can be 
determined using maximum entropy image reconstruction techniques \cite{bro91}.  The field orientation can also be inferred within 
the magnetic regions \cite{don97a}.  After the initial success of ZDI in recovering the first magnetic maps of a star other than
the Sun \cite{don92}, the technique has been applied to construct magnetic surface maps of stars of different spectral type at various evolutionary stages
(for example, \cite{hus01,hus02,don06,don08a,don08b,don08c,mor08}).  In the latest incarnation of ZDI the field topology at the stellar
surface is expressed as a spherical harmonic decomposition \cite{don06}.  The surface field is described as the sum of a poloidal plus a toroidal
component, which allows us to determine how much the field departs from a purely potential (poloidal) state.  For ZDI
of accreting T Tauri stars polarisation signals in both accretion related emission lines and photospheric absorption lines are considered when constructing
magnetic maps \cite{don07,don08c,don10}.  Photospheric absorption lines form across the entire star, while accretion related emission lines
form locally where accreting gas impacts the stellar surface.  Thus magnetic maps constructed from photospheric absorption lines only miss strong field
regions which contain the foot points of the large scale field lines that interact with the disc \cite{don07}.  The resolution of such magnetic 
surface maps is limited by the stellar rotation period and inclination, with a finer spatial resolution at the surface of the star achievable 
for faster rotators, and by the flux cancellation effect of two nearby opposite polarity regions giving rise to oppositely
polarised signals, resulting in a zero net polarisation signal \cite{val04}.  As a result, the smallest scale magnetic features, such as 
stellar equivalents of the small bipolar groups observed on the Sun (for example, \cite{dew08}), remain unresolved.  Instead, spectropolarimetric Stokes V (circular polarisation) 
studies are limited to probing the medium and larger scale fields, and likely miss a large fraction of the total magnetic flux \cite{rei09}.
  None-the-less, significant advances in the study of stellar magnetism have been made
over the past few years using optical spectropolarimeters, in particular in the mapping of the magnetic fields of forming solar-like stars, as we discuss in the following subsection.  

Zeeman-Doppler imaging is not the only method that has been developed to map stellar surface fields.  Magnetic Doppler imaging (MDI), which can trace its origins back to work that pre-dates 
the development of ZDI \cite{pis85,gla85}, is an alternative technique that incorporates polarised radiative transfer, and can also include linear polarisation diagnostics 
(Stokes Q and U) in the field reconstruction \cite{pis02,koc02,koc02b,pis03,koc09}.  However, MDI has thus far only been applied to construct maps of a few chemically peculiar stars \cite{koc04,luf10,koc10}.  Furthermore, as argued by the authors of \cite{don09}, MDI is currently limited to only the brightest and most magnetic stars.  None-the-less, development of MDI 
will continue to be scientifically fruitful in future years, and will provide important comparison tests with the results of ZDI studies.


\subsection{Accreting T Tauri stars and observations of their magnetic fields}
Classical T Tauri stars (CTTS) represent a key transitional period in the life of a star, between the embedded protostellar phase of spherical accretion and the main sequence stage. 
They are low mass pre-main sequence stars which accrete material from dusty circumstellar discs. They possess strong magnetic fields, of order a few kG \cite{joh07}, which truncate 
the disc and force in-falling gas to flow along the field lines. Material rains down on to the stellar surface, where it shocks and produces hot spots that emit in the optical, ultraviolet (UV), and X-ray wavebands.  
CTTS are observed typically to rotate well below break-up speed, and are more slowly rotating (on average) than older pre-main sequence weak-line T Tauri stars whose discs have largely 
dispersed (for example, \cite{bou93}).  CTTS can be in excess of 1000 times more active in X-rays than the Sun is presently. X-rays from the central star may influence the dynamics and chemistry of the circumstellar disc, which will in turn 
set the initial conditions for planet formation \cite{fei02,pas07}.  Understanding the final stage of formation of CTTS, and how they interact with their discs, is crucial 
if we wish to understand the formation of the Sun and our own Solar System.  Many accreting T Tauri stars will eventually evolve along the main sequence surrounded by planetary systems much
like our own.

There is an abundance of observational evidence which supports the basic scenario of magnetically controlled accretion from truncated circumstellar discs.  Excess IR emission is consistent
with CTTS being surrounded by dusty discs, while the shapes of 
spectral energy distributions (SEDs) in the near-infrared (nIR) are consistent with magnetospheric cavities (for example, \cite{rob07}).\footnote{Inner dust disc 
radii derived from interferometric measurements have often been found to be larger than that derived from SED fitting, see \cite{mil07} and references therein.  This 
discrepancy may arise from the crudeness of the disc models used to convert from interferometric visibilities to disc inner radii \cite{pin08}.  For CTTS, however, it is the location of the 
inner gas disc, which extends beyond the dust disc, for example \cite{ake05}, that is important.  Interferometric studies are just beginning to probe gas on such a small scale, for example \cite{eis10} 
(see also the CO transition spectroscopy work in \cite{naj03} and \cite{car07}).}
Inverse P-Cygni profiles are commonly detected in many emission lines, with broad
redshifted absorption components indicating gas infall at approximately free-fall velocities, consistent with columns of gas being magnetically channelled on to the stellar 
surface \cite{edw94,fis08}.  Blue shifted absorption is also commonly detected, indicating that strong outflows are common from CTTS, although whether such outflows originate mainly from the star,
or from the disc surface remains an open question \cite{kwa07}.  Excess continuum emission at optical and in particular UV wavelengths is consistent with the existence of accretion shocks
at the stellar surface, formed due to the high velocity impact of accreting gas.  This excess emission, the veiling continuum, makes absorption lines shallower that they would
appear in non-accreting stars of the same spectral type \cite{har91}.  Estimates of the amount of veiling provides a method of determining the mass accretion rate on to the star \cite{gul98}.  CTTS
are highly variable at all wavelengths, over a variety of timescales.  Such variability is thought to arise due to a complex mixture of hot (accretion related) and cool (magnetic flux
emergence related) spots distributed across the stellar surface, time variable mass accretion and outflows, as well as obscuration effects such as warped inner discs, and columns
of accreting gas rotating across the line-of-sight \cite{bou95}.  Meanwhile copious amounts of X-ray emission, thought to arise due to particle acceleration along field lines following reconnection events, 
indicates that CTTS are extremely magnetically active (see \cite{bou07pp} and \cite{men99} for comprehensive reviews).    

The magnetospheric accretion scenario requires that T Tauri stars possess strong magnetic fields that are sufficiently globally ordered to 
truncate the disc.  Measuring their magnetic fields, however, remains difficult as the stars are faint, and subject to high levels of spectral 
variability.  Initial spectropolarimetric studies at optical wavelengths failed at directly detecting magnetic fields \cite{bro81,joh86,joh87}, 
however, due to the flux cancellation effect whereby signal from regions of opposite polarity cancel, these
initial failures were in fact early indications of the complex nature of T Tauri magnetic fields.  The first field detections were made by 
estimating the increase in line equivalent width that arises due to the saturation of the Zeeman components.  Reference \cite{gue99}, through careful analysis of photospheric Fe I absorption
lines, found the product of magnetic field strength and (magnetic) filling factor of order $\sim$kG on a few accreting T Tauri stars (reference \cite{bas92}
had previously used the same technique to make similar field detections on non-accreting T Tauri stars).  The most successful method 
of measuring T Tauri field strengths has been through the analysis of magnetically sensitive lines at IR wavelengths, as  
Zeeman broadening increases more rapidly with increasing wavelength compared to rotational broadening.  Strong average fields 
of order 1-3$\,{\rm kG}$ have now been detected on T Tauri stars in Taurus, the Orion Nebula Cluster (ONC) and the TW Hydrae Association
\cite{joh99b,yan05,yan08,yan09,val04,joh07}.  Typically magnetically sensitive IR lines of Ti I are used, their shape being best described if the stellar surface contains a distribution of field strengths 
(up to $\sim6-7\,{\rm kG}$). Averaging over the distribution yields photospheric surface fields of $\sim1-3\,{\rm kG}$, with similar values for 
both accreting and non-accreting (i.e. disc-less) T Tauri stars \cite{joh04,val04}.   
Average field strengths of this magnitude are strong enough to disrupt the disc (as discussed in the following section), however, reference \cite{saf98} points out 
that such strong fields are not necessarily sufficiently globally ordered to truncate the disc at several stellar radii.  

Zeeman broadening measurements of lines in intensity spectra have yielded several intriguing results.  The average field strengths of T Tauri stars are a few kG.
The authors of \cite{joh07} and \cite{yan09} argue that  such strong fields may inhibit the coronal X-ray emission.  The quiescent X-ray emission is thought to be due to many small flares triggered by
by reconnection events arising from the release of magnetic stresses built up due to the motion of field line foot points.  Strong fields may inhibit the foot point motions
and the consequent tangling of the field.  This may explain why T Tauri stars appear to be less luminous in X-rays than predicted from a correlation between X-ray luminosity and (unsigned) magnetic flux
\cite{pev03,joh07,yan08}.  Another major finding from Zeeman broadening studies of stars in different star forming regions is the apparent decrease in 
unsigned magnetic flux ($4\pi R_{\ast}^2\bar{B}$) with age \cite{yan09}.  Such a trend remains unexplained.  

 T Tauri stars, the majority of which are completely convective, host magnetic fields that are most likely to be dynamo generated, see for example \cite{sch75,kuk99},
 and \cite{dob06,bro08}, and references therein, for some of the recent work on magnetic field generation in completely convective stars in general.  However, it is occasionally 
 suggested that their magnetic fields may possess a fossil component 
 \cite{tay87,dud95,mos03,joh07}.  Fossil magnetic fields are fields that have survived from the initial collapse of the magnetised cloud core during the earliest stages of the formation of the star.  
The arguments against the existence of fossil fields at typical T Tauri ages ($\sim$ few Myr) have been succinctly summarised by the authors of \cite{don09}.  
Firstly, the onset of convection is thought to rapidly destroy any remnant fossil field on a timescale of not more than 1000 years \cite{cha06}.  Secondly, indicators such as flares 
(commonly observed on T Tauri stars) suggests reordering of their magnetic fields, and thus they cannot be linked to evolutionary processes from millions of years in the past.  
Thirdly, the similarity between the large scale magnetospheric properties of T Tauri stars and those of M-dwarfs (with an age of order Gyr Ð i.e. so old that their fields are certainly not fossil), 
that we discuss at the end of this section, is further evidence that fields are dynamo generated.   Although dynamo magnetic field generation models for partially and fully convective 
stars are still debated, the current generation of spectropolarimeters is providing the community with crucial information on how stellar field topologies vary with quantities such as 
stellar mass and rotation period (see the review article \cite{don09}).  

Zeeman broadening measurements do not yield information about the magnetic geometry of accreting T Tauri stars.  However, small wavelength
shifts in spectral lines observed in right and left circularly polarised light provide another method of diagnosing stellar magnetic fields.  As previously mentioned,
the earliest spectropolarimetric studies failed at detecting T Tauri magnetic fields.  A major break through was the detection of strong circular polarisation
in the He I 5876{\AA } line \cite{joh99a}.  This line of He I has a high excitation potential and is thus believed to form at the accretion shock, where
columns of magnetically channelled gas impact the stellar surface \cite{ber01}.  Polarisation detections in this line are thus tracing the field on the star
where the large scale field lines that interact with the disc are anchored.  The polarisation signal is often found to be rotationally modulated, with the variation
in the derived line-of-sight (or longitudinal) field component with rotation phase well fitted by a simple model with a single accreting spot in the visible hemisphere
\cite{val04}.  This is fully consistent with the findings from ZDI studies, discussed below, where evidence for single dominant accretion 
spots at high latitudes is found.  However, despite arguments that variations in the longitudinal field component, derived from polarisation detections in the 
He I 5876{\AA } line, were attributable to rotational modulation (for example, \cite{val03,val04}), \cite{chu07} refutes such suggestions and argues that the field 
in the line formation region is constantly evolving and restructuring on a 
timescale of only a few hours. The ESPaDOnS/NARVAL spectropolarimetric data presented in \cite{don08c}, however, clearly show 
that although the He I 5876{\AA } line is subject to intrinsic variability, its temporal evolution is dominated by rotational modulation.  
This suggests the magnetic field in the He I line formation region remains stable on timescales of longer than a rotation cycle, and that T Tauri magnetospheres
remain stable, at least over a few rotation periods, consistent with earlier line profile variability studies of individual stars (for example, \cite{gia93,joh95,joh97,bou07aatau}).
Strong polarisation signals in He I 5876{\AA } and 
other accretion related emission lines have now been reported on a number of accreting T Tauri stars \cite{val03,val04,smi04,sym05,yan07,don07,don08c,don10,don10aatau}.  
However, polarisation measurements made using magnetically
sensitive photospheric absorption lines, which presumably form uniformly across the entire stellar surface, yield small longitudinal field strengths, well below
the average fields obtained from Zeeman broadening measurements \cite{smi03,smi04,smi05,dao06}.  Commonly the net polarisation signal is zero \cite{val04}. 
 This suggests that accreting T Tauri stars host complex
surface magnetic fields.  In contrast, the strong (and rotationally modulated) polarisation signal detected in accretion related emission lines suggests 
that the bulk (although perhaps not all) of gas accreting on to stars from their discs, lands on single polarity regions of the stellar surface.  However, it is 
only in the past three years that the geometry of T Tauri magnetic fields have been revealed.  

ZDI studies, combined with tomographic imaging techniques, have now revealed the true complexity of the magnetic fields of accreting T Tauri stars.
At the time of writing surface magnetic maps of six stars have been published, namely V2129~Oph \cite{don07}, BP~Tau \cite{don08c},
CR Cha, CV Cha \cite{hus09}, V2247~Oph \cite{don10} and AA~Tau \cite{don10aatau}.  All have been found to have magnetic fields consisting of many high order 
components.  At 1.35$\,{\rm M}_{\odot}$ V2129~Oph is believed to have already developed a small radiative core, despite its young age (where the stellar mass has 
been derived using the Siess \etal pre-main sequence evolutionary models \cite{sie00}, as with the other stars discussed below).  The magnetic energy
was found to concentrate mainly in a strong octupole component of polar strength 1.2$\,{\rm kG}$ tilted by $\sim20\,^{\circ}$ relative to the stellar rotation axis.
The dipole component was found to be weak, only 0.35$\,{\rm kG}$ at the pole and tilted by $\sim30\,^{\circ}$ relative to the stellar rotation axis, but in a different plane from the 
octupole component.  The surface field in the visible hemisphere was dominated by a 2$\,{\rm kG}$ positive radial field spot at high latitude, with the footpoints 
of the accretion funnel rooted in the same region, but differs significantly from a dipole \cite{don07}.  Evidence for high latitude (polar) cool spots \cite{joh97} and 
for high latitude accretion hot spots \cite{str05} had already been found previously through Doppler imaging of other CTTS. The lower mass and completely convective
star BP~Tau (0.7$\,{\rm M}_{\odot}$) is found to have a much simpler field topology with the magnetic energy shared between strong dipole (of polar strength 1.2$\,{\rm kG}$) and octupole 
(1.6$\,{\rm kG}$ at the pole) field components \cite{don08c}.  Two surface magnetic maps were derived for BP~Tau from circularly polarised spectra taken about 10 months apart, 
but little variation in the large scale field topology was detected.  A similar result was found for AA~Tau, from magnetic maps derived from data taken about one year apart \cite{don10aatau}.
AA~Tau is of similar mass, radius, and rotation rate as BP~Tau, although its magnetic field is even simpler, consisting of strong ($\sim$2-3$\,{\rm kG}$) dipole, almost
anti-parallel with respect to the angular momentum vector of the star, with an octupole component five times weaker.

The initial ZDI results suggest that the field complexity is intimately related to the depth of the convection zone, with completely convective pre-main sequence stars
hosting simpler dominantly poloidal large scale magnetic fields with strong dipole components.  These results are consistent with the ZDI study of the massive accreting
T Tauri stars CR Cha ($1.9\,{\rm M}_{\odot}$) and CV Cha ($2\,{\rm M}_{\odot}$) presented by \cite{hus09}, as both stars have significant radiative cores and have 
particularly complex large scale field topologies.  It is also mirrors the findings from ZDI studies of low-mass main sequence M-dwarfs \cite{mor08,don08a}.    
M-dwarfs which are completely convective (those below 
$\sim0.35\,{\rm M}_{\odot}$ \cite{cha97}) were found to host simple dominantly poloidal large scale magnetic fields, with strong dipole components \cite{mor08}
(the exception being stars below $\sim0.2\,{\rm M}_{\odot}$; only some of which host such simple large scale fields, see below and \cite{mor10}). 
In contrast to the findings for mid M-dwarfs, early M-dwarfs (spectral types M0-M3) with small radiative cores have more complex large scale fields with strong 
toroidal and weak dipole components \cite{don08a}.  Of course, due to the flux cancellation that effects
circular polarisation studies, discussed in the previous section, the dominant and strongest magnetic field components are likely be the highest order multipole
components that constitute the very small scale field close to the stellar surface.  This suggestion is emphasised by the authors of $\cite{rei09}$ who argue that 
the bulk of the total magnetic flux is missed by polarisation studies, with Zeeman broadening measurements indicating the presence of small scale field components 
far stronger than those detected by Stokes V studies alone.  The work of \cite{pha09} is also consistent with the bulk of the magnetic energy being stored in the 
small scale features that remain unresolved in stellar magnetic maps.

Zeeman signals are also suppressed 
within dark (cool) surface spots due to the low surface brightness.  Cool spots, which on T Tauri stars are believed cover a far larger fraction of the stellar
surface when compared to sun spots, for example \cite{don07}, thus represent a potential source of additional missing flux in stellar magnetic maps \cite{joh10}. 
The flux at the stellar surface is the result of several different processes: the dynamo that generated the flux to begin with, the processes that took place during 
the buoyant rise of that flux through the convective zone (and its interaction with the convective cells) and finally the surface effects as the flux emerges into the 
low-density region of the photosphere.  T Tauri stars are known to have average field strengths of a few kG \cite{joh07}, in excess of the mean solar field strength, 
although in sun spots the field can reach $\sim3-4\,{\rm kG}$.
It may be the case that the fields in T Tauri cool spots are similarly large compared to the mean photospheric field strengths. 
It is therefore interesting to ask the questions: i) what is the relative contribution to the total magnetic flux through the surface of T Tauri stars from the dark spots, 
the flux that is resolved in the ZDI maps, and the unresolved flux missing from the ZDI maps due to the flux cancellation effect? and ii) are these contributions in the same ratios
as we see on the Sun?   Untangling the different contributions to the total flux through the surface of T Tauri stars promises to be a challenge for future theories.      

The picture of completely convective T Tauri stars having simple dominantly poloidal large scale fields with strong dipole components may not be valid for the lowest mass
T Tauri stars.  For low mass accreting T Tauri stars (below 0.5$\,{\rm M}_{\odot}$) the picture may
be more complicated.  Recently the authors of \cite{don10} have presented magnetic maps of the low mass T Tauri star V2247~Oph ($0.35\,{\rm M}_{\odot}$), which has 
a faster rotation rate ($P_{rot}=3.5\,{\rm d}$) in 
comparison with the more massive stars BP~Tau ($P_{rot}=7.6\,{\rm d}$) and V2129~Oph ($P_{rot}=6.53\,{\rm d}$). Various accretion related emission lines are detected in the optical spectra,
indicating that mass accretion is ongoing in this system, despite little IR excess evident from Spitzer satellite data (indicating that the dust component of the disc, but not necessarily the gas component, 
has larger dispersed).  The magnetic field of V2247~Oph
is found to be particularly complex with a very weak dipole component compared to that of BP~Tau.  However, this also appears to be consistent with new results for late-type 
M-dwarfs (below $0.2\,{\rm M}_{\odot}$, or spectral types M5-M8), where individual stars are found to host a mixture of complex non-axisymmetric magnetic fields that are very different from the strong 
and simple large scale fields of mid M-dwarfs, and strong axisymmetric dipoles which are more consistent with the large scale topologies of mid M-dwarfs \cite{mor10}.
The V2247~Oph results demonstrate that more spectropolarimetric data for a larger sample of stars are required in order to disentangle the effects of differing stellar masses,
rotation periods, and mass accretion rates.  What is clear, however, is that the magnetic fields of accreting T Tauri stars can be significantly more complex
than a dipole. Before considering how analytic models of multipolar stellar magnetospheres can be constructed, we briefly overview 
the development of T Tauri magnetospheric accretion models with dipole magnetic fields.


\section{Magnetospheric accretion models with dipolar magnetic fields}\label{dipole_accn}
Although it had been suggested by various authors that T Tauri magnetospheres would disrupt circumstellar discs and channel columns of gas on to the 
star (for example, \cite{uch81,uch84,cam90}), it was the inspirational paper of K{\"o}nigl \cite{kon91} that
demonstrated that a multitude of observational features could be explained through the magnetospheric accretion scenario.  By adapting the Ghosh and Lamb
model of accretion on to neutron stars \cite{gho77,gho79a,gho79b}, K{\"o}nigl argued that provided T Tauri stars had magnetic fields of 
order a kG that could effectively couple to the disc, discs could be disrupted at several stellar radii, the alignment of accretion columns with the line-of-sight could 
explain the development of inverse P-Cygni profiles, while shocks at the base of funnels of accreting gas could naturally explain the observed UV excess.    
K{\"o}nigl argued that the observed slow rotation of accreting T Tauri stars could be explained provided that the spin-up torque exerted on the star due to
the accretion of high angular momentum material, and the magnetic connection to regions of the disc rotating more quickly than the star, 
was exactly balanced by a spin-down torque transmitted by the field lines threading the disc exterior to 
corotation.  The corotation radius $R_{co}$ is an important point for models of magnetospheric accretion.  In the stellar equatorial plane,    
\begin{equation}
R_{co} = \left(\frac{GM_{\ast}}{\omega_{\ast}^2}\right)^{1/3},
\end{equation}
which is the radius at which the Keplerian rotation rate of the disc material is equal to that of the star ($\omega_{\ast}=2\pi/P_{rot}$).  Interior (exterior) to this radius, the disc
material is spinning faster (slower) than the star. At radii interior to corotation, material would naturally accrete on to the star, while stellar field lines threading the disc
at corotation would rotate as a solid body and would not be stretched due to differential rotation.    
	
Magnetospheric accretion models, such as those proposed in \cite{kon91,col93,paa96}, provide magnetic links between the star and 
regions of the disc beyond $R_{co}$ which are rotating more slowly than the 
star. By having field lines threading the disc at a range of radii the star is able to accrete material without experiencing a net spin-up torque, which would
act to slowly increase the stellar rotation rate.  However, field lines threading the disc beyond the corotation radius 
may quickly become wrapped up, inflate, and be torn open after only a few rotation periods (see for example \cite{aly90,van94,lov95}).  
The Shu X-wind model, developed through a series of papers, gets around this problem \cite{shu94a,shu94b,shu95,shu96,naj94,ost95}. 
The model introduces the idea of trapped flux, where the closed field lines connecting the 
star and the disc are pinched together in a small interaction region about the corotation radius (called the X-region). In such a way the strong dipolar field of the star rotates 
as a solid body with material from the inner part of the X-region accreting onto the star. Torques in the funnel flow deposit excess angular momentum into the X-region which 
is then removed by a wind that carries material away from the outer portion of the X-region.  References \cite{li96a} and \cite{li96b} consider the funnel flow of gas on to the star, and assuming that
accretion occurs at a steady rate, also find that the matter angular momentum in the funnel flow is transferred to the disc, not to the star, in agreement with the X-wind model.
Thus, the Shu X-wind model allows accretion to occur without spinning up the 
star, whilst also providing a connection between the accretion process and outflows. Models which combine accretion and outflows often predict a correlation between the 
mass accretion and mass outflow rates (for example \cite{ost95,fer97}), which has been observed (for example \cite{ard02}).  However, there is no reason to
expect that discs will always be truncated at the corotation radius.  Based on IR spectroscopy of CO transitions references \cite{naj03} and \cite{car07} conclude that gas in the inner disc 
extends to well within the corotation radius, suggesting that there is nothing special about corotation (in terms of the location of the disc truncation radius).\footnote{The 
author of \cite{cai10}, who reviews recent progress with the X-wind model,
refutes this by pointing out that part of the CO emission may come from the accretion funnel itself.} The process of accretion, considered alone, should thus 
act to spin-up the star, in the absence of an additional angular momentum loss mechanism, or significant magnetic connections to regions of the disc beyond corotation. 

The author of \cite{wan97} argues that it is physically impossible in a steady state scenario for angular momentum to be transported from the funnel flow 
region back to the outer disc, and thus the material torque must be transferred to the star.  Disc-locking models (where the stellar rotation rate matches 
the Keplerian rotation rate at the disc truncation radius) have been criticised as being physically \cite{mat04} unfounded, and is often observationally 
controversial \cite{sta99,sta01} (we note, however, that other observational studies do find good evidence linking the presence of discs, and/or accretion, 
with slow stellar rotation, for example \cite{cie07,fal06}).  The authors of \cite{mat05a} argue that for typical T Tauri accretion rates, the large scale magnetic
field threading the disc is opened to such an extent that the star will receive no spin-down torque at all (see also \cite{zwe06} who consider the effects of time varying stellar 
magnetic fields).  A recent T Tauri spin-evolution model, the first to combine the opening of the large scale magnetosphere due to the interaction with the disc 
with variations in stellar radius and mass accretion rate with time, find that all stars experience a net spin-up torque \cite{mat10}.  For their preferred case of strong 
disc coupling, stars are spun-up and end up with rotation periods of less than 3 days by the end of their simulations at 3 Myr.  However, their model only accounts for 
spin-down torques provided by a small (in terms of radial extent) connection to the disc outside of corotation (with spin-up torques arising from the small connection 
to disc interior to corotation, and from matter accreting on to the star); it does not consider additional spin-down
torques arising from disc winds, or from stellar winds, that appear to be required in order to explain the observed spread in T Tauri rotation periods \cite{mat10}.  
Clearly there remain many unanswered questions regarding the balance of torques in the star-disc system.

From an observational perspective, reference \cite{har02} argues that stellar spin-down 
cannot occur faster than the rate
at which angular momentum can be removed by a disc wind, or through viscous processes.  For stars in the youngest star forming regions, such as the ONC studied by the authors of \cite{sta99},
the angular momentum loss time scale is comparable to the age of the stars, and thus disc braking may not have had sufficient time to slow the stellar rotation rates.  Whether-or-not
stars can be efficiently braked depends crucially on how well the stellar field couples to the disc (for example \cite{arm96,mat05a,bax08,mat10}).   
Furthermore, as argued by the authors of \cite{fer00} (see their section 2.1), stars must undergo strong braking during the initial optically embedded phase of evolution.  They \cite{fer00} argue
that the Shu X-wind cannot explain such strong and efficient braking, and is thus unable to explain the observed slow rotation of accreting T Tauri stars.  
An alternative magnetospheric accretion/outflow model is the reconnection X-wind 
of Ferreira \etal \cite{fer00,fer06} where angular momentum is removed by a wind launched from the entire surface of the disc.  A unique feature of this model is that the outflow
is powered by the rotational energy of the star itself, and thus the reconnection X-wind provides a torque that can efficiently brake the star (see the review in \cite{fer08}).  Another alternative 
is the accretion powered stellar wind model developed by Matt and Pudritz \cite{mat05b,mat08a,mat08b}, which assumes that the spin-up torque due to accretion is balanced by the spin-down torque from the 
stellar wind.  However, it is not yet clear how accretion can power a stellar outflow (see the discussion in \cite{mat07}).  A possible suggestion is that gas accreting on to the stellar surface in accretion columns
excites magnetohydrodynamic (MHD) surface waves which drive the stellar outflow \cite{cra08}.  Unfortunately, the derived mass lose rates 
are an order of magnitude below what is required to extract enough angular momentum 
in the wind to explain the observed slow rotation of accreting T Tauri stars.  Ultimately the angular momentum removal mechanism may be some combination of disc, and accretion powered stellar,
winds, for example \cite{cra09}, although this is an open question.
  
The star-disc interaction, and the process of accretion and outflows, is likely to be highly time dependent.
In order to incorporate time dependent effects, MHD simulations are required.  To date, a myriad of MHD models have been constructed of both funnel 
flows and the star-disc interaction, which vary in their 
assumptions regarding the disc physics \cite{hay96,goo97,mil97,kol02a,rom02,rom03,rom04,kuk03,kuk04,von04,von06a,lon05,von06b,yel06,bes08,zan09}.  Some models predict 
episodic accretion, periodic inflation, opening, and reconnection of the magnetosphere, plasmoid ejection, the launching of winds from the disc, 
field line collimation into jet-like structures,
and variable epochs of stellar spin-up and spin-down.  The next generation of spectropolarimeters, which will be able to detect magnetic fields in the inner disc, will provide
data to discriminate between the various MHD simulations.  A common feature of such models is the assumption that the star possesses a simple dipolar magnetic field.  The 
recent ZDI studies discussed in the previous section have now demonstrated conclusively that T Tauri magnetic fields are multipolar, with complex surface field regions distorting the structure 
of the large scale accreting field in the regions close to the star \cite{gre08}.     

The authors of \cite{joh02} took the first steps towards investigating accretion models with non-dipolar magnetic fields. They 
demonstrated that if the dipole field assumption is removed from the Shu X-wind model, and under the assumption that the field strength does not vary from star to star, 
there should be a correlation between the stellar and accretion parameters of the form $R_{\ast}^2f_{acc}\propto (M_{\ast}\dot{M}P_{rot})^{1/2}$ (where $\dot{M}$ is the mass accretion 
rate and $f_{acc}$ the accretion filling factor).  Such a correlation
agrees reasonably well with the observational data.  Over the past few years the first models of magnetospheric accretion that consider 
multipolar T Tauri magnetic fields have been developed  \cite{gre05,gre06a,gre07,gre08,lon07,lon08,lon10,moh08,rom10}.  
Before reviewing such models in \S\ref{mag_models}, we explore results developed for multipole field expansions in classical electromagnetism and molecular physics in 
order to demonstrate how simple analytic models of multipolar stellar magnetic fields can be constructed.  


\section{Multipole magnetic fields}\label{derivations}
Multipole expansions for the potential of a finite static charge distribution in electrostatics, and 
of a continuous current distribution in magnetostatics, are commonly encountered
in the electromagnetism literature (for example \cite{buc59,gra84,raa05}).  The practical applications of such expansions, however, have been 
most exploited by molecular physicists and chemists, where the multipole moments of molecules with various symmetries
have been measured for decades (for example \cite{sto66,gra84}), and by geophysicists, with models of the magnetic fields of the planets within the Solar System 
readily found in the literature (for example \cite{wil82}).    

In this section we derive the magnetic field components of an arbitrary multipole of order $l$ in spherical tensor form directly from the magnetostatic potential
(where $l=1,2,3,4,5,\ldots$ represent the dipole, the quadrupole, the octupole, the hexadecapole, the dotriacontapole, and so on).  We show how the 
field components can be expressed in terms of either the polar, or the equatorial, field strength of the particular multipole being considered.
We extend the analysis to demonstrate how the field components can be modified via the inclusion of an additional boundary condition designed 
to mimic the effects of plasma opening field lines to form a stellar wind, and show how the field can be written in co-ordinate free form.
For stellar or planetary applications it is natural to employ spherical tensors, and later choose spherical polar components of the magnetic field vector $\mathbf{B}$.  
However, in light of recently published work by the authors of \cite{lon07} and \cite{lon08}, who follow 
\cite{lan75} by adopting a Cartesian tensor approach for the multipole expansion, we conclude this section by demonstrating how an alternative approach to the multipole expansion 
can lead to differing expressions for the large scale magnetic field components.

Throughout this work we assume a standard spherical polar coordinate system $(r,\theta,\phi)$ with the coordinate origin ($r=0$) taken 
to be the centre of the star (or equivalently a planet), $0\le\theta\le\pi$ and $0\le\phi\le2\pi$, with $\theta=0$ corresponding to the stellar rotation pole.
When Cartesian coordinates are considered, the stellar rotation axis is assumed to be aligned with the $z$-axis and $\phi=0$ the $x$-axis. 
 
 
\subsection{Magnetostatic expansion}\label{magn_static}
\begin{figure}[t]
   \centering
   \includegraphics[width=65mm]{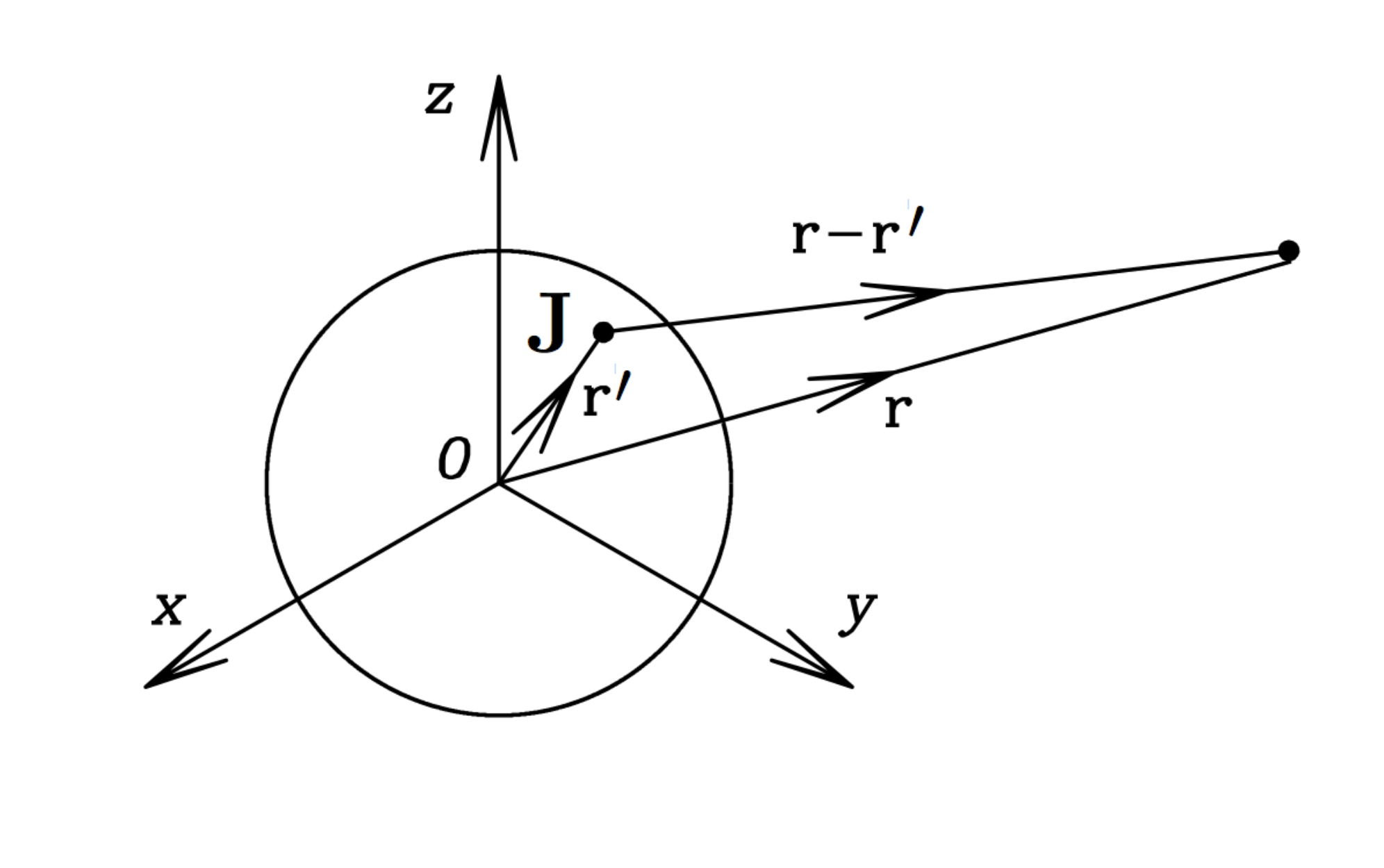} 
   \caption{A current source $\mathbf{J}$ internal to a star which also contains the origin of the coordinate system $0$.  
                $\mathbf{r}$ denotes a distant field point external to the star, where the potential due to the current source at 
                source point $\mathbf{r'}$ is to be calculated.}
   \label{fieldsource}
\end{figure}

We are interested in deriving expressions for the multipolar field components that can be used to describe the large scale magnetospheres of stars 
(or equivalently planets).  Their external magnetic fields are generated due to the dynamo action and the distribution of current sources internal to the star/planet. 
Expansions of the magnetostatic scalar potential, which we denote $\Psi$, for a finite continuous current distribution
can be carried out in several ways - by introducing Debye potentials \cite{gra78a}, using spinors \cite{tor02}, or most elegantly, directly 
from Maxwell's equations for a magnetostatic field \cite{gra78b} (see also \cite{bro71}).  In cgs units the magnetostatic Maxwell equations are,
\begin{eqnarray}
\nabla \cdot \mathbf{B} &=& 0 \label{divmax}\\
\nabla \times \mathbf{B} &=& \frac{4\pi}{c}\mathbf{J},\label{maxwell}
\end{eqnarray}
where $\mathbf{J}$ is the source current density, assumed localised near the coordinate origin in figure \ref{fieldsource}.  By taking 
the curl of both sides of (\ref{maxwell}) and using a vector identity for $\nabla \times (\nabla \times \mathbf{B})$ and
(\ref{divmax}) it is straightforward to show that
\begin{equation}
\nabla^2 \mathbf{B} = -\frac{4\pi}{c}\nabla \times \mathbf{J}.
\label{current}
\end{equation}
In a region which is source-free (for example in the region external to the star/planet) the complete form of the field $\mathbf{B}$
can be determined purely from its radial component $\mathbf{r}\cdot\mathbf{B}$ (see appendix B of \cite{gra78a}, 
and \cite{gra78c}, for general proofs).  By operating on both sides of (\ref{current}) 
with $\mathbf{r}\cdot$ and using another vector identity and (\ref{divmax}) it can be shown that,
\begin{equation}
\nabla^2(\mathbf{r}\cdot\mathbf{B}) = - \frac{4\pi}{c}\mathbf{r}\cdot\nabla \times \mathbf{J}.
\end{equation}     
This is Poisson's equation, which has a solution in terms of the static Green's function for the Laplacian, $|\mathbf{r}-\mathbf{r'}|^{-1}$,
\begin{equation}
\mathbf{r}\cdot\mathbf{B} = \frac{1}{c}\int \frac{\mathbf{r'}\cdot \nabla ' \times \mathbf{J'}}{|\mathbf{r}-\mathbf{r'}|} d\mathbf{r'},
\label{rdotB}
\end{equation}
where $\mathbf{r}$ denotes a field point external to the star, $\mathbf{r'}$ denotes a source point internal to the star, while 
$\mathbf{J'}=\mathbf{J}(\mathbf{r'})$ and the source $\mathbf{J}$ is also assumed to be to internal to the star, as illustrated in 
figure \ref{fieldsource}.

By applying the cosine rule to the triangle in figure \ref{fieldsource} the $|\mathbf{r}-\mathbf{r'}|^{-1}$ term in (\ref{rdotB}) can be
re-written as
\begin{equation}
|\mathbf{r}-\mathbf{r'}|^{-1} = \frac{1}{r}\left[1+\left(\frac{r'}{r}\right)^2-2\left(\frac{r'}{r}\right)\cos{\bar{\theta}}\right]^{-1/2}
\label{sphere3}
\end{equation}
where $\bar{\theta}$ is the angle between $\mathbf{r}$ and $\mathbf{r'}$.  The term in the square brackets is the 
generating function for Legendre polynomials, which allows (\ref{sphere3}) to be be re-written as
\begin{equation}
|\mathbf{r}-\mathbf{r'}|^{-1}=\sum_l \frac{r'^l}{r^{l+1}}P_l(\cos{\bar{\theta}}).
\end{equation}
Using the addition theorem for spherical harmonics, which expresses the Legendre polynomials
$P_{l}(\cos{\bar{\theta}})$ as the sum of the product of the spherical harmonics $Y_{lm}(\theta',\phi')$ and $Y_{lm}^{\ast}(\theta,\phi)$
over the range $m=-l,\ldots,l$, equation (\ref{rdotB}) can be rewritten as,
\begin{equation}
\mathbf{r}\cdot\mathbf{B} = \frac{1}{c}\int d\mathbf{r'} \mathbf{r'}\cdot\nabla ' \times \mathbf{J'} \sum_l \sum_m \left( \frac{4\pi}{2l+1}\right)\frac{r'^l}{r^{l+1}}
                                         Y_{lm}(\theta ', \phi ')Y^{\ast}_{lm}(\theta, \phi).
\label{rdotB2}
\end{equation} 
The spherical harmonics are given by,
\begin{equation}
Y_{lm}(\theta,\phi) = (-1)^m \left(\frac{2l+1}{4\pi}\right)^{1/2}\left(\frac{(l-m)!}{(l+m)!}\right)^{1/2}P_{lm}(\cos{\theta}){\rm e}^{{\rm i}m\phi},
\label{sph_ham}
\end{equation}
for $m \ge 0$ [while for $m<0$, $Y_{l(-m)}(\theta,\phi)=(-1)^m Y_{lm}^{\ast}(\theta,\phi)$].  The definition of the spherical harmonics 
differs between research areas via the inclusion or omission of 
the first two bracketed terms, or parts thereof.  The definition we use here includes the Condon-Shortley phase [the $(-1)^m$ term].  
Other definitions lead to differing expressions for the associated Legendre
functions $P_{lm}(\cos{\theta})$, defined below by (\ref{plm}).  Defining the non-primitive spherical magnetostatic multipole moments as\footnote{The 
difference between primitive and non-primitive
multipole moments is discussed in \ref{electro_cart}.},
\begin{equation}
M_{lm} = \frac{1}{c(l+1)} \int d\mathbf{r'} r'^l Y_{lm}(\theta ', \phi ')\mathbf{r'}\cdot \nabla ' \times \mathbf{J'},
\label{mag_multi}
\end{equation}
allows (\ref{rdotB2}) to be written as,
\begin{equation} 
\mathbf{r}\cdot\mathbf{B}=\sum_l\sum_m (l+1)\left(\frac{4\pi}{2l+1}\right)M_{lm}Y_{lm}^{\ast}(\theta,\phi)/r^{l+1},
\label{rdotB3}
\end{equation}
where the reason for the inclusion of the additional $(l+1)$ factor in (\ref{rdotB3}) will become obvious during the integration of the 
separable differential equation (\ref{pde}) below.  
We note that he non-primitive multipole moments, defined by (\ref{mag_multi}) can be re-written in several equivalent ways using various vector 
identities \cite{gra78a}.  Alternatively, the $|\mathbf{r}-\mathbf{r'}|^{-1}$ term in (\ref{rdotB}) can be expanded in a Taylor series 
and then written in traceless form, in analogy with the electrostatic Cartesian tensor derivation in \ref{electro_cart}.  Interested readers can find 
details of this in \cite{gra79} and \cite{gra80}.   

External to the star, in the source free region, (\ref{maxwell}) further reduces to $\nabla \times \mathbf{B} = \mathbf{0}$.  This condition can be
satisfied by writing the field $\mathbf{B}$ in terms of a magnetostatic scalar potential $\Psi$,
\begin{equation}
\mathbf{B} = -\nabla \Psi.
\label{BPsi}
\end{equation}  
In order to determine the field components $B_r$ and $B_\theta$, required to describe the large scale structure of a stellar (or equivalently a planetary)
magnetosphere, we first need to derive an expression for $\Psi$.  Operating on both sides of (\ref{BPsi}) with $\mathbf{r}\cdot$ gives,
\begin{equation}
\mathbf{r}\cdot\mathbf{B} = -r \frac{\partial \Psi}{\partial r}.
\label{Bradial}
\end{equation}
By equating (\ref{Bradial}) and (\ref{rdotB3}) a separable partial differential
equation is created that can be solved for $\Psi$,
\begin{equation}
-r\frac{\partial \Psi}{\partial r} = \sum_l \sum_m (l+1) \left( \frac{4\pi}{2l+1}\right)M_{lm}Y^{\ast}_{lm}(\theta,\phi)/r^{l+1}.
\label{pde}
\end{equation}
With the assumption that the potential $\Psi \rightarrow 0$ as $r \rightarrow \infty$, (\ref{pde}) can be integrated to give,
\begin{equation}
\Psi(\mathbf{r}) = \sum_l \sum_m \left(\frac{4\pi}{2l+1}\right)M_{lm}Y_{lm}^{\ast}(\theta,\phi)/r^{l+1},
\label{33}
\end{equation}
where the integration constant is zero (see \cite{gra79} for further discussion on this subtle point).  Equation (\ref{33}) gives the 
general form of the multipole expansion of the magnetostatic potential in spherical tensor form. Note that the correct number of components 
(i.e., $2l+ 1$) of the non-primitive multipole moment of order $l$, $M_{lm}$ with $m = -l,\ldots,+l$, occurs automatically using spherical tensors (compare this with the 
Cartesian tensor method briefly discussed in \S\ref{electro_cart}). The corresponding expression for the spherical tensor form of the non-primitive electric multipole moment 
$Q_{lm}$ is obtained from $M_{lm}$ in (\ref{mag_multi}) by replacing $\mathbf{r'}\cdot \nabla ' \times \mathbf{J'}/[c(l+1)]$ with the charge density $\rho(\mathbf{r'})$.

In this paper we are interested in deriving expressions for the axial multipole field components, which correspond to the $m=0$ terms
of (\ref{33}) when we choose space-fixed axes with $z$ along the multipole moment symmetry axis.  The
potential then cannot depend on the azimuthal angle $\phi$, so that only terms with $m=0$ can contribute to (\ref{33}).
By substituting for the spherical harmonics using 
(\ref{sph_ham}) the potential becomes,
\begin{equation}
\Psi(\mathbf{r}) = \sum_l \left( \frac{4\pi}{2l+1}\right)^{1/2}M_{l0}P_l(\cos{\theta})/r^{l+1}.
\end{equation}
The $M_{l0}$ terms are determined directly from (\ref{mag_multi}),
\begin{eqnarray}
M_{l0} &=& \frac{1}{c(l+1)} \int d\mathbf{r'}r'^l \left(\frac{2l+1}{4\pi}\right)^{1/2}P_l(\cos{\theta '})\mathbf{r'}\cdot \nabla ' \times \mathbf{J'} \nonumber \\
           &=& \left( \frac{2l+1}{4\pi}\right)^{1/2}M_l,
\end{eqnarray}  
where the quantities $M_l$ are defined as \emph{the} magnetic multipole moments, with $M_1 \equiv \mu$ the dipole moment, $M_2 \equiv Q$ the quadrupole moment, 
$M_3 \equiv \Omega$ the octupole moment, and so on.  The $l$th component of the scalar potential of the large scale stellar magnetosphere is therefore given by,
\begin{equation}
\Psi_l = \frac{M_l}{r^{l+1}}P_l(\cos{\theta}).
\label{Psi_mag}
\end{equation}

We note that in general $M_{lm}$ is a complicated function of the orientation of the current source distribution, but for axial distributions it is a simple function of the 
orientation ($\theta$, $\phi$) of the symmetry axis \cite{gra76}, i.e.,
\begin{equation}
M_{lm} = M_l Y_{lm}(\theta, \phi).
\label{newMlm}
\end{equation}
A corresponding expression can be derived for the Cartesian form of the non-primitive multipole moment \cite{gra84,gra10}, and   
analogous spherical and Cartesian tensor formulae exist for the non-primitive electric multipole moments \cite{gra84,gra09}.


\subsection{The magnetic field components}\label{field} 
\begin{figure*}
        \def\subfigtopskip{4pt}
        \def\subfigbottomskip{4pt}
        \def\subfigcapskip{2pt}
        \centering
        \begin{tabular}{cc}
                \subfigure{
                        \includegraphics[width=55mm]{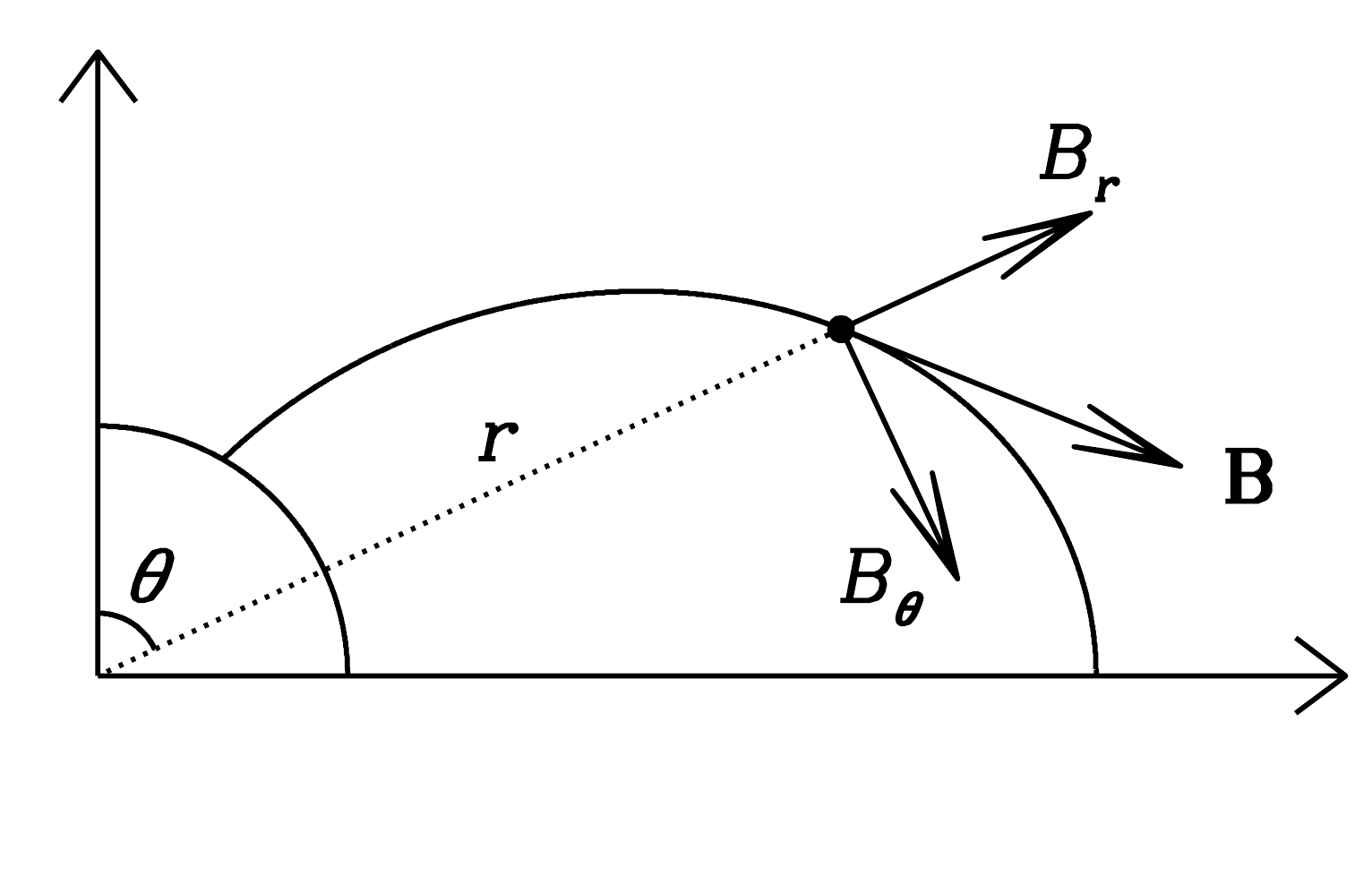}  \hspace{-3mm}   
                        } &
                \subfigure{
                        \includegraphics[width=55mm]{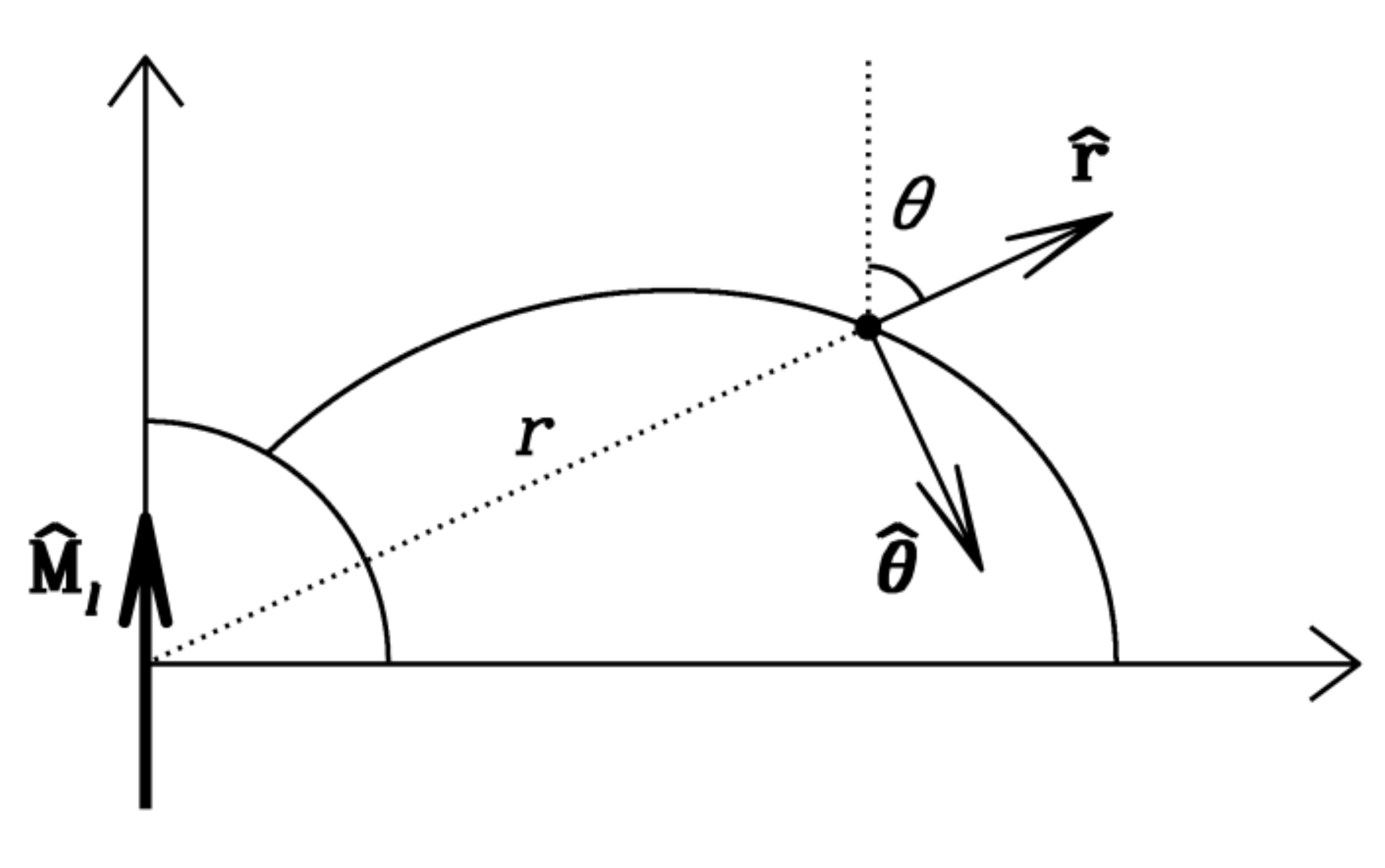}  \hspace{-3mm}  
                        } \\ 
        \end{tabular}
        \caption[]{The left hand panel shows a field vector $\mathbf{B}$ decomposed into the radial $B_r$ and polar $B_{\theta}$ components at a point along a 
           field line a distance $r$ from the centre of the star, at a co-latitude of $\theta$.  The field components are 
           used to illustrate their definitions and are not to scale.  The right hand panel shows the unit vectors in the radial and polar directions and $\mathbf{\hat{M}}_l$ is
           a unit vector along the symmetry axis of axial multipole $l$.}
        \label{field_fig}
\end{figure*}

As the stellar magnetic field is described by the gradient of the magnetostatic scalar potential, 
$\mathbf{B}=-\nabla \Psi$, the field components in spherical coordinates of a multipole of order $l$ are obtained via,
\begin{equation}
B_r = -\frac{\partial \Psi_l}{\partial r} \hspace{5mm} B_\theta = -\frac{1}{r}\frac{\partial \Psi_l}{\partial \theta} \label{brdef}
\end{equation}
while for the axial multipoles that we consider in this paper $B_\phi = 0$.  Figure \ref{field_fig} illustrates a field
vector $\mathbf{B}$ at a point along a field line, decomposed into the $B_r$ and $B_{\theta}$ components.  
Noting that the associated Legendre functions can be written as
\begin{equation}
P_{lm}(x) = (1-x^2)^{m/2}\frac{\rmd^m}{\rmd x^m}P_l(x)
\label{plm}
\end{equation}
where $x \equiv \cos{\theta}$ and $P_l(x)$ are the Legendre polynomials\footnote{We remind readers that the Condon-Shortley phase,
a term of the form $(-1)^m$, has already been included in our definition of the spherical harmonics, and
is therefore not included as a pre-factor in the associated Legendre functions.  Users of IDL should note that the
legendre package of IDL does include the $(-1)^m$ term.  It is also worth noting that the geophysics community uses Schmidt 
quasi-normalised associated Legendre functions, by convention, which leads to slightly different expressions for $B_r$ and $B_{\theta}$, see \cite{win05}.  
No such convention exists in the astrophysics community.}, 
\begin{equation}
P_l(x) = \frac{1}{2^ll!}\frac{\rmd^l}{\rmd x^l}(x^2-1)^l,
\label{pl}
\end{equation}
the magnetic field components of an axial multipole of order $l$ can be obtained from (\ref{brdef}) using (\ref{Psi_mag}) and (\ref{plm}),
\begin{equation}
B_r =  \frac{(l+1)}{r^{l+2}}M_l P_l(\cos{\theta}) \hspace{5mm}
B_\theta = \frac{M_l}{r^{l+2}}P_{l1}(\cos{\theta}), \label{B_general1}
\end{equation}
where we have made use of the fact that $P_{l0}(\cos{\theta})=P_l(\cos{\theta})$ and $M_l$ is
the multipole moment.

The field components for the lower order multipoles can then be obtained. 
For the dipole magnetic field,
\begin{equation} 
B_r = \frac{2\mu}{r^3} \cos{\theta} \hspace{5mm} B_\theta = \frac{\mu}{r^3} \sin{\theta}.
\end{equation}
The axial quadrupole magnetic field components are given by,
\begin{equation} 
B_r = \frac{3Q}{2r^4}(3\cos^2{\theta}-1) \hspace{5mm} B_\theta = \frac{3Q}{r^4}\cos{\theta}\sin{\theta},
\end{equation}
and the axial octupole field components are,
\begin{equation} 
B_r = \frac{2\Omega}{r^5}(5\cos^2{\theta}-3)\cos{\theta} \hspace{5mm}
B_\theta = \frac{3\Omega}{2r^5}(5\cos^2{\theta}-1)\sin{\theta}.
\end{equation}   
Expressions for higher order multipoles can be derived from (\ref{B_general1}).

For some applications, including lunar magnetism, the multipolar $\mathbf{B}$ fields within the source region are required,
which can be conveniently represented in terms of contact fields involving the Dirac delta function $\delta(\mathbf{r})$ \cite{gra10}.
Because $\nabla \times \mathbf{B} \ne 0$ in the source region we cannot use the usual magnetic scalar potential $\Psi$ in this region.
Instead $\mathbf{B}$ can be represented by the two scalar Debye potentials $\psi$ and $\chi$ \cite{gra10},
$\mathbf{B} = \mathbf{L}\psi + \nabla \times \mathbf{L}\chi$, where $\mathbf{L} = -i \mathbf{r} \times \nabla$ is the angular momentum operator.  
The two terms are the toroidal and poloidal components, respectively, of $\mathbf{B}$.  The coefficients of contact 
field terms for $\psi$, $\chi$ and $\mathbf{B}$ involve the primitive magnetic multipole moments.  An application of contact multipolar fields 
to lunar magnetism is described in reference \cite{gra09}.  The corresponding electrostatic multipolar contact
fields \cite{gra09}, which {\it can} be derived using the electric scalar potential since $\nabla\times\mathbf{E}=\mathbf{0}$ in the source regions, 
are mentioned in \S\ref{electro_cart}.
 
 
\subsection{General expressions for magnetospheres}\label{fields2}

\subsubsection{Polar field strength}
Rather than specifying the strength of the various multipole moments, it is more 
convenient to discuss the polar strength of each component, $B^{l,pole}_{\ast}$, i.e. the strength of the particular
multipole component at the stellar rotation pole.  At the rotation pole, $r=R_{\ast}$ and $\theta = 0$, $P_l(\cos{\theta}) = 1$ 
for all values of $l$, and the field of all the axial multipoles is purely radial.  
From (\ref{B_general1}) any multipole moment can be written as 
\begin{equation}
M_l = R_{\ast}^{l+2}B^{l,pole}_{\ast}/(l+1)
\label{moment}
\end{equation}
where $B^{l,pole}_{\ast}$ is the polar field strength of the $l$th order multipole.
The dipole, quadrupole and octupole moments can be written as
$\mu = B^{1,pole}_{\ast}R_{\ast}^3/2$, $Q=B^{2,pole}_{\ast}R_{\ast}^4/3$ and 
$\Omega = B^{3,pole}_{\ast}R_{\ast}^5/4$ respectively, which allows the field components to be re-expressed
in a more convenient form,
\begin{eqnarray}
B_{r,dip} &=& B^{1,pole}_{\ast}\left(\frac{R_{\ast}}{r}\right)^3\cos{\theta}, \label{dipBr} \\
B_{\theta,dip} &=& \frac{1}{2}B^{1,pole}_{\ast}\left(\frac{R_{\ast}}{r}\right)^3\sin{\theta} \label{dipBt}\\
B_{r,quad} &=& \frac{1}{2}B^{2,pole}_{\ast}\left(\frac{R_{\ast}}{r}\right)^4(3\cos^2{\theta}-1), \label{quadBr}\\
B_{\theta,quad} &=& B^{2,pole}_{\ast}\left(\frac{R_{\ast}}{r}\right)^4\cos{\theta}\sin{\theta},\label{quadBt}\\
B_{r,oct} &=& \frac{1}{2}B^{3,pole}_{\ast}\left(\frac{R_{\ast}}{r}\right)^5(5\cos^2{\theta}-3)\cos{\theta},\label{octBr} \\
B_{\theta,oct} &=& \frac{3}{8}B^{3,pole}_{\ast}\left(\frac{R_{\ast}}{r}\right)^5(5\cos^2{\theta}-1)\sin{\theta} \label{octBt},
\end{eqnarray}
For an $l$th order multipole the general field components can be derived from (\ref{B_general1}) using (\ref{moment}),
\begin{equation}
B_{r} = B^{l,pole}_{\ast} \left(\frac{R_{\ast}}{r}\right)^{l+2}P_l(\cos{\theta}) \hspace{5mm}
B_{\theta} = \frac{B^{l,pole}_{\ast}}{l+1}\left(\frac{R_{\ast}}{r}\right)^{l+2}P_{l1}(\cos{\theta}). \label{mag_gen1}
\end{equation}
These simple expressions for the magnetic field components of axial stellar (or planetary) magnetospheres
are straightforward to adapt as inputs to numerical simulations.  They can be used to derive the components
of an individual $l$th order multipole, while linear combinations of the various multipoles may be used to
construct expressions for more complex multipolar fields.  The use of (\ref{mag_gen1}) requires knowledge 
of $P_{l}(\cos{\theta})$ and $P_{l1}(\cos{\theta})$ as well as the polar field strength of each multipole component being
considered.  The Legendre polynomials and associated Legendre functions can be looked up in tables, or derived
through a combination of (\ref{plm}) and (\ref{pl}) and use of Bonnet's recursion formula.  For models of stellar 
magnetospheres the polar field strength of each multipole component is determined observationally by decomposing the 
field into poloidal and toroidal components, each of which is then expressed as a spherical harmonic expansion.  The 
coefficients of such a fit to the data contain information on the strength of the individual field components \cite{don06}.  


\subsubsection{Equatorial field strength}
It is also possible to derive expressions for $B_r$ and $B_{\theta}$ which include the stellar
equatorial, rather than the polar, field strength.  Various authors define the strength $B_{\ast}$ 
of a low order multipole as being the field strength at the stellar equator (for example \cite{arm96,cla95} for
a dipole and \cite{mat08b} for a quadrupole) while others follow the convention used in the previous subsection, and 
define the strength as the polar value (for example \cite{mci06,gre08}).  
With $\theta = \pi/2$ and $r=R_{\ast}$ it can be seen from (\ref{dipBr}) and (\ref{dipBt}) that the 
equatorial field strength of a dipole is $1/2$ of the polar value.  For a quadrupole, it can be shown from (\ref{quadBr}) and 
(\ref{quadBt}) that $B^{2,equ}_{\ast}= B^{2,pole}_{\ast}/2$.\footnote{Note that for the quadrupole, in the equatorial plane,
$B_{\theta}=0$ and $B_r = -B^{2,pole}_{\ast}/2$.  This then gives an equatorial field strength of 
$B^{2,equ}_{\ast}=(B_r^2 + B_{\theta}^2)^{1/2} = B^{2,pole}_{\ast}/2$.}  However, this is not a general result for
higher order multipoles.  For example, using the expressions for the octupole field components, (\ref{octBr}) and (\ref{octBt}),
we find that $B^{3,equ}_{\ast}=3B^{3,pole}_{\ast}/8$.  In \ref{equ_appendix} we derive a general relationship between
$B^{l,equ}_{\ast}$ and $B^{l,pole}_{\ast}$ for a multipole of arbitrary $l$. 

At the stellar rotation pole the field is purely radial, and we therefore end up with single expressions for $B_r$ and $B_{\theta}$ and the particular
multipole moment $M_l$ [see (\ref{moment}) and (\ref{mag_gen1})] that are valid for all multipoles when expressed in terms of $B^{l,pole}_{\ast}$
irrespective of the $l$ value.  This is not the case if the expressions for $B_r$, $B_{\theta}$ and $M_l$ are re-expressed
in terms of the equatorial field strength $B^{l,equ}_{\ast}$.  For odd $l$ number axial multipoles the field in the stellar equatorial plane
has only a polar ($\theta$) component ($B_r=0, B_{\theta} \ne 0$), while for even $l$ number axial multipoles the field only
has a radial component in the equatorial plane ($B_r \ne 0, B_{\theta}=0$).  Therefore the expressions for $B_r$, $B_{\theta}$
and $M_l$, when written in terms of the equatorial field strength, are different for odd and even $l$ number multipoles.     
For odd $l$ number multipoles, such as the dipole, the octupole and the dotriacontapole, our expressions (\ref{mag_gen1})
can be re-written in terms of the equatorial field strength $B_{l,equ}^{\ast}$ rather than the polar field strength $B^{l,pole}_{\ast}$ using the results developed
in \ref{equ_appendix},
\begin{equation}
B_r = \frac{2^l[(l-1)/2]![(l+1)/2]!}{l!}B^{l,equ}_{\ast}\left(\frac{R_{\ast}}{r}\right)^{l+2}P_l(\cos{\theta})
\end{equation}
\begin{equation}
B_{\theta} = \frac{2^l[(l-1)/2]![(l+1)/2]!}{(l+1)!}B^{l,equ}_{\ast}\left(\frac{R_{\ast}}{r}\right)^{l+2}P_{l1}(\cos{\theta}),
\end{equation}
while for even $l$ number multipoles, such as the quadrupole and the hexadecapole, the corresponding expressions are,
\begin{equation}
B_r = \frac{2^l[(l/2)!]^2}{l!}B^{l,equ}_{\ast}\left(\frac{R_{\ast}}{r}\right)^{l+2}P_l(\cos{\theta})
\end{equation}
\begin{equation}
B_{\theta} = \frac{2^l[(l/2)!]^2}{(l+1)!}B^{l,equ}_{\ast}\left(\frac{R_{\ast}}{r}\right)^{l+2}P_{l1}(\cos{\theta}).
\end{equation}
The constant terms can be modified
by introducing double factorials, however, this does not lead to a significant simplification.
Clearly the expressions for $B_r$ and $B_{\theta}$ when written in terms of the equatorial field strength, lack the 
simplicity and elegance of the corresponding expressions based on the polar field strength (\ref{mag_gen1}).


\subsubsection{Incorporating open field by using a source surface boundary condition}
References \cite{alt69} and \cite{sch69} introduced the source surface boundary
condition in order to produce global potential (i.e. current free) field extrapolations of the Sun's coronal magnetic field from maps of the photospheric field.  
This outer boundary condition of the potential field source surface (PFSS) model mimics the effect of the solar wind dragging and distorting the field lines of the solar corona,
and gives a simple way to incorporate open field into global magnetospheric models.  In reality the distortion 
of the field by the coronal plasma will induce a current system, and therefore a proper solution for the field structure requires
a solution to the equations of MHD.  None-the-less PFSS models have been used extensively in the study
of solar magnetism.  At some height above the solar surface, the source surface $R_S$, the plasma pressure in the corona pulls open the field lines forming 
a wind.  Above the source surface the field is purely radial and is often described by a Parker spiral \cite{par58}.   The source surface
represents the radius at which all of the field becomes radial, but there is no reason why it may not do so closer to the surface.  
For the Sun $R_S$ is typically taken to be $2.5\,{\rm R}_{\odot}$, a value consistent with satellite observations of the interplanetary magnetic field 
(see the discussion in \cite{ril06}).  
This same boundary condition (with differing values of $R_S$) has since been applied to stellar field extrapolations for young rapid rotators \cite{jar02a}
and pre-main sequence stars \cite{gre06a}, providing a simple method of incorporating open field
regions along which a wind could be launched.  In this section we show how our general expressions for $B_r$ 
and $B_{\theta}$, (\ref{mag_gen1}), are modified for a stellar magnetosphere 
with a source surface (note that PFSS models have also been applied to planetary magnetospheres, for example, \cite{sch96}).  
In \S\ref{mag_models} we compare the results of PFSS extrapolation models with more complex MHD simulations.

\begin{figure*}
        \def\subfigtopskip{4pt}
        \def\subfigbottomskip{4pt}
        \def\subfigcapskip{2pt}
        \centering
        \begin{tabular}{ccc}
                \subfigure{
                        \includegraphics[width=40mm]{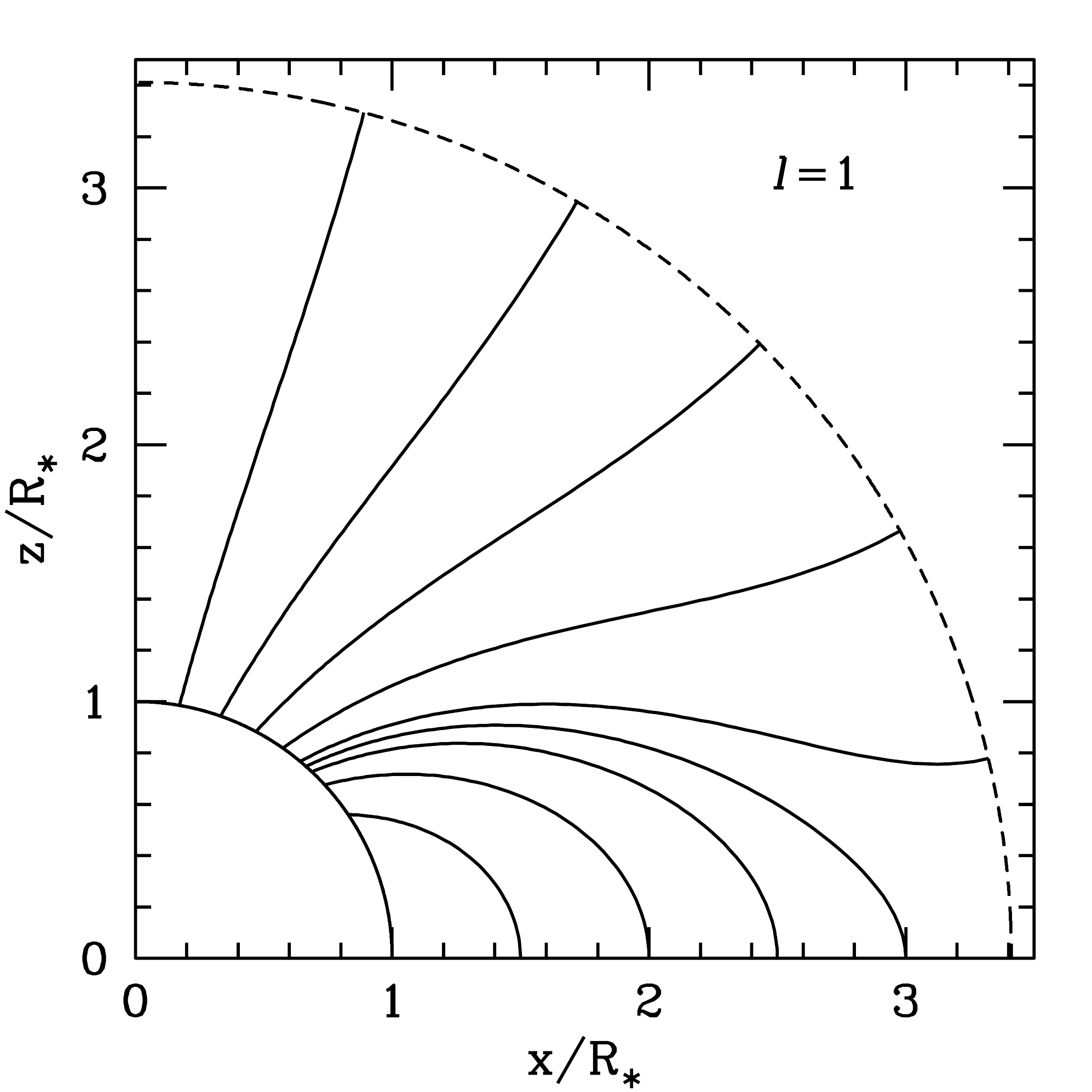}  \hspace{-3mm}   
                        } &
                \subfigure{
                        \includegraphics[width=40mm]{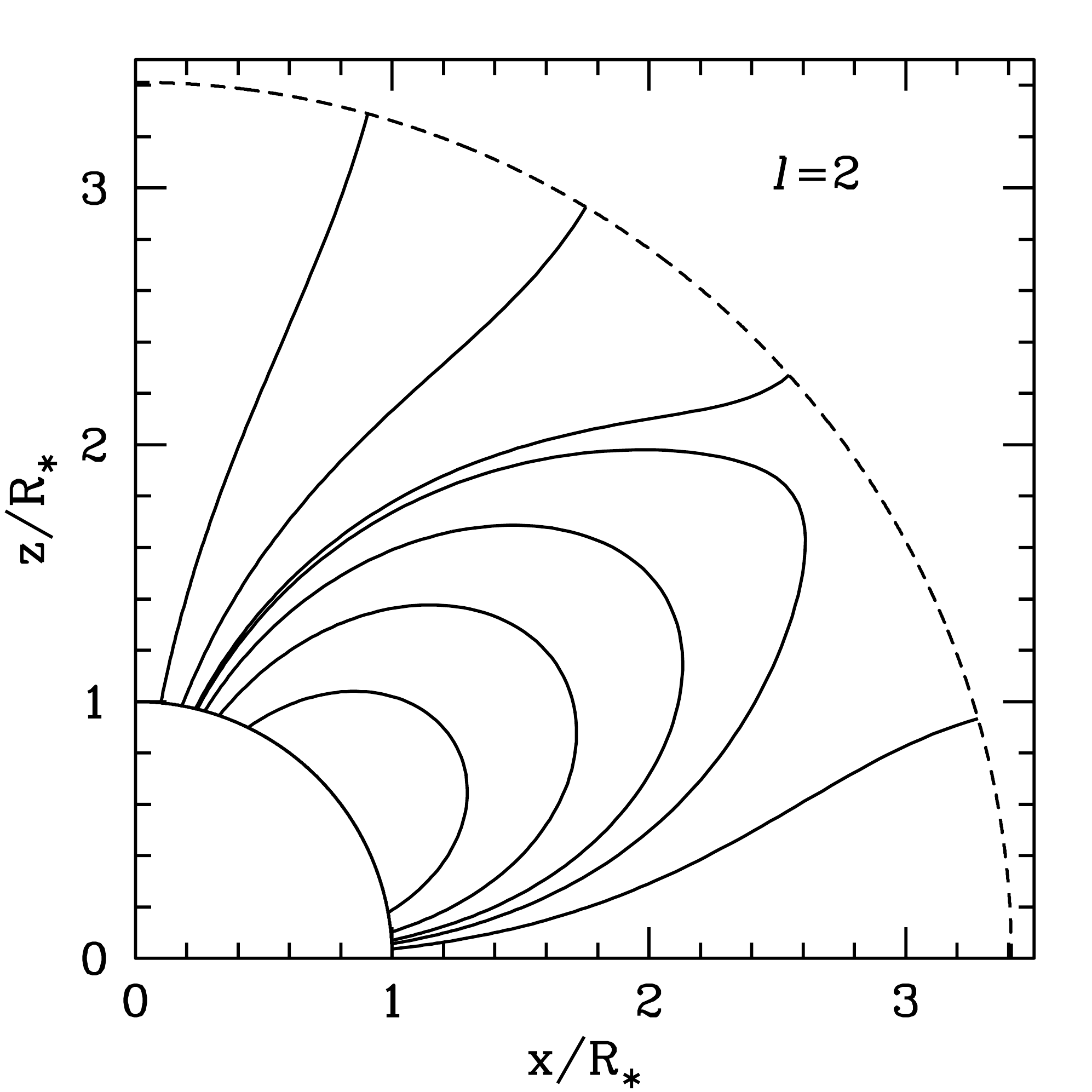}  \hspace{-3mm}  
                        } &
                \subfigure{
                        \includegraphics[width=40mm]{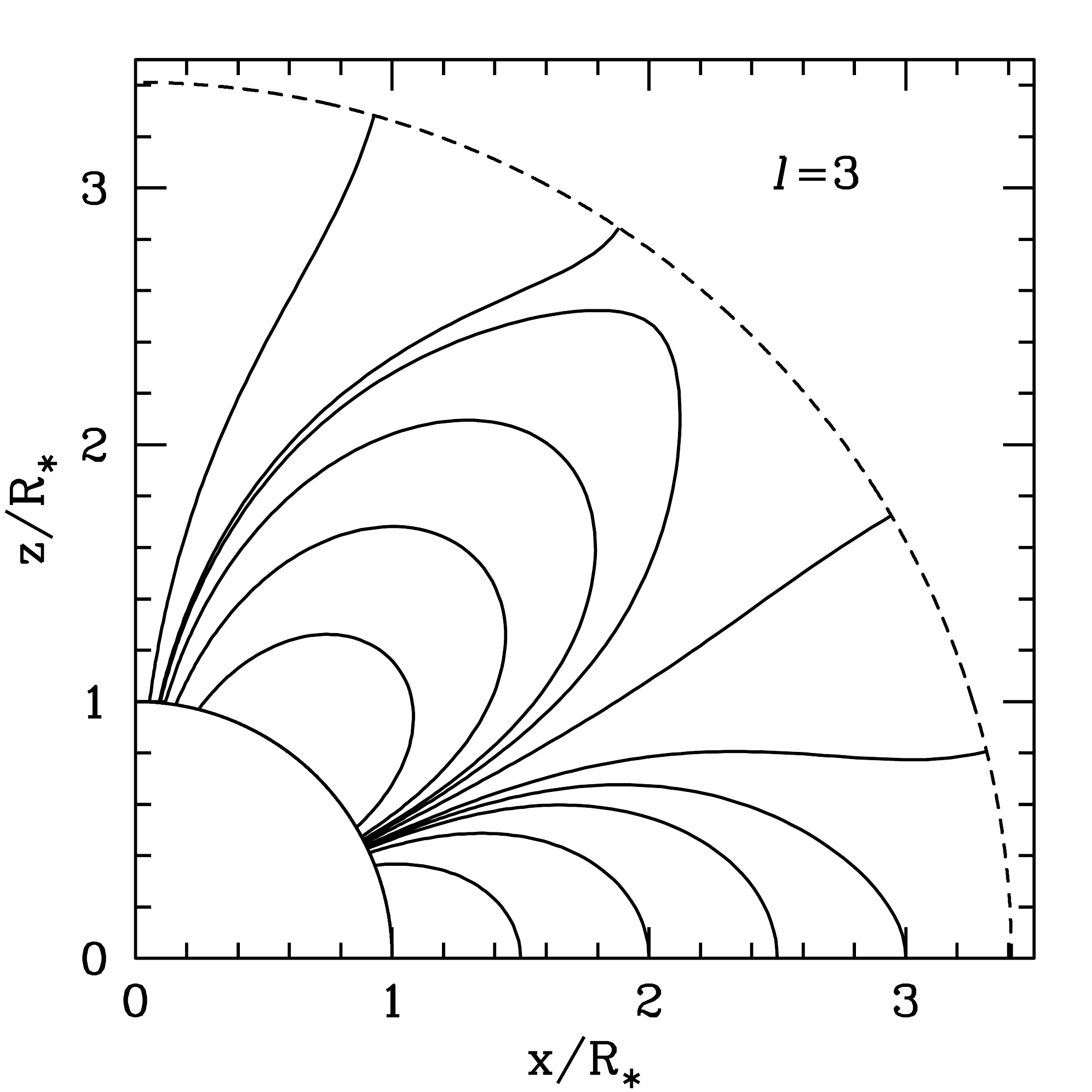}  \hspace{-3mm}  
                        } \\ 
        \end{tabular}
        \caption[]{Field lines for the lowest order multipoles (a dipole, $l=1$, a quadrupole, $l=2$, and an octupole, $l=3$) with a source surface set to 
                   $R_S \sim 3.4R_{\ast}$ plotted as the dashed line.  The field line shape is calculated using the components $B_r$ and $B_{\theta}$ 
                  obtained from (\ref{br_source}) and (\ref{bt_source}).  For higher order multipoles there are regions of open field at lower latitudes along 
                  which a stellar wind could be launched.  The magnetic fields are symmetric about the $x$ and $z$ axes, while a multipole of order $l$ has $2l$ 
                  shells of closed loops around the entire star (for example \cite{ass02}).}
        \label{pictures}
\end{figure*}

The large scale magnetospheric field must satisfy Maxwell's equation that the field be divergence free, (\ref{divmax}), and this,
combined with (\ref{BPsi}) (the PFSS model assumes that $\mathbf{B}$ is source free), means that the magnetic scalar potential must satisfy Laplace's equation,  
\begin{equation}
\nabla^2 \Psi = 0.
\end{equation}
This has a separable solution in spherical coordinates of the form
\begin{equation}
\Psi = \sum_l \sum_m \left[ a_{lm}r^l + b_{lm} r^{-(l+1)}\right]P_{lm}(\cos{\theta}){\rm e}^{{\rm i}m\phi}
\label{Psi_extrap}
\end{equation}
where here we change from the normalised spherical harmonics in (\ref{sph_ham}) to conform with our previously published models of stellar 
magnetosphere \cite{jar02a,gre06a,gre08}, and where the coefficients $a_{lm}$ and $b_{lm}$ are determined from the boundary conditions.  The first boundary condition is to specify the 
radial component at the stellar surface.  For field extrapolation models, this is determined directly from the observationally derived 
magnetic surface maps.  In our case, as we are considering axial multipoles, the radial field at the stellar surface ($r=R_{\ast}$) is given by (\ref{mag_gen1}),   
\begin{equation}
B_{r}(R_{\ast}) = B^{l,pole}_{\ast} P_l(\cos{\theta}),
\label{bound1}
\end{equation}
for a multipole of order $l$.  The second boundary condition is that at the source surface $R_S$ the field becomes purely radial,
\begin{equation}
B_{\theta}(R_S) = B_{\phi}(R_S) = 0.
\label{bound2}
\end{equation}
The magnetic field components themselves can be derived from (\ref{BPsi}) using (\ref{Psi_extrap}),
\begin{equation}
B_r  =  -\sum_l \sum_m
               [la_{lm}r^{l-1} - (l+1)b_{lm}r^{-(l+2)}]
               P_{lm}(\cos{\theta}) {\rm e}^{{\rm i} m \phi}
\label{br_extrap}
\end{equation}
\begin{equation}
B_\theta  =  -\sum_l \sum_m 
               [a_{lm}r^{l-1} + b_{lm}r^{-(l+2)}]
               \frac{\rmd}{\rmd\theta}P_{lm}(\cos{\theta}) {\rm e}^{{\rm i} m \phi}
\label{bt_extrap}
\end{equation} 
\begin{equation} 
B_\phi  =  -\sum_l\sum_m 
               [a_{lm}r^{l-1} + b_{lm}r^{-(l+2)}]
               \frac{P_{lm}(\cos{\theta})}{\sin{\theta}} {\rm i}m{\rm e}^{{\rm i} m \phi}.
\label{bp_extrap}
\end{equation}
From (\ref{bt_extrap}) and (\ref{bp_extrap}) it is clear that boundary condition (\ref{bound2}) is satisfied if
\begin{equation}
b_{lm} = -a_{lm}R_S^{2l+1},
\label{almblm}
\end{equation}
while for the axial multipoles ($m=0$) it can be seen from (\ref{bp_extrap}) that $B_{\phi} = 0$.  From (\ref{Psi_extrap}) it can be seen that (\ref{almblm}) is equivalent to
the assumption that $\Psi(r=R_S)$ is an equipotential surface.  Substituting (\ref{almblm}) 
into (\ref{br_extrap}), with $m=0$, and applying boundary condition (\ref{bound1}) and noting that $P_{l0}(\cos{\theta})=P_l(\cos{\theta})$, an expression 
for $a_{l0}$ in terms of $R_{\ast}$ and $R_S$ can be derived for a particular multipole of order $l$,
\begin{equation}
a_{l0} = -\frac{B^{l,pole}_{\ast}}{lR_{\ast}^{l-1}+(l+1)R_S^{2l+1}R_{\ast}^{-(l+2)}}.
\end{equation} 
Substituting this result and (\ref{almblm}) into (\ref{br_extrap}) and (\ref{bt_extrap}) with $m=0$ gives, for a particular 
$l$ value, general expressions for $B_r$ and $B_{\theta}$ (valid in the region $R_\ast\le r \le R_S$) for the large scale magnetosphere with a source surface,    
\begin{eqnarray}
B_r &=& B^{l,pole}_{\ast} \left(\frac{R_{\ast}}{r}\right)^{l+2}P_l(\cos{\theta}) \left[ \frac{lr^{2l+1}+(l+1)R_S^{2l+1}}{lR_{\ast}^{2l+1} + (l+1)R_S^{2l+1}}\right],\label{br_source} \\
B_{\theta} &=& \frac{B^{l,pole}_{\ast}}{l+1}\left(\frac{R_{\ast}}{r}\right)^{l+2}P_{l1}(\cos{\theta}) \times \nonumber \\
    && \left[ \frac{-(l+1)r^{2l+1}+(l+1)R_S^{2l+1}}{lR_{\ast}^{2l+1} + (l+1)R_S^{2l+1}}\right] \hspace{5mm} (R_{\ast}\le r\le R_S),\label{bt_source}
\end{eqnarray}
where in deriving (\ref{bt_source}) we have used the fact that $P_{l1}(\cos{\theta})=-\rmd P_l(\cos{\theta})/\rmd\theta$ [see (\ref{plm}) with $m=1$]. 
The field lines of the lowest order multipoles, with a source surface, are sketched in figure \ref{pictures}.  
These are found by numerical integration of $dr/B_r=rd\theta/B_\theta$ with $B_r$ and $B_{\theta}$ given by (\ref{br_source}) and (\ref{bt_source}), although an analytic
solution can also be found, but we do not discuss this here.
Note that the magnetic field components with the imposed source surface boundary condition are the same as (\ref{mag_gen1}) multiplied by correction terms.  Changing the radius
of the source surface $R_S$ modifies the structure of the entire magnetosphere, with more open field, along which a wind could be launched, available for 
smaller values of $R_S$.  Equations (\ref{mag_gen1}) are recovered in the limit of $R_S\rightarrow \infty$.   


\subsubsection{Coordinate free field components and tilted magnetospheres}
The initial ZDI results on V2129~Oph and BP~Tau have suggested that the octupole field component of the magnetospheres of 
accreting T Tauri stars contains a significant fraction of the magnetic energy \cite{don07,don08c}.  The dipole components of their magnetospheres, however, 
remain the most dominant at typical disc truncation radii, despite the large scale dipole-like field being distorted close to the surface of the star \cite{gre08}. 
Composite magnetic fields consisting dipole plus octupole field components have been used for many years by the solar physics community in the study
of coronal mass ejections (for example \cite{ant99,dev05}).  For stellar magnetism, \cite{lon10} and \cite{rom10} have recently 
presented MHD simulations of accretion to stars with composite dipole-octupole fields, the latter comparing their model directly with observations of V2129~Oph \cite{don07}.  
In their 3D models the octupole and dipole moment symmetry axes are tilted relative to each other, and to the stellar rotation axis, and the three axes do not lie in one plane.  
Their prescription for the total field $\mathbf{B}^l$ of axial multipole $l$ is presented in coordinate free form.  The total field $\mathbf{B}^l$ can be written using our equations 
(\ref{B_general1}) as
\begin{equation}
\mathbf{B}^l = \frac{M_l}{r^{l+2}}\left[(l+1)P_l(\cos{\theta})\mathbf{\hat{r}} + P_{l1}(\cos{\theta})\mathbf{\hat{\uptheta}} \right].
\label{total_B}
\end{equation}
Let $\mathbf{\hat{M}}_l$ be a unit vector along the symmetry axis of axial multipole $l$.  It can be seen from figure \ref{field_fig} (right panel) that 
$\mathbf{\hat{M}}_l\cdot\mathbf{\hat{r}}=\cos{\theta}$ and $\mathbf{\hat{M}}_l\cdot\mathbf{\hat{\uptheta}}=-\sin{\theta}$.  To make the derivation 
easier to follow we have assumed in figure \ref{field_fig}
that $\mathbf{\hat{M}}_l$ is aligned with the stellar rotation axis, and that both lie in the same stellar meridional plane (for example, the $xz$-plane).  
However, the results derived in this section apply generally to tilted multipole 
symmetry axes, and with appropriate coordinate and vector frame transformations are equally applicable to the case of two (or more) axial moments with arbitrary tilts with respect to the 
stellar rotation axis.  As $\mathbf{\hat{M}}_l$ can be written generally as $\mathbf{\hat{M}}_l=\cos{\theta}\mathbf{\hat{r}}-\sin{\theta}\mathbf{\hat{\uptheta}}$ then 
$(\mathbf{\hat{M}}_l\cdot\mathbf{\hat{\uptheta}})\mathbf{\hat{\uptheta}}=\mathbf{\hat{M}}_l-(\mathbf{\hat{M}}_l\cdot\mathbf{\hat{r}})\mathbf{\hat{r}}$.  Thus a simple expression for
$\mathbf{\hat{\uptheta}}$ can be derived, $\mathbf{\hat{\uptheta}} = -{\rm cosec}\,{\theta}\mathbf{\hat{M}}_l + {\rm cot}\,{\theta}\mathbf{\hat{r}}$.
Using this result to eliminate $\mathbf{\hat{\uptheta}}$ in (\ref{total_B}) then the total field may be written as
\begin{eqnarray}
\mathbf{B}^l = \frac{M_l}{r^{l+2}}\big\{[(l+1)P_l(\cos{\theta)}&+&{\rm cot}\,\theta P_{l1}(\cos{\theta})]\mathbf{\hat{r}}- \nonumber \\
&&{\rm cosec}\,\theta P_{l1}(\cos{\theta})\mathbf{\hat{M}}_l\big\}.
\label{total_B2}
\end{eqnarray}
In \ref{equ_appendix} the associated Legendre functions (with $m=1$) and the Legendre polynomials are written as series expansions.  Using (\ref{l_series}) and (\ref{l_series3})
to replace $P_l(\cos\theta)$ and $P_{l1}(\cos\theta)$ in (\ref{total_B2}), and using (\ref{moment}) to replace the $l$-th order multipole moment with the polar
field strength, then after some manipulation, a general result for the total field of an arbitrary titled axial multipole of order $l$ in coordinate free form can be obtained
\begin{eqnarray}
\mathbf{B}^l = \frac{B^{l,pole}_{\ast}}{(l+1)}\left(\frac{R_\ast}{r}\right)^{l+2} &&
         \sum_{k=0}^{N}\Big\{\frac{(-1)^k(2l-2k)!}{2^lk!(l-k)!(l-2k)!}(\mathbf{\hat{M}}_l\cdot\mathbf{\hat{r}})^{l-2k}\times \nonumber \\
         && \left[(2l-2k+1)\mathbf{\hat{r}}-(l-2k)(\mathbf{\hat{M}}_l\cdot\mathbf{\hat{r}})^{-1}\mathbf{\hat{M}}_l\right]\Big\}
\label{coord_free_B}
\end{eqnarray}
where $N=l/2$ or $N=(l-1)/2$, whichever is an integer.  This general result can be used to construct models of composite magnetic fields consisting
of the fields of two or more axial multipoles, arbitrarily tilted with respect to the stellar rotation axis, such as presented by \cite{lon10}.  Analogous to (\ref{coord_free_B}) 
a similar expression can be derived for a magnetosphere with regions of open field introduced by applying the source surface boundary condition.  Starting from (\ref{br_source}) and 
(\ref{bt_source}) and following the same argument used in deriving (\ref{coord_free_B}) we obtain, 
\begin{eqnarray}
\mathbf{B}^l &=& \frac{B^{l,pole}_{\ast}}{lR_\ast^{2l+1}+(l+1)R_S^{2l+1}}\left(\frac{R_\ast}{r}\right)^{l+2}\sum_{k=0}^N \Big\{\frac{(-1)^k(2l-2k)!}{2^lk!(l-k)!(l-2k)!}\times \nonumber \\
&& (\mathbf{\hat{M}}_l\cdot\mathbf{\hat{r}})^{l-2k}\big[(2kr^{2l+1}+(2l-2k+1)R_S^{2l+1})\mathbf{\hat{r}} + \nonumber \\ 
&& (l-2k)(r^{2l+1}+R_S^{2l+1})(\mathbf{\hat{M}}_l\cdot\mathbf{\hat{r}})^{-1}\mathbf{\hat{M}}_l\big]\Big\}.
\end{eqnarray}

 
\subsection{Difference between a spherical and Cartesian tensor approach}\label{quad_danger} 
The authors of \cite{lon07,lon08} have recently constructed models of composite dipole-quadrupole
magnetic fields, however, the expressions that they derive for the quadrupole field components are a factor
of $1/2$ smaller than those derived in this paper (see \S\ref{field}).  In their work they follow the approach of
\cite{lan75} who develop an expression for the magnetostatic potential exterior to the star
due to a pseudo magnetic ``charge'' distribution interior to the star.  In their approach the magnetostatic potential
is derived by analogy with the electrostatic potential expanded using Cartesian tensors.  In \ref{electro_cart}
we consider the electrostatic case by expanding the potential for a finite static charge distribution in Cartesian coordinates.  
As part of that derivation the non-primitive quadrupole moment must be defined (see equation (\ref{quad_moment})).   
There are three different definitions of the traceless quadrupole moment tensor used in 
the literature (for example \cite{kom03,lan75,kie72}), which ultimately leads to different expressions for $B_r$ and $B_\theta$,
and explains why the expressions used in \cite{lon07} and \cite{lon08} are a factor of $1/2$ smaller.     
However, as demonstrated in \cite{gra84}, the factor of $1/2$ arises naturally when the potential is expanded using spherical 
harmonics.  Given that stars and their circumstellar environments, and planets, are most straightforwardly described using a spherical 
coordinate system, (\ref{quad_moment}) is the most convenient definition of the non-primitive quadrupole moment.  Furthermore, as we now show,
our equations (\ref{mag_gen1}) represent the simplest way of expressing high order field components, and do not suffer from any 
ambiguity that can arise due to the differing definitions of multipole moments in use.

As pointed out in \ref{electro_cart}, three different definitions for the non-primitive (traceless) electrostatic quadrupole moment 
tensor $\mathbf{Q}$ are used\footnote{Further definitions of $\mathbf{Q}_1$ are possible and are discussed in \cite{gra78a} for both 
the electric and magnetic multipoles in terms of the equivalent spherical tensor $Q_{2m}$.  In \cite{gra80} equivalent 
Cartesian forms of the traceless magnetic quadrupole moment corresponding to $\mathbf{Q}_1$ are discussed.},
\begin{eqnarray}
\mathbf{Q}_1 &=& \frac{1}{2}\sum_i q_i (3\mathbf{r}_i\mathbf{r}_i - r_i^2 \mathbf{I}) \nonumber \\
\mathbf{Q}_2 &=& \sum_i q_i (3\mathbf{r}_i\mathbf{r}_i - r_i^2 \mathbf{I}) \nonumber \\
\mathbf{Q}_3 &=& \sum_i q_i \left(\mathbf{r}_i\mathbf{r}_i - \frac{1}{3}r_i^2 \mathbf{I}\right), \nonumber
\end{eqnarray}
where $\mathbf{r}_i\mathbf{r}_i$ is the tensor product of the vectors $\mathbf{r}_i$, $\mathbf{I}$ is the second rank identity tensor, and 
$\mathbf{Q}\hspace{1mm} \mathbf{:} \hspace{1mm} \mathbf{T}^{(2)}(\mathbf{r})$ used below is the double dot product
representing the full contraction of the tensors $\mathbf{Q}$ and the gradient tensor $\mathbf{T}^{(2)}(\mathbf{r})=\nabla\nabla(1/r)$
(see \ref{electro_cart} for full details).  Following \cite{lan75}, the authors of \cite{lon07,lon08} adopt the $\mathbf{Q}_2$ definition of the 
non-primitive quadrupole moment.  As a result of this, it is trivial to show
that the quadrupole potential, and therefore the field components $B_{r,quad}$ and $B_{\theta,quad}$ derived
in \cite{lon07,lon08}, are a factor of $1/2$ smaller than we have derived in this paper, based on the $\mathbf{Q}_1$ definition.  
Adoption of the $\mathbf{Q}_3$ definition results in a third different form for the potential
and field components for the quadrupole.  However, it is easy to demonstrate that our general expressions (\ref{mag_gen1}), 
based on the polar field strength, produce quadrupole field components that are independent of how $\mathbf{Q}$ 
is defined.  

We define the non-primitive quadrupole moment for the electrostatic Cartesian tensor case between 
(\ref{quaddef}) and (\ref{quaddef2}).  The (electrostatic) quadrupole potential can clearly be written as
$\Phi_2(\mathbf{r})=\eta \mathbf{Q}\hspace{1mm} \mathbf{:} \hspace{1mm} \mathbf{T}^{(2)}(\mathbf{r})$, where the constant $\eta$ depends on whether the
$\mathbf{Q}_1$, $\mathbf{Q}_2$ or $\mathbf{Q}_3$ definition is used (with $\mathbf{Q}_1, \eta=1/3$, with $\mathbf{Q}_2, \eta=1/6$
and with $\mathbf{Q}_3, \eta = 1/2$).  Following the argument in \ref{appendix_B2}
and carrying out the double dot product to determine the electrostatic potential $\Phi_2$ (which has an identical form to the 
magnetostatic potential $\Psi_2$ derived from equation (\ref{Psi_mag})), and then using (\ref{brdef}) 
to determine $B_r$ and $B_{\theta}$, we find that,
\begin{equation}
B_r   = \frac{9Q}{2r^4}\eta (3\cos^2{\theta}-1) \hspace{5mm}
B_{\theta} = \frac{9Q}{r^4} \eta \cos{\theta}\sin{\theta}.
\end{equation}   
At the stellar rotation pole, where the field only has a $B_r$ component, it is clear that 
\begin{equation}
Q=\frac{B^{2,pole}_{\ast}R_{\ast}^4}{9\eta}. \label{getitright}
\end{equation}
As both $B_r$ and $B_{\theta}$ are directly proportional to both $\eta$ and $Q$, and as $Q$ itself is inversely
proportional to $\eta$, then the field components are independent of $\eta$ and consequently independent
of the chosen definition of the non-primitive quadrupole moment tensor.  Thus, our equations (\ref{mag_gen1}) 
represent the simplest way of expressing the field components, and do not suffer from the ambiguities encountered in the 
literature due to differing definitions of the various multipole moments.


\section{Magnetospheric accretion models with multipolar magnetic fields} \label{mag_models}
Models of the magnetospheres of accreting T Tauri stars have traditionally assumed dipolar magnetic fields.  In the past few years however, 
motivated in part by the availability of stellar magnetic surface maps, new models have been developed which consider fields with an 
observed degree of complexity.  Two types of models which incorporate multipolar stellar magnetic fields have been developed.  Firstly, those which 
assume that the large scale magnetosphere can be modelled as a potential field, and then include effects of the coronal plasma, and secondly, 3D magnetohydrodynamic simulations.


\subsection{Development of PFSS models and comparison with MHD field extrapolations}
Over the past 40 years the PFSS model has undergone several modifications.  Reference \cite{pne71} considered how the coronal
gas pressure distorted the large scale field.  In order to account for the effects of the coronal plasma, the authors of \cite{sch78} introduced an analytic field
extrapolation model with a non-spherical source surface, but only considered the dipole component of the solar field.  The resulting source surface
was a prolate spheroid with major axis aligned with the solar rotation axis.  Reference \cite{lev82} subsequently extended the ideas of Schulz to incorporate more
generalised fields (see also \cite{sch97}).  It is interesting to note, however, that the prolate spheroid source surface also arises in more 
complex MHD field extrapolation models, at least at solar minimum \cite{ril06}.  At solar maximum, when the photospheric field is more complex, the 
source surface is, on average, more spherical.  

Various other studies have examined the effects of including current sheets and other volume currents into 
solar magnetic field extrapolation models.  Such models better reproduce the topologies of coronal streamers detected during solar eclipse observations, as 
well as satellite observations of the large scale heliospheric field (for example \cite{sch71,wol85,aly93,zha94,zha95,gar99}).  As such detailed 
observations are not available for forming solar-like stars we do not discuss the various models here.  Interested 
readers can find up-to-date discussions comparing PFSS models with the results of more complex current sheet source surface (CSSS) models
in \cite{sch06} and \cite{jia10}.

\begin{figure}[t]
  \centering
  \includegraphics[width=95mm]{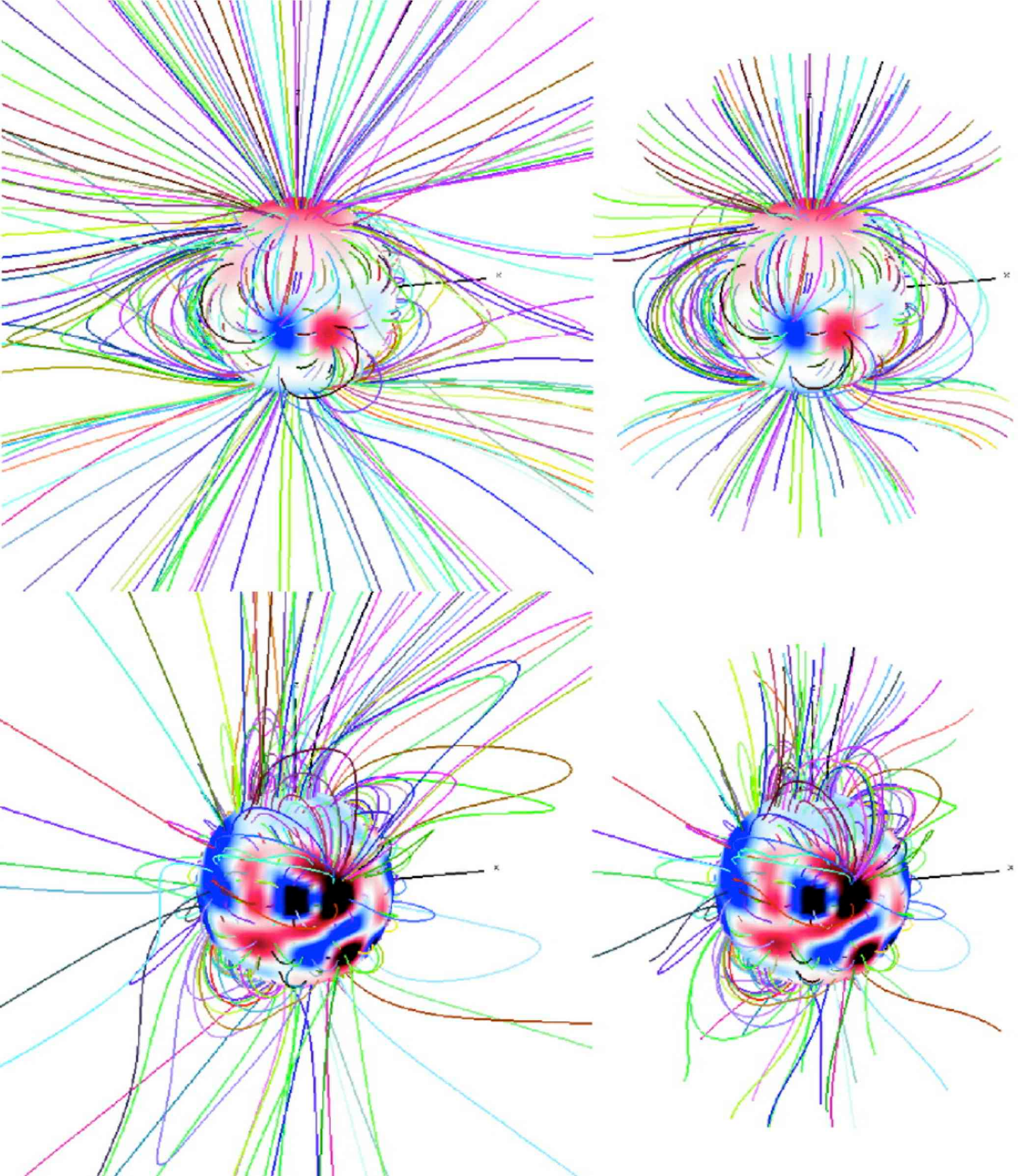}
  \caption{A comparison of the field structure of the solar corona obtained using the potential field source surface method (PFSS; right hand panel) with a MHD solution (left hand panel) 
               at solar minimum (top panel) and close to solar maximum (bottom panel).  For both the PFSS and MHD models the field is extrapolated using
               the same starting positions at the photosphere with field lines assigned the same (arbitrary) colour to aid comparison.  
               Reproduced by permission of the AAS, and Dr. P. Riley, from \cite{ril06}.}
  \label{riley_extrap}
\end{figure}

MHD field extrapolation models are more desirable as they not only include information on the magnetic field structure, but also about how fields interact with, and are influenced by, outflows and the 
coronal plasma.  For the solar corona, the first global MHD simulations to incorporate magnetic maps of the photospheric field were developed in the late 1990s (for example \cite{usm96} and \cite{lin99}).  
A thorough and balanced overview of the advantages and disadvantages of the use of the PFSS model compared to MHD models is presented in
\cite{ril06} (see figure \ref{riley_extrap}).
The PFSS allows 3D global field extrapolations to be produced quickly with moderate computing resources.  In contrast, MHD 
models require powerful computing resources.  The main advantage of MHD models is their ability to include the non-potentiality of the field and time dependent 
surface transport effects (meridional circulation, differential rotation, supergranular flows etc).  PFSS models produce static field configurations and cannot incorporate time 
dependent effects, such as magnetic reconnection events.  It remains an open question, however, how important surface effects are on T Tauri stars, 
although the authors of \cite{don10} have recently reported the detection of surface differential rotation on the low mass CTTS V2247~Oph.     

MHD models require the inclusion of a number of free parameters, particularly with regards the thermodynamic properties of the coronal plasma.   
In the MHD model of \cite{ril06} a simple polytropic equation of state for the coronal plasma is considered, with the temperature and density at the base of the 
solar surface treated as free parameters.  Other authors have considered more complex implementations of the energy equation; reference \cite{dow10} for example, considers
heating and cooling terms in the MHD energy equation.  Unfortunately, such models also introduce more unknown physical parameters, the choice of which
directly influences the resulting field topology.  In contrast to MHD models, the PFSS model produces a unique solution for the coronal field \cite{aly87}.  
Although reference \cite{ril06} concludes that the PFSS model often produces results close to the more physically realistic MHD models, when considering the large scale coronal structure, 
the authors anticipated that discrepancies between the models would increase once new solar vector magnetograms became available (readers are referred to 
\cite{rua08} and \cite{tad09} for some of the latest work on solar field extrapolations).
Extrapolations of the local magnetic field in active regions have recently been compared to the observed 3D field structure derived from the STEREO satellite data.  
Reference \cite{san09} concludes that the potential field approximation is poor at reproducing the observed fields, while 
reference \cite{liu08} concludes that provided the active region under consideration is relatively stable then the potential field approximation is adequate.  

It is important to remember
however that small scale magnetic features, such as bipolar groups, that are easily detected on the Sun, remain well below the achievable resolution of 
stellar magnetic maps.  A star in the Taurus star forming region, at a distance of $\sim 140\,{\rm pc}$ \cite{ken94}, is some 29 million times further from the Earth than the Sun.  
Stellar spectropolarimetric measurements can only probe the large scale properties of the magnetosphere, and not the small
scale field regions where the PFSS model most frequently breaks down; thus the PFSS model has since been extensively applied to produce coronal 
field extrapolations from stellar magnetograms derived through ZDI (for CTTS, however, there is additional non-potentiality induced due to the 
interaction of the large scale magnetosphere with the disc, as we discuss in \S\ref{MHDsection}).        
  
In addition to the finite achievable resolution Zeeman-Doppler maps of stellar surface fields are also subject to missing information due to inclination effects, whereby
for a given stellar inclination, much of one hemisphere is hidden to an observer.  All of these limitations, and their effects on the derived coronal
structure and X-ray emission properties, have been thoroughly examined in \cite{joh10}.  Of the three limitations (finite resolution, the suppression of the Zeeman
signal in dark spots, and the unobservable hemisphere) they find that the largest effect on the global coronal structure is caused by stellar inclination effects.  What theoretical
models assume for the flux distribution in the hidden surface area changes the way the larger scale field lines connect to opposite polarity regions between the 
hemispheres.  However, the authors of \cite{joh10} conclude that the field in the visible hemisphere is reliably reproduced by field extrapolation models, and in particular,
the quantities calculated from PFSS models of T Tauri magnetospheres (the location of hot spots, accretion filling factors, disc truncation radii etc) are unaffected by missing information.
For CTTS the detection of high latitude accretion spots strongly suggests field configurations that are antisymmetric with respect to the centre of the star.  If it were the 
case that the field in the hidden hemisphere was such that stellar magnetosphere was symmetric with respect to the centre of the star, high latitude accretion spots
would be difficult to explain.  The authors of \cite{joh10} also conclude, however, that missing flux in magnetic surface maps causes PFSS models to overestimate the 
amount of open flux.  This may have important implications for T Tauri angular momentum loss (see \S\ref{summary}; in the following section we discuss the 
PFSS model applied specifically to T Tauri coronae).  
          
A problem for PFSS models of stellar coronae is where to locate the source surface.  Unlike solar observations, in-situ satellite measurements 
of stellar heliospheric fields are not available.  Indirect indicators of coronal structure have therefore been used to estimate the maximum extent of 
the closed field regions of stellar coronae.  For AB~Dor, a star for which several magnetic maps have been derived \cite{don97b,don99}, 
slingshot prominences have been detected extending to $\sim4.5{\rm R}_{\ast}$, in excess
of the equatorial corotation radius of $\sim2.7{\rm R}_{\ast}$ \cite{col89,don99}.
Thus, in the first PFSS field extrapolation models to be applied to a stellar magnetic map (of AB~Dor), \cite{jar02a} made a conservative estimate of
$R_S = 3.4{\rm R}_{\ast}$.  However, changing the location of the source surface influences the structure of the 
entire corona, in particular the extent of open field relative to closed field regions.  Using the AB~Dor magnetic maps \cite{jar02b} extended the 
PFSS model to include the effects of coronal plasma.  Assuming that stellar coronae are isothermal and that the plasma trapped along the closed loops is in
hydrostatic equilibrium, enables the gas density at each point within the corona to be determined.  

In the Jardine coronal model if the gas pressure exceeds the 
magnetic pressure at any point along a field line loop, that field line is assumed to be torn open, and the coronal plasma lost in a wind.  Such loops would therefore
be dark in X-rays.  In such a way the model allows the stellar X-ray emission properties such as the global X-ray emission measure, or the amount of rotational modulation
of X-ray emission, to be estimated using the coronal fields extrapolated from magnetic surface maps.  With a large source surface, however, often the 
simulated values of the coronal density are too low, suggesting that more compact coronae, with small values of $R_S$ are required.  At first, this appears
to contradict the requirement of large extended magnetic structures for prominence support.  However, even if the closed field regions are confined close 
to the surface of the star, extended stable magnetic loops may form due to the reconnection of wind-bearing open field lines \cite{jar05}.
This model provides a mechanism for supporting prominences beyond corotation, and beyond any reasonable estimate of the source
surface radius.  The location of the source surface in stellar models remains a somewhat free parameter, although changing its value affects the amount of modulation
of X-ray emission and the X-ray luminosity predicted.  The coronal model does, however, include an additional free parameter.  It is
assumed that the gas pressure at the field line foot points (equivalently the density) scales with the magnetic pressure ($p_0\propto B_0^2$).  Adjusting the free parameter,
the constant of proportionality between these two pressures, affects the total amount of X-ray emission predicted.  For smaller values the gas pressure 
is reduced at each point along the coronal loops and 
therefore the magnetic field is able to contain more of the coronal plasma.  For larger values, the opposite is the case, and more loops are unable to contain 
the plasma and are assumed to be blown open.  For T Tauri stars specifically the authors of \cite{jar06} developed a method of constraining the free parameter by comparing the model 
predicted X-ray emission measures with those derived from a large sample of stars in the Orion Nebula Cluster, for each assumed surface magnetic field distribution. 
We note that such free parameters are not unique to stellar coronal models, and are commonly employed for model of solar loops, for example \cite{wan97_2}.

Independently obtained X-ray observations of stellar coronae can provide a test 
of the coronal fields derived via field extrapolation.  In order to test the model properly, however, contemporaneous spectropolarimetric (from which
the magnetic surface maps are derived) and X-ray satellite observations are required, see \cite{hus07}.  The simultaneity of the multiwavelength observations is
crucial as if there is a large separation between the ground based spectropolarimetry and the space based X-ray observations there is a danger that the
magnetic field of the star will have evolved significantly during the delay.  Thus, the X-ray properties predicted from the derived magnetic map may be
significantly different from those observed.  This of course would not provide a true test of the theoretical models, as it would be impossible to 
ascertain if the difference was due to magnetic evolution of the stellar corona, or inadequacies of the model, or both. 
However, the lack of significant change in the large scale field topology of the accreting 
T Tauri stars BP~Tau and AA~Tau apparent from spectropolarimetric data taken years apart, suggests that a strictly
simultaneous observing strategy, although highly desirable, may not be required \cite{don08c,don10aatau}.
A contemporaneous observing strategy was successfully employed in \cite{hus07} on AB~Dor and is currently
being used in order to test the ability of PFSS models to capture the true magnetospheric geometry of accreting T Tauri stars \cite{gre09}, 
which we now discuss.


\subsection{Potential field models of T Tauri magnetospheres with complex fields}
The PFSS model was extended through a series of papers in order to construct models of the magnetospheres of accreting T Tauri stars 
\cite{gre05,gre06a,gre07,gre08,jar06}.  The initial simulations used magnetic fields extrapolated from surface magnetograms
of young rapidly rotating zero-age main sequence stars, AB~Dor and LQ~Hya, as pre-main sequence maps were not available at the time \cite{gre05,gre06a}.
By adjusting the stellar parameters to typical T Tauri values the extrapolated fields were divided into three distinct regions.  Close to the star the complex and 
loopy field lines contained the X-ray emitting corona, while regions of open field (typically at high latitude) were available to carry a stellar wind.  The larger scale 
field, that which interacts with the disc, is simpler in structure and more ``dipole-like''.  This simple larger scale field is, however, distorted close to the star by the 
complex surface field regions \cite{gre08}.  This model for T Tauri magnetospheres assumed that any of the large scale field lines that passed through
the stellar equatorial plane would support gas accretion from the disc on to the star, provided they passed through the disc interior to the corotation radius down to 
some inner disc truncation radius.  Interior to corotation the 
effective gravity points inwards towards the star and accretion would naturally take place.  This simple model, the first to incorporate multipolar magnetic
fields into models of the accretion process on to T Tauri stars, successfully reproduced the observed correlation between X-ray luminosity and stellar mass 
\cite{pre05,jar06}, the observed rotational modulation of X-ray emission \cite{fla05,gre06b}, the observed reduction
in X-ray luminosity of accreting stars relative to non-accretors \cite{gre07},
and the correlation between disc mass accretion rate and stellar mass \cite{gre06a}.  However, it is worth noting that despite several 
publications describing the possible origin of the accretion rate - stellar mass correlation (originally discovered by, although thus-far not credited to, reference \cite{hil92}), it may
be nothing more than an artefact of observational selection and detection limitations \cite{cla06}.

\begin{figure}[t]
  \centering
  \includegraphics[width=115mm]{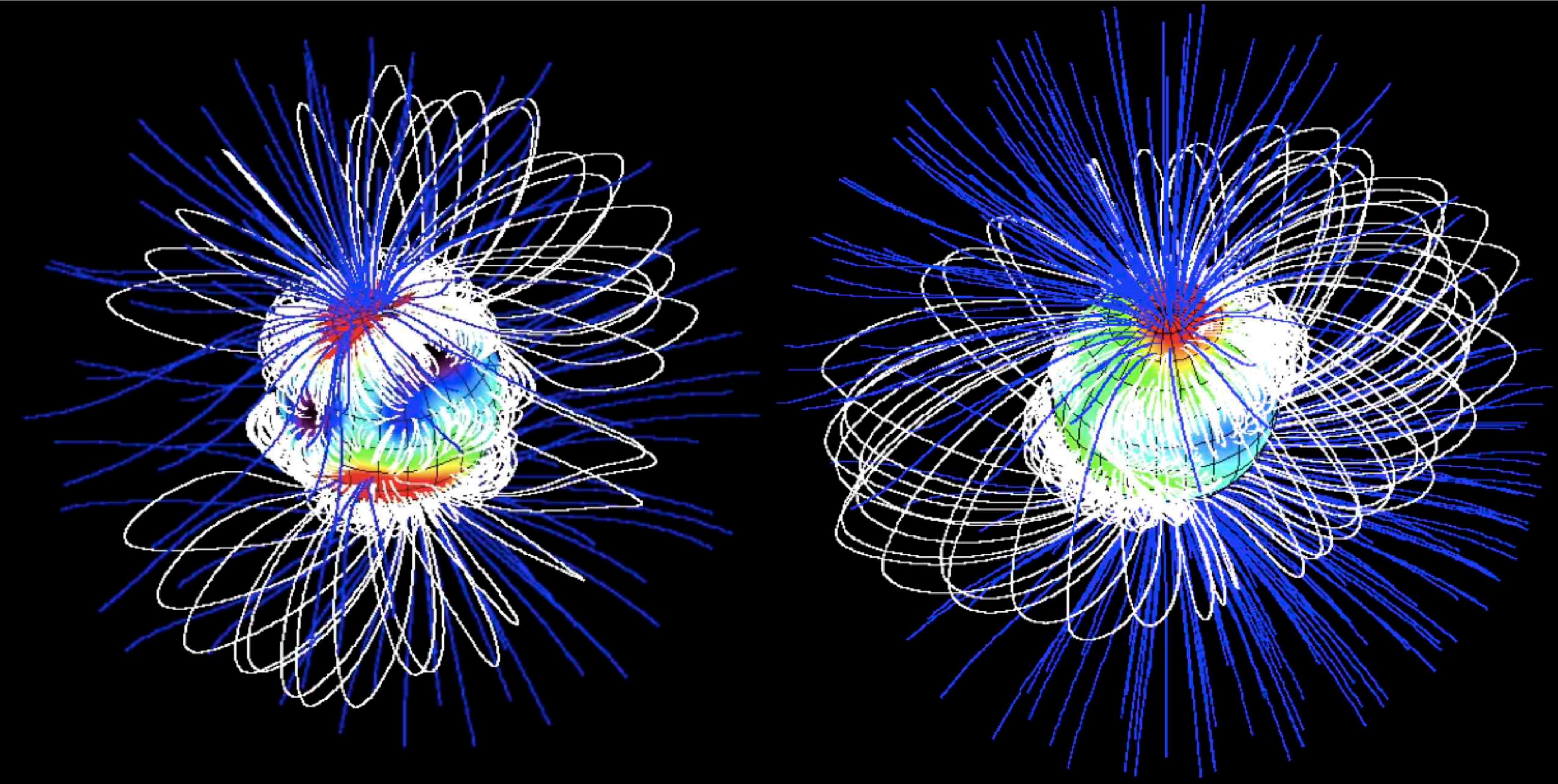}
  \caption{Numerical field extrapolations of the magnetospheres of the accreting T Tauri stars V2129~Oph and BP~Tau (left and right panels respectively).  Open field lines 
               are coloured blue, with closed fields line in white.}
  \label{field_extraps}
\end{figure}

Figure \ref{field_extraps} shows potential field extrapolations from photospheric maps of the magnetic fields of the accreting T Tauri stars V2129~Oph and BP~Tau.  
There is a clear distinction between the simple large scale field, and the more complex surface field. 
From the rotational modulation of unpolarised and circularly polarised profiles of the CaII infrared triplet emission (and in particular the contribution 
from accretion spots at the foot points of accretion flows), \cite{don07} and \cite{don08c} infer that the majority of accreting gas flows into spots at high latitude (see \cite{don08c} for details
of the construction process for excess CaII emission maps, which has recently been improved and further developed by the authors of \cite{don10aatau}).
In order to ensure that the large scale field is able to reach high latitudes, the source surface in such models must
be set to at least the equatorial corotation radius $R_{co}$.  This does not mean, however, that the disc need be truncated at corotation.  By assuming that circumstellar discs are 
truncated where the differential magnetic torque due the stellar magnetosphere is balanced by the differential viscous torque in the disc, reference \cite{gre08} calculated disc
truncation radii using the extrapolated fields of BP~Tau and V2129~Oph.  For BP~Tau, a completely convective star with strong dipole and octupole components, the disc was found
to be truncated at $\sim0.7R_{co}$.  The disc of V2129~Oph, a star which despite its young age has already developed a small radiative core and which has a dipole component four
times weaker than that of BP~Tau, the truncation radius was $\sim0.5R_{co}$.  The process of accretion of gas from well within corotation should exert a 
spin-up torque upon the star in the absence of an efficient angular momentum removal mechanism.  We note, however, that
based on a re-evaluation of the accretion related emission lines in the optical spectrum of V2129~Oph its mass accretion rate may be an order of magnitude lower than previously 
considered by both \cite{don07} and \cite{gre08}.  With a lower accretion rate the torque balance calculation would result in a larger disc truncation radius.
A stronger dipole component and/or a weaker octupole component would have the same effect.    

The incorporation of complex multipolar magnetic fields into magnetospheric accretion models naturally provides an explanation for small accretion filling factors (the fractional
surface area of the star covered in accretion hot spots).  Magnetospheric accretion with an aligned dipole field tends to produce filling factors that are typically an order
of magnitude larger than observationally inferred (for example \cite{val04}).  For magnetospheric accretion with non-dipolar magnetic fields 
accreting gas is funnelled on to discrete regions of the stellar surface \cite{gre05,gre06a}.     
The prediction of small accretion hot spots spanning a range of latitudes is not unique to PFSS models of T Tauri magnetospheres.  Hot spots at low latitudes 
are a natural consequence of considering non-dipolar, as well as tilted dipolar, magnetospheres.  They also arise in MHD models of magnetospheric accretion (as we
discuss in the following section) and in a recent generalisation of the Shu X-wind model \cite{moh08}.    

The multipolar X-wind model of reference \cite{moh08} is constructed from 2D axisymmetric potential stellar magnetic fields.  Only the odd $l$-number multipoles,
which have $B_r=0$ in the equatorial plane, are considered.\footnote{Reference \cite{moh08} 
erroneously and repeatedly refers to the quadrupole as being an odd multipole and the $l=3$ and $l=5$ multipoles as the quadrupole and the octupole.  In fact, 
such $l$-numbers represent the octupole ($l=3$) and the dotriacontapole ($l=5$).}  As with the dipolar X-wind model, all field 
lines initially originate from the star.\footnote{The model has recently been extended via the inclusion of a separate disc magnetic field \cite{kra09}.} 
During the initial phases of evolution it is assumed that matter somehow opens the large scale closed field threading the disc.  
This produces a disconnected region of open field at the corotation radius from where an outflow is launched.  Although a non-dipolar magnetosphere results in funnel
flows from the disc arriving at the star at different latitudes, the basic properties of the X-wind model remain unchanged, namely the assumption of disc-locking and 
that of trapped flux within a small region (the X-region) at the corotation radius.  The field lines both interior and exterior to corotation are pinched together,
while angular momentum from the funnel flow is transferred backwards through the X-region and removed by the X-wind.  
Thus, the multipolar X-wind model, as well as the dipole version \cite{shu94a,ost95}, allows accretion to take place without a net transfer of angular
momentum to the star.  The process of accretion proceeds without a spin-up torque being exerted on the star, which would have slowly increased the stellar rotation rate towards break-up speed.  
This is desirable as most accreting T Tauri stars have rotation rates about an order of magnitude below the break-up rate (for example \cite{bou93}).     
The model is admirable in that it provides a simultaneous description of both accretion and outflows, processes which are intimately
linked in accreting T Tauri stars (large collimated jets, for example, are not observed in non-accreting systems).
However, the multipolar X-wind model suffers from the same limitations as its dipolar counterpart (see \S\ref{dipole_accn}).
Furthermore, the multipolar X-wind model, nor the PFSS field extrapolation models, incorporate time dependent effects, such as the 
field evolution due to the star-disc interaction, or magnetic reconnection events.   The modelling of such complex processes requires a full MHD solution.  


\subsection{3D MHD models of T Tauri magnetospheres with non-dipolar fields} \label{MHDsection}
Reference \cite{rom10} presents 3D MHD simulations of V2129~Oph.  This represents the culmination (thus far) of a series of papers by the same group of
authors examining the star-disc interaction with ever increasing physical reality - from 2D MHD simulations \cite{rom02}, to 3D models with tilted 
dipole magnetospheres \cite{rom03,rom04} (see also \cite{kol02b}), simulations with composite dipole-quadrupole magnetic fields \cite{lon07}, 
then with composite magnetic fields where the dipole and quadrupole moments are tilted by different amounts relative to the stellar rotation axis, and in different 
planes \cite{lon08}, and finally similar 3D models but with composite fields consisting of a dipole plus an octupole field component \cite{lon10}.  

\begin{figure}[t]
  \centering
  \includegraphics[width=130mm]{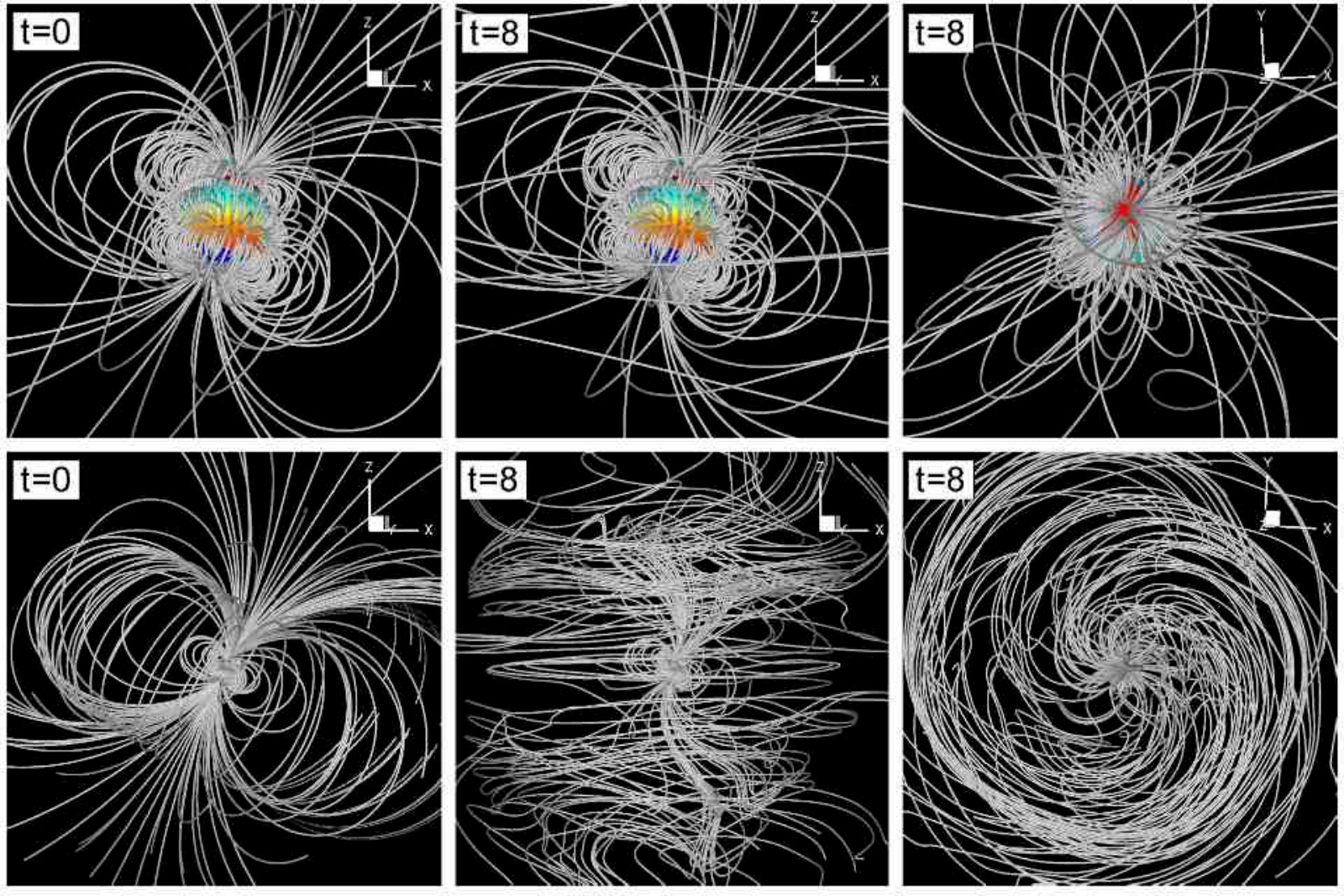}
  \caption{A MHD simulation of the magnetic field of the CTTS V2129~Oph showing the distortion of an initially potential field ($t=0$, left hand panel) due to the star-disc interaction.  
               The middle and right panels show two different orientations of the field structure after 14.4 days ($t=8$ in simulation units), a little more than 
               two stellar rotation periods.  The upper panel is the field closest to the star, which shows 
               little departure from the initial potential configuration.  The lower panel shows the formation of a magnetic tower, which arises due to the differential rotation of the field line foot points 
               anchored in the disc and on the stellar surface, and being wrapped around the rotation axis.  Figure reproduced, with permission, from \cite{rom10}.}
  \label{romanova_extrap}
\end{figure}

The most recent simulations have shown that the potential field approximation remains valid in regions where the magnetic stresses dominate over the material stresses
of the accreting gas \cite{lon10,rom10}, as was argued previously \cite{jar08b}.  The 3D MHD models can, however, include the time dependent
effects of the star-disc interaction.  The larger scale field is found to quickly depart from a 
potential field configuration due to the shearing of the field lines caused by the differential rotation between where they are anchored on the stellar surface and where they thread the disc, see figure \ref{romanova_extrap}, which leads to inflation of the field.  The field lines wrap around the rotation axis forming a magnetic tower (\cite{rom10}, and references therein).        

The shape and distribution of accretion hot spots at the base of flows of accreting gas are particularly sensitive to the field topology.  The inclusion of multipole moments 
of different strength, tilt relative to the rotation axis and the other moments, as well as the plane of the tilt, are all important in determining the location of hot spots.  For example,
in the case of accretion to dipole-octupole composite fields, a strong dipole component dominates the kinematics of the in-falling gas, leading to the formation of two 
discrete funnel flows for a slightly tilted dipole moment \cite{lon10}.  In other simulations with a stronger octupole component, the more complex
field regions dominate leading to the formation of lower latitude spots.  Intriguingly, the position and shapes of photospheric hot spots determined from the simulations
appear to remain fairly constant in time \cite{lon08}.  This immediately suggests that the commonly observed complex variability of accreting T Tauri stars, evident
from their photometric light curves, is due to unsteady (clumpy) accretion flows, with a variable mass accretion rate.  Alternatively this complex variability may be 
due to the restructuring of the stellar magnetic field due to dynamo processes interior to the star (see the discussion in \cite{lon10}).\footnote[1]{It is also worth noting
that the authors of \cite{lon08} conclude that stars which show simple sinusoidal light curves can also host complex non-dipolar magnetic fields.}

The authors of \cite{lon07} and \cite{lon08} have also calculated the angular momentum flux on to the star due to accretion and the outward angular momentum flux along (polar) 
stellar open field lines.  They find that the main spin-up torque on the star is due to the magnetic connection with regions of the disc interior to corotation, with a smaller spin-down 
torque arising from the outward angular momentum flow along the open field.  The spin-up torque from the accreting matter itself is a factor of $20-100$ times smaller than the spin-up 
torque from the magnetic connection to the inner disc for a purely dipolar stellar magnetic field, but only $10-20$ times smaller for a purely quadrupolar field \cite{lon07}.  In all cases 
the MHD models presented in references \cite{lon07} and \cite{lon08} feature stars that are experiencing a net spin-up torque.  For stars with magnetic fields that are more complex than a dipole, 
the spin-up torque is less than for the dipole models despite the disc being truncated closer to the star.  This may initially seem contradictory, however the authors of \cite{lon08} argue that it arises 
due to the smaller connectivity with the inner disc than they find for dipolar fields.  Thus the authors of \cite{lon08} conclude that the spin-up torque experienced by a star with a non-dipole 
magnetic field may not be as severe as for stars with dipole magnetic fields.  Further investigations incorporating generalised multipolar stellar fields, simulated for many rotations 
of the star would be welcomed to explore these suggestions further.

The MHD models of accretion to non-dipolar magnetic fields do not yet incorporate restructuring of the stellar magnetosphere due to surface transport effects, or time 
variable accretion flows (but see \cite{rom08} and \cite{kul08}).  However, it is not yet clear observationally whether or not 
the surface fields of accreting T Tauri stars vary significantly with time.  Reference \cite{don08c} derived two magnetic maps of the accreting T Tauri star BP~Tau 
from circularly polarised spectra taken approximately 10 months apart.  This corresponds to almost 39 stellar rotations, and while the maps appeared
to show an apparent quarter phase shift in the field (most likely related to a small error in the assumed rotation period \cite{gre08}), the large scale field structure   
is remarkably similar at both epochs.  The same result was found for AA~Tau where magnetic maps derived from spectropolarimetric data taken about one year apart also show
little difference in the large scale field topology \cite{don10aatau}.  This would then suggest that time variable mass accretion has a larger role to play in T Tauri photometric variability; however, the authors 
of \cite{ngu09}, who search for variations in accretion rates over short ($\sim$ hours - days) and long ($\sim$ months) timescales conclude that accretion hot spots rotating 
across the star is the dominant cause of variability.  This result also appears to be consistent with earlier line profile variability studies of individual stars that suggested that 
accretion flows are modulated with the stellar rotation (for example, \cite{gia93,joh95,joh97,bou07aatau}).  However, it should also be noted that
the common detection of large X-ray flares believed to extend to several stellar radii may indicate large scale field evolution \cite{get08a,get08b}.  This, coupled 
with the recent detection of strong differential rotation on the completely convective T Tauri star V2247~Oph \cite{don10}, suggests that   
more spectropolarimetric data on a greater number of stars is required to thoroughly investigate the issue of evolving magnetic fields. 

An important finding from the MHD simulations is that accretion hot spots are inhomogeneous, being hotter 
and denser in their cores compared to their peripheries, as most of the kinetic energy of the accreting gas is carried in the centre of the funnel flow 
(a finding first discussed for accretion along tilted dipole magnetospheres - \cite{rom04}).  Thus, hot spots appear smaller when viewed at higher 
temperature and densities (i.e. at shorter wavelengths).  This has important implications as it suggests that UV observations
used to estimate accretion filling factors under estimate the true size of the spots (being sensitive only to the hotter regions).  This also appears to explain why accretion filling
factors and mass accretion rates derived from density sensitive line triplets detected in high resolution X-ray spectra are found to be an order of magnitude smaller
than those derived from optical/UV spectra, with the X-ray emission arising from only the hottest regions of the spot \cite{gun07,arg07}.  
However, it must also be remembered that unsteady accretion flows may alter the accretion filling factor, \cite{ard00}, for example, find a factor
of more than 40 difference in the accretion filling factor for BP~Tau.  This is of course a variability study of a single star, and long term observations targeting 
the same stars repeatedly would be welcome to search for variations in accretion spots sizes and distributions.
The non-uniform variability of accreting T Tauri stars clearly presents a formidable challenge for 
theoretical models (see the discussion in \cite{joh02}).  Future models that incorporate magnetic surface maps as a boundary condition, and which
consider stellar surface transports effects, will drive further progress in this field. 


\section{Summary and applications to outstanding problems} \label{summary}
Over the past few years instrumentation has advanced to the stage where the magnetic fields of stars across the Hertzsprung-Russell diagram can be probed
in unprecedented detail.  Of particular interest are the magnetic fields of forming solar-like stars, as they allow us to study the history of the Sun at the epoch of the formation of the 
Solar System.  The ability to produce surface maps of their magnetic fields is a remarkable achievement, given that stars (with the exception of the Sun) are mere points of 
light in the night sky.  However, those points of light are rotating, and by monitoring the distortion of Zeeman signatures as stars rotate, ZDI studies have revealed the complex
nature of T Tauri magnetic fields.  In this work, we have demonstrated how results developed in the classical electromagnetism and molecular physics literature can be used to derive 
analytic expressions for the axial multipole ($m=0$) magnetic field components of a stellar, or equivalently a planetary, magnetosphere.  The resulting expressions 
(\ref{mag_gen1}) depend only on the polar field strength of the particular multipole component and on the associated Legendre function $P_{l1}(\cos{\theta})$ 
and Legendre polynomial $P_l(\cos{\theta})$.  Our general expressions for $B_r$ and $B_{\theta}$ in terms of the polar field strength are valid regardless of the definition of the 
non-primitive (traceless) multipole moments.

Complex magnetic fields can be created by considering linear combinations of the multipole field components, for example, by following a similar
approach to that used by the authors of \cite{lan98,koc02,lon07,lon08,kol09}, who have 
all considered a dipole plus a quadrupole composite field, or the approach of the authors of \cite{moh08} and \cite{lon10}, who have considered higher order multipoles.  Our 
expressions for $B_r$ and $B_{\theta}$ provide the most straightforward method of constructing complex magnetospheric geometries consisting of axial multipoles.

The analytic descriptions of axial multipoles derived in this paper, when used as inputs to new models, will increase our understanding of many important astrophysical 
processes.  Applications include models of the magnetic star-disc interaction, wind launching on both the pre-main sequence, and main sequence, and studies of the
magnetic interaction between stars and orbiting close-in giant planets.

The interaction between T Tauri magnetospheres and the disc is believed to control the rotational evolution of the star \cite{bou07iau} and may also be responsible
for the collimation and launching of outflowing winds and jets \cite{fer06}. 
This process may determine the mass accretion rate, setting the lifetime of the disc, and consequently the brief window of time in which planets may form.
The star-disc interaction depends on the stellar magnetic field topology, with the disc truncation radius being particularly sensitive
to the relative polar strength of each of the multipole field components \cite{gre08}.  Recent observations of the classical T Tauri star V2129~Oph
revealed a complex magnetic topology with field modes up to $l=15$ required to fit the data \cite{don07}.  The magnetic energy was concentrated 
dominantly in the octupole field component, with a weak dipole component dominating at larger radii \cite{gre08}.  As an alternative to numerical field 
extrapolations (see \cite{jar08b} and \cite{gre08}) the large scale magnetosphere of V2129~Oph may be (rather crudely) approximated analytically by
a combination of the $l=1$ and $l=3$ multipoles, giving $B_r = B_{r,dip} + B_{r,oct}$ and $B_{\theta}=B_{\theta,dip} + B_{\theta,oct}$,
where ``dip'' and ``oct'' refer to the dipole and octupole components respectively.  Each of the terms in such linear combinations
are easily derived from (\ref{mag_gen1}).  Such composite expressions for the radial and polar field components do not violate Maxwell's equation 
that the field be solenoidal ($\nabla\cdot \mathbf{B}=0$).   

MHD simulations of the star-disc interaction which incorporate observationally derived magnetic surface maps will represent the next major advancement in the field.  However, this will
require significant computational resources.  The magnetic surface maps contain information about many high order field components, thus any simulations will
require small grid resolutions at the stellar surface to handle the steep gradients in field strength, and this must be coupled with a large enough computational domain to capture the behaviour
at the disc interaction region.  It is questionable whether or not the resulting simulations will yield new insights not already gained from simulations with
tilted dipole plus octupole composite fields \cite{lon10,rom10}.  Such composite fields broadly match the magnetic topologies obtained to date through ZDI of 
accreting T Tauri stars \cite{don07,don08c}.  A more fruitful line of research may be the inclusion of surface transport effects, differential rotation, supergranular diffusion and meridional 
circulation, into MHD simulations, as well as a more complete consideration of accretion shocks, and surface waves, generated by the high velocity impact of the dense accreting gas
\cite{kol08,cra08,cra09}.  

Another anticipated application for our general expressions for $B_r$ and $B_{\theta}$, is in models of stellar winds.  Understanding the rotational 
evolution of stars at all evolutionary phases requires knowledge of how the stellar wind torque 
varies with the stellar parameters (for example \cite{iva03}). Recently, the authors of \cite{mat09} have considered how the stellar wind torque for main sequence solar-like stars 
depends on quantities such as the stellar radius, mass, and rotation rate, the mass outflow rate, and the 
equatorial field strength of the stellar magnetosphere.  By considering both dipolar and quadrupolar magnetospheres, they find that
the stellar wind torque is particularly sensitive to the assumed field geometry.  However, as pointed out \cite{mat09}, further simulations are
required to fully quantify the effects of varying the field topology.  Our analytic expressions for the magnetic field components provide an easy way of 
incorporating more complex field geometries in to stellar wind models.           

Exoplanet migration may be influenced by the stellar wind plasma, as well as by how the stellar magnetosphere interacts with the disc. 
Simulations \cite{rom06}, and analytic work \cite{lin96,fle08}, suggest that the inner disc hole, cleared by 
the star-disc interaction, may provide a natural barrier that decreases the rate of inward migration of forming planets.  However, if a multipolar magnetosphere
were to be considered, the structure of columns of accreting gas within the magnetospheric gap may alter the migration rate of planets \cite{rom06}.  By 
considering the azimuthal ram pressure of a stellar wind, the authors of \cite{lov08} find that a planet will migrate inwards or outwards depending 
on whether the stellar rotation period is greater or less than the planets orbital period.  The current picture is confused, however, as the authors of \cite{vid09} have presented 
3D MHD simulations of stellar winds and conclude that they have little influence on the migration of exoplanets.  But, as pointed out \cite{lov08}, 
multipolar stellar magnetic fields are expected to have a non-uniform distribution of open field, along which a stellar wind could be launched.  A migrating 
planet, in the stellar equatorial plane, may therefore be immersed directly in the stellar outflow.  This potentially will have a larger influence upon planet migration than would be expected
from dipole stellar field models where the wind is launched from the star in a direction away from the equatorial plane.  Such arguments, however, are highly speculative and more 
quantitative work in this area is required.  

Other anticipated applications for our analytic expressions for multipolar magnetic fields include models of magnetic star-planet interaction, and of planetary magnetospheres.  The magnetic fields of 
Jupiter and Saturn, for example, are well known to be multipolar with estimates of their multipole moments readily found in the literature (for example \cite{wil82}). 
Outside the Solar System, it has been argued that orbiting close-in giant exoplanets can lead to stellar activity enhancements. 
This is an emerging research area, which may provide a method for characterising exoplanetary magnetic fields, and consequently their internal 
structure \cite{shk09}.  Variations in the activity of host stars, synchronised with the planetary orbital period, have been observed, although sometimes 
with a short phase lag, perhaps due to the planet perturbing the stellar field lines (for example \cite{wal08}). A possible explanation 
for this is that a planet ploughing through a stellar magnetosphere triggers a release of energy stored-up in the coronal field by decreasing its relative helicity 
(for example \cite{lan08,lan09}).  However, such models have yet to consider multipolar stellar magnetic fields. Realistic multipolar fields have significantly less 
magnetic flux at the position of orbiting close-in planets, even if the large scale topology is ``dipole-like'' in structure \cite{gre08}. 
More observations and theoretical modelling are required to confirm if
magnetic star-planet interaction can indeed lead to stellar activity enhancements.  A more promising method of investigating exoplanetary magnetic fields is 
through the detection of planetary radio emission, arising due to the electron-cyclotron maser instability, where electrons are accelerated along the field lines of the 
planetary magnetosphere due to the electric field generated from the reconnection of the stellar and planetary magnetic fields \cite{jar08b}.      
 
The current generation of spectropolarimeters are providing the community with unrivalled new datasets with which to probe stellar magnetism as a 
function of stellar age and spectral type.  Such datasets can provide crucial guidance for new theoretical models designed to deepen and broaden our 
understanding of all of the above physical processes.  In turn, such new models may be used to make predictions that can be tested with future 
instrumentation.  The HARPS spectropolarimeter will provide better access to astrophysical objects in the southern hemisphere sky \cite{sni08}, and,
as with the PEPSI (Potsdam Echelle Polarimetric and Spectroscopic Instrument) at the Large Binocular Telescope, will allow Stokes Q and U linear polarisation studies,
important for the investigation of magnetic fields at the inner edge of circumstellar discs \cite{str04,lly09}.  In the longer term, SPIRou, a nIR 
spectropolarimeter for the Canada-France-Hawai'i telescope to be implemented in 2014, will provide simultaneous Zeeman broadening measurements and 
magnetic topology information.  


\ack
We thank Dr Aline Vidotto (St Andrews) for comments on how stellar winds may effect planet migration, and both of the anonymous referees 
to whom this manuscript was sent simultaneously for their constructive and detailed comments. This work was supported by the Science and 
Technology Facilities Council [grant number ST/G006261/1].   


\appendix

%
%
\section[]{Relations between the equatorial and polar field strength for a multipole of arbitrary order $l$}\label{equ_appendix}
The Legendre polynomials are defined by (\ref{pl}), however, an alternative definition is 
\begin{equation}
P_l(x) = \sum_{k=0}^N (-1)^k \frac{(2l-2k)!}{2^l k! (l-k)! (l-2k)!}x^{l-2k}
\label{l_series}
\end{equation}
where $N=l/2$ or $N=(l-1)/2$, whichever is an integer \cite{kre06}.  Using this alternative definition 
expressions for $B_r$ and $B_{\theta}$ based on the stellar equatorial, rather than the polar, 
field strength, can be derived. 


\subsection[]{Odd-order multipoles}       
For multipoles of odd-order $l$, for example, a dipole ($l=1$) or an octupole ($l=3$),  $N=(l-1)/2$.  In
the stellar equatorial plane $B_r = 0$, and the field only has a $B_{\theta}$ component.  From 
(\ref{B_general1}) the $B_{\theta}$ component depends on the associated Legendre function $P_{l1}(x)$.  Calculating
$P_{l1}(x)$ requires the Legendre polynomials to be differentiated [see (\ref{plm})].  From (\ref{l_series}) and (\ref{plm})
we obtain,
\begin{eqnarray}
P_{l1}(x)&=&(1-x^2)^{1/2} \Big[\sum_{k=0}^{N-1} (-1)^k \frac{(2l-2k)!(l-2k)}{2^l k! (l-k)! (l-2k)!}x^{l-2k-1}\nonumber \\
         &+& (-1)^{(l-1)/2} \frac{(l+1)!}{2^l [(l-1)/2]![(l+1)/2]!} \Big],
\label{l_series3}
\end{eqnarray}  
where we have substituted for $N=(l-1)/2$ in the final term.  As $x \equiv \cos{\theta}$ then at the equator of the star, where $\theta = \pi/2$, 
the only non-zero term of (\ref{l_series3}) is the final term and therefore, from (\ref{B_general1}), 
\begin{equation}
B_{\theta} = \frac{M_l}{R_{\ast}^{l+2}} \frac{(-1)^{(l-1)/2}(l+1)!}{2^l [(l-1)/2]! [(l+1)/2]!}.
\label{Bequ2}
\end{equation} 
Therefore the field strength at the stellar equator for an odd $l$ number multipole is,
\begin{equation}
B^{l,equ}_{\ast} = (B_r^2 + B_{\theta}^2)^{1/2} = \frac{M_l}{R_{\ast}^{l+2}} \bigg |\frac{(-1)^{(l-1)/2}(l+1)!}{2^l [(l-1)/2]! [(l+1)/2]!}\bigg |.
\label{Bequ3}
\end{equation}
By noting that the only possible negative term in (\ref{Bequ3}) is the $(-1)^{(l-1)/2}$ term, 
and that $|(-1)^{(l-1)/2}| = 1$ regardless of the odd $l$ value, (\ref{Bequ3}) can be re-arranged for the multipole
moment $M_l$.  Comparing the result to (\ref{moment}), a relationship 
between the polar and equatorial field strengths can be determined for an odd $l$-number multipole,
\begin{equation}
B^{l,pole}_{\ast} = \frac{2^l [(l-1)/2]! [(l+1)/2]!}{l!}B^{l,equ}_{\ast}.
\end{equation}
These result is not valid for even $l$-number multipoles.


\subsection[]{Even-order multipoles}
For multipoles of even-order $l$, for example, a quadrupole ($l=2$) or a hexadecapole ($l=4$),  $N=l/2$ and 
(\ref{l_series}) can be re-written as
\begin{equation}
P_l(x) = \sum_{k=0}^{N-1} (-1)^k \frac{(2l-2k)!}{2^l k! (l-k)! (l-2k)!}x^{l-2k} + \frac{(-1)^{l/2} l!}{2^l [(l/2)!]^2}
\label{l_series2}
\end{equation}
where we have substituted for $N=l/2$ in the final term.
As $x \equiv \cos{\theta}$ then at the equator of the star, where $\theta = \pi/2$, the only non-zero term of 
(\ref{l_series2}) is the final term.  For even $l$ number multipoles the field is purely radial in the equatorial plane
($B_{\theta} = 0$).  Therefore, from (\ref{B_general1}), 
\begin{equation}
B_r = \frac{(l+1)}{R_{\ast}^{l+2}}M_l \frac{(-1)^{l/2}l!}{2^l [(l/2)!]^2}
\label{Bequ}
\end{equation}
The equatorial field strength for an even $l$ number multipole is then,
\begin{equation}
B^{l,equ}_{\ast} = (B_r^2 + B_{\theta}^2)^{1/2} = \frac{(l+1)}{R_{\ast}^{l+2}}M_l \bigg | \frac{(-1)^{l/2}l!}{2^l [(l/2)!]^2} \bigg |.
\label{Bequ4}
\end{equation}
Noting that the only possible negative term in (\ref{Bequ4}) is the $(-1)^{l/2}$ term, and that $|(-1)^{l/2}|=1$ regardless of
the even $l$ value, (\ref{Bequ4}) can be re-arranged for the multipole
moment $M_l$.  By then comparing the result to (\ref{moment}), a relationship 
between the polar and equatorial field strengths for an even $l$-number multipole can be determined,
\begin{equation}
B^{l,pole}_{\ast} = \frac{2^l[(l/2)!]^2}{l!}B^{l,equ}_{\ast}.
\end{equation}
This expression is not valid for odd $l$-number multipoles.  


\section[]{Electrostatic expansion using Cartesian tensors}\label{electro_cart}
The authors of \cite{lon07} have recently presented a derivation of the quadrupole component of the potential 
expansion in Cartesian coordinates, based on the electrostatic approach presented in \cite{lan75}, but using pseudo 
magnetic ``charges''.  In this Appendix we derive an expression for the electrostatic potential $\Phi$ external
to a volume (for example, the star) containing the ``charges'' and the coordinate system origin.  As part of the derivation,
for terms in the potential expansion corresponding to the quadrupole and the higher order multipoles, a choice must be made 
regarding the definition of the traceless multipole moment.  Various different definitions are used in the literature,
which ultimately leads to different expressions for the magnetic field components (see the discussion in \S\ref{quad_danger}).   

We consider a finite discrete charge distribution with $N$ charges $q_i$ contained within a volume
which also contains the coordinate origin.  The electrostatic
potential $\Phi$ at a distant field point $\mathbf{r}$ due to a charges at source points 
$\mathbf{r}_1, \mathbf{r}_2, \ldots$ is given by
\begin{equation}
\Phi(\mathbf{r}) = \sum_i \frac{q_i}{|\mathbf{r} - \mathbf{r}_i|}.
\label{expansion}
\end{equation}
This, and the other results below, are straightforward to adapt for a continuous charge distribution
by replacing the sum over the individual charges with an integral over volume of the charge density
(i.e. $\sum_i q_i \rightarrow \int \rho(\mathbf{r})d\mathbf{r}$).  The multipole expansion in terms of Cartesian tensors can then be obtained by 
expanding $|\mathbf{r} - \mathbf{r}_i|^{-1}$ as a Taylor series in $\mathbf{r}_i$,
\begin{eqnarray}
\Phi(\mathbf{r}) &=& \sum_i q_i \Big[ \frac{1}{r} + (-\mathbf{r}_i)\cdot \mathbf{\nabla} \left(\frac{1}{r}\right)
                 + \frac{1}{2!}(-\mathbf{r}_i)(-\mathbf{r}_i) \hspace{1mm}\mathbf{:}\hspace{1mm} \mathbf{\nabla} \mathbf{\nabla} \left(\frac{1}{r}\right) \nonumber \\
&+& \frac{1}{3!}(-\mathbf{r}_i)(-\mathbf{r}_i)(-\mathbf{r}_i) \hspace{1mm}\mathbf{\vdots}\hspace{1mm} \mathbf{\nabla} \mathbf{\nabla} \mathbf{\nabla}\left(\frac{1}{r}\right)+ \ldots \Big], 
\end{eqnarray}
which can be more conveniently written as
\begin{equation}
\Phi (\mathbf{r}) =qT^{(0)}(\mathbf{r}) - \mathbf{\upmu}\cdot\mathbf{T}^{(1)}(\mathbf{r}) + 
                   \frac{1}{2}\mathbf{\uptheta}\hspace{1mm}\mathbf{:}\hspace{1mm}\mathbf{T}^{(2)}(\mathbf{r}) - 
                   \frac{1}{6} {\rm \mathbf{O}}\hspace{1mm}\mathbf{\vdots}\hspace{1mm} \mathbf{T}^{(3)}(\mathbf{r}) + \ldots
\label{general_Psi}
\end{equation}
where $T^{(0)}=1/r$, $\mathbf{T}^{(1)}(\mathbf{r})=\mathbf{\nabla}(1/r)$, $\mathbf{T}^{(2)}(\mathbf{r})=\mathbf{\nabla}\mathbf{\nabla}(1/r)$, $\ldots$ are
gradient tensors.  The primitive multipole moments relative to the origin are given by $q = \sum_i q_i$, $\mathbf{\upmu} = \sum_i q_i \mathbf{r}_i$, 
$\mathbf{\uptheta} = \sum_i q_i \mathbf{r}_i \mathbf{r}_i$, ${\rm \mathbf{O}} = \sum_i q_i \mathbf{r}_i \mathbf{r}_i \mathbf{r}_i$ and so on,
where $q$ is the total charge, and $\mathbf{\upmu}, \mathbf{\uptheta}$ and $\mathbf{O}$ are the dipole, quadrupole and octupole primitive moments 
respectively.  The term primitive moment \cite{raa05} is used here to distinguish these multipole moments from the traceless multipole moments (which are referred
to in the literature as \emph{the} multipole moments) discussed below in \S\ref{appendix_B2}.  
The Cartesian tensors which define the primitive moments are symmetric (e.g. $\theta_{\alpha\beta} = \theta_{\beta\alpha}$) but not 
traceless (e.g. $\Sigma_\alpha \theta_{\alpha\alpha} \ne 0$).  In this notation quantities such
as $\mathbf{r}_i\mathbf{r}_i$ represents the tensor product of the vectors $\mathbf{r}_i$ (in this 
case specifically the dyadic product).   Note that a tensor $\mathbf{rr}$ has components $(\mathbf{rr})_{\alpha\beta}=r_{\alpha}r_{\beta}$ and 
similarly $(\mathbf{rrr})_{\alpha\beta\gamma} = r_{\alpha}r_{\beta}r_{\gamma}$ etc.  The double and triple
dot products are represented by $\mathbf{:}$ and $\mathbf{\vdots}$ respectively and represent full
contractions of the appropriate rank tensors.\footnote{The tensor product is also denoted in the literature
by $\mathbf{r}_i \otimes \mathbf{r}_i \equiv \mathbf{r}_i \mathbf{r}_i$, while a full contraction
of two rank-$n$ tensors $\mathbf{A}$ and $\mathbf{B}$ can also be written as $\mathbf{A}\cdot n \cdot \mathbf{B}$
or $\mathbf{A}[n]\mathbf{B}$.  For example,
$\mathbf{O}\hspace{1mm}\vdots\hspace{1mm} \mathbf{T}^{(3)} \equiv \mathbf{O}\cdot 3 \cdot \mathbf{T}^{(3)}
\equiv \mathbf{O}[3]\mathbf{T}^{(3)}$.  Partial contractions of tensor products can be defined similarly \cite{gra84} but will not be needed here.}  For example, 
$\mathbf{\uptheta}\hspace{1mm}\mathbf{:}\hspace{1mm}\mathbf{T}^{(2)}(\mathbf{r}) = \sum_\alpha \sum_\beta \theta_{\alpha\beta}T^{(2)}_{\beta\alpha}$
represents a full contraction, yielding a scalar, of the second-rank tensors $\mathbf{\uptheta}$ and $\mathbf{T}^{(2)}$.   
(Since the tensors used in this paper are all symmetric we can also write 
$\mathbf{\uptheta}\hspace{1mm}\mathbf{:}\hspace{1mm}\mathbf{T}^{(2)}(\mathbf{r}) = \sum_\alpha \sum_\beta \theta_{\alpha\beta}T^{(2)}_{\alpha\beta}$.)
Equation (\ref{general_Psi}) is valid for an electrostatic potential generated by true electric charges \cite{gra84}, or for a magnetostatic potential generated by
pseudo magnetic charges.  

The multipole expansion (\ref{general_Psi}) is valid for field points $\mathbf{r}$ outside the source region. For some applications \cite{gra09} the multipole 
contributions to the electric field inside the source region can be expressed in terms of so-called contact terms, involving the delta function 
$\delta(\mathbf{r})$. It is interesting to note that the primitive multipole moments are necessary to fully describe these contact field terms; the traceless moments 
are sufficient for the external (long-range) multipole fields (as we discuss below for the electric case and in \S\ref{magn_static} for the magnetic case) but not for 
the internal (contact) multipole fields.  Magnetostatic multipolar contact fields \cite{gra10} are mentioned in \S\ref{field}.

\subsection{The dipole term}
The dipole term of the multipole expansion, the second term in (\ref{general_Psi}),
is straight-forward to derive.  The first gradient tensor
can be written as $\mathbf{T}^{(1)}(\mathbf{r})=\nabla(1/r)=-\mathbf{\hat{r}}r^{-2}$ 
where $\mathbf{\hat{r}}$ is a unit vector along the direction of $\mathbf{r}$.  The dipole moment vector $\mathbf{\upmu}$ has three
components, however, by transforming to the principal axes frame such that the $z$-axis lies along the direction of $\mathbf{\upmu}$, gives
a dipole moment of $\mu = \mu_z$ (principal axes frames are discussed in greater detail below for the quadrupole moment).  By performing the 
dot product contained in the second term of (\ref{general_Psi}) we obtain,  
\begin{equation}
\Phi_1 = \frac{\mu}{r^2}\cos{\theta}.
\label{Psi_dip}
\end{equation}
where we have used the fact that $z/r = \cos{\theta}$, and where $\Phi_1$ is the dipole component of the potential expansion.

\subsection{The quadrupole term}\label{appendix_B2}
As a symmetric second rank tensor, the primitive Cartesian quadrupole moment $\mathbf{\uptheta}$ has six independent
components.  However, the independent components can be reduced to five as terms of the form $\lambda \mathbf{I}$ can
be added to $\mathbf{\uptheta}$, where $\lambda$ is an arbitrary scalar and $\mathbf{I}$ the second rank identity tensor, since
\begin{equation}
\lambda \mathbf{I:T}^{(2)}(\mathbf{r})=\lambda \nabla^2\left(\frac{1}{r}\right)=0 \hspace{10mm} (r>0),
\end{equation}
i.e. $\lambda \mathbf{I}$ contributes nothing to the field external to the distribution.  Note that
in general, a primitive multipole moment of order $l$, which is symmetric, has $(l+1)(l+2)/2$ independent 
components, and $l(l-1)/2$ traces \cite{mel91}.  Generally the non-primitive moment has $(2l+1)$ independent components,
which emerge automatically in the spherical tensor method (see the discussion of equation (\ref{mag_multi}) in \S\ref{magn_static}).  
The form of $\lambda$ is chosen to be (see for example \cite{lei67}),
\begin{equation}
\lambda = -\frac{1}{3}\sum_i q_i r_i^2,
\end{equation} 
from which it is straight forward to show that $\mathbf{\uptheta}+\lambda\mathbf{I}$ is traceless, thus reducing
the number of independent components to five.  The quadrupole term in (\ref{general_Psi}) is then,
\begin{eqnarray}
\Phi_2 (\mathbf{r}) &=& \frac{1}{2}\mathbf{\uptheta :T}^{(2)}(\mathbf{r}) = \frac{1}{2}\sum_i q_i \left(\mathbf{r}_i \mathbf{r}_i - \frac{1}{3}r_i^2\mathbf{I}\right)\mathbf{:T}^{(2)}(\mathbf{r}) \label{quaddef}\\
                    &=& \frac{1}{3}\mathbf{Q:T}^{(2)}(\mathbf{r}) \label{quaddef2},
\label{fullPsi_quad}
\end{eqnarray}
where the traceless (or non-primitive) quadrupole moment is defined as,
\begin{equation}
\mathbf{Q} = \frac{1}{2}\sum_i q_i (3\mathbf{r}_i\mathbf{r}_i - r_i^2 \mathbf{I}).
\label{quad_moment}
\end{equation}
The derivation of the corresponding octupolar term $\Phi_3(\mathbf{r})$ in terms of the non-primitive or traceless Cartesian octupole moment is discussed in 
reference \cite{gra84} and in the arXiv version of this paper.  At this stage it is worth pointing out that there are three different definitions of the non-primitive quadrupole moment used in 
the literature (for example \cite{kom03,lan75,kie72}), which ultimately leads to different expressions for $B_r$ and $B_\theta$
(see \S\ref{quad_danger}).  However, the definition we chose here ensures that the Cartesian tensor approach produces expressions for
the electrostatic case, and equivalently the magnetostatic potential case (see below) expanded using a spherical tensor approach, that are entirely equivalent.
The components of the traceless quadrupole moment tensor are written as,
\begin{equation}
Q_{\alpha\beta} = \frac{1}{2}\sum_i q_i (3r_{i\alpha}r_{i\beta}-r_i^2\delta_{\alpha\beta}),
\end{equation}
where $\delta_{\alpha\beta}$ the Kronecker delta ($\delta_{\alpha\beta}=1$ if $\alpha=\beta$ or =0 if $\alpha \ne \beta$). 

Multipole moments are mathematical constructions, the values of which, in general, depend upon the choice of origin \cite{mel91,raa05}.
The first non-vanishing multipole moment is independent of the origin \cite{gra84}.  
For a body of general shape three of the quadrupole moment components represent the orientation of the body-fixed axes (of the star or planet) relative to the space-fixed 
axes of the coordinate system being used to define the components \cite{gra84}.  As the multipole moment tensors are real and symmetric it is always possible to
transform to the so-called principal axes frame, which has the effect of reducing all off-diagonal elements of 
(\ref{quad_moment}) to zero.  This is equivalent to calculating the principal moments of inertia (moments of a mass distribution)
in mechanics problems (see \cite{gol50} for a thorough discussion of the principal axis transformation).  Choosing the body-fixed 
axes to coincide with the principal axes reduces the number of independent components of $\mathbf{Q}$ to two (recall that as
$\mathbf{Q}$ is traceless, $Q_{xx}+Q_{yy}+Q_{zz}=0$).  In the axisymmetric case the number of independent principal axes components 
reduces to one.  To see this choose the body-fixed $z$-axis of the star (the rotation axis) as the symmetry axis so that $Q_{xx}=Q_{yy}$ by
symmetry and both are equal to $-Q_{zz}/2$ since $\mathbf{Q}$ is traceless.  For axial distributions the body-fixed component $Q_{zz} = Q$ 
is referred to as {\it the} quadrupole moment.  

In order to derive the final expression for the quadrupole term of the potential expansion, we also require
the components of the second rank gradient tensor.  Noting that $\mathbf{T}^{(2)}(\mathbf{r}) = \mathbf{\nabla \nabla} 
(1/r) = (3\mathbf{\hat{r}\hat{r}}-\mathbf{I})r^{-3}$ ($r \ne 0$) then the components of $\mathbf{T}^{(2)}(\mathbf{r})$ can be written as,
\begin{equation}
T_{\alpha\beta}^{(2)}=(3r_{\alpha}r_{\beta}-r^2\delta_{\alpha\beta})r^{-5}.
\end{equation}  
Calculating the double dot product (where $\mathbf{Q:T}^{(2)}=\sum_\alpha \sum_\beta Q_{\alpha\beta}T^{(2)}_{\beta\alpha}$)
in (\ref{fullPsi_quad}) gives the quadrupole component of the multipole expansion for the axial case,
\begin{equation}
\Phi_2 = \frac{Q}{2r^3}(3\cos^2{\theta}-1).
\label{Psi_quad}
\end{equation}

The expressions above for the dipole and quadrupole terms of the electrostatic potential expansion derived from a 
Cartesian tensor approach, $\Phi_1$ and $\Phi_2$ as given by equations (\ref{Psi_dip}) and (\ref{Psi_quad}), are entirely analogous  
to the equivalent terms obtained form the spherical tensor approach to the magnetostatic expansion, $\Psi_1$ and $\Psi_2$ as derived
from equation (\ref{Psi_mag}).  Derivation of the traceless octupole and higher order traceless multipole terms of the potential expansion using the Cartesian tensor approach
is more cumbersome and time consuming; however, the components of the non-primitive octupole and hexadecapole
moments ($l=3$ and $l=4$ respectively) can be found in the literature (for example \cite{sto66}), and    
reference \cite{gra84} also provides a general expression for the $l$-th traceless moment of such an expansion.
Electrostatic multipole expansions can also be carried out using spherical tensors by expanding the $|\mathbf{r}-\mathbf{r'}|^{-1}$ 
term in equation (\ref{expansion}) and using the addition theorem for spherical harmonics following a similar
argument to that discussed in \S\ref{magn_static}.  The resulting expression for
$\Phi_l$, the general term of electrostatic expansion, is entirely analogous to $\Psi_l$, the general term of the magnetostatic expansion given by 
equation (\ref{Psi_mag}).  A detailed first principles derivation can be found in \cite{gra84}.


\section{Octupole term of the scalar potential}\label{octu_appendix}
{\bf (This appendix is included in the astro-ph version only and will not appear in the published version of the article)}.  

\noindent The octupole moment tensor $\mathbf{O}$, as a third rank tensor, has three possible traces
($\sum_\beta O_{\alpha\beta\beta}$).  
In general, a primitive multipole moment of order $l$, which is symmetric, has $(l+1)(l+2)/2$ independent 
components, and $l(l-1)/2$ traces \cite{mel91}.  The non-primitive moment has $(2l+1)$ independent components in general
[as seen automatically using spherical tensors - see (\ref{qlm_def})], or seven for the octupole moment ($l=3$) as we show below.  
Following a similar argument to that of the quadrupole case, see \ref{appendix_B2}, 
a term of the form $(\mathbf{aI})^{sym}$, where $\mathbf{a}$ is an arbitrary vector and $(\mathbf{aI})^{sym}$ denotes
a symmetrised tensor with components 
$(\mathbf{aI})^{sym}_{\alpha\beta\gamma} = (a_\alpha\delta_{\beta\gamma}+a_\beta\delta_{\gamma\alpha}+a_\gamma\delta_{\alpha\beta})$, 
can be added to the primitive tensor since
\begin{equation}
(\mathbf{aI})^{sym} \hspace{1mm} \mathbf{\vdots} \hspace{1mm} \mathbf{T}^{(3)}(\mathbf{r}) = 0, \hspace{10mm} (r>0).
\end{equation}
Because $\mathbf{a}$ is arbitrary we can add a sum of such terms of the form
\begin{equation}
-\frac{1}{5}\sum_i q_i r_i^2 (\mathbf{r}_i\mathbf{I})^{sym}
\end{equation}
such that all three traces of the new octupole moment tensor vanish.  The octupole term in (\ref{general_Psi}) is then,
\begin{eqnarray}
\Phi_3 (\mathbf{r}) &=& -\frac{1}{6}\mathbf{O} \hspace{1mm} \mathbf{\vdots} \hspace{1mm} \mathbf{T}^{(3)}(\mathbf{r})
                    = -\frac{1}{6}\sum_i q_i\left[\mathbf{r}_i\mathbf{r}_i\mathbf{r}_i-\frac{1}{5}r_i^2(\mathbf{r}_i\mathbf{I})^{sym}\right]
                        \hspace{1mm} \mathbf{\vdots} \hspace{1mm} \mathbf{T}^{(3)}(\mathbf{r}) \nonumber \\
                    &=& -\frac{1}{15} \mathbf{\Omega} \hspace{1mm} \mathbf{\vdots} \hspace{1mm} \mathbf{T}^{(3)}(\mathbf{r}),
\label{oct_moment}
\end{eqnarray}
where $\mathbf{\Omega}$, the traceless octupole moment tensor, is defined as,
\begin{equation}
\mathbf{\Omega}=\frac{1}{2}\sum_i q_i(5\mathbf{r}_i\mathbf{r}_i\mathbf{r}_i - r_i^2(\mathbf{r}_i\mathbf{I})^{sym}).
\label{A4}
\end{equation}
This has components,
\begin{equation}
\Omega_{\alpha\beta\gamma} = \frac{1}{2}\int \rho(\mathbf{r})d\mathbf{r}(5r_\alpha r_\beta r_\gamma - 
                             r^2[r_\alpha\delta_{\beta\gamma}+r_\beta\delta_{\gamma\alpha}+r_\gamma\delta_{\alpha\beta}])
\label{app_oct}
\end{equation}
where we have switched to considering a continuous charge distribution rather than a summation over individual 
charges $q_i$ ($\sum_i q_i \rightarrow \int \rho(\mathbf{r})d\mathbf{r}$ where $\rho(\mathbf{r})$ is the charge
density) for ease of notation.  We have chosen the numerical prefactors in (\ref{oct_moment}) and (\ref{A4}) so that (\ref{A4}) agrees with the 
definition using spherical coordinates.  For an arbitrary charge distribution (\ref{app_oct}) has seven independent components 
in a general reference frame.  In the axisymmetric case, where there is one independent principal axes component, called {\it the} 
octupole moment $\Omega = \Omega_{zzz}$, there are seven non-vanishing principal axes components, i.e.    
$\Omega_{xxz}=\Omega_{zxx}=\Omega_{xzx}=\Omega_{yyz}=\Omega_{zyy}=\Omega_{yzy}$ and $\Omega_{zzz}$ with $\Omega_{zzz}=-(\Omega_{xxz}+\Omega_{yyz})$.

The components of the third rank gradient tensor, 
$\mathbf{T}^{(3)}(\mathbf{r})=\nabla\mathbf{T}^{(2)}(\mathbf{r})=\nabla[(3\mathbf{\hat{r}}\mathbf{\hat{r}}-\mathbf{I})r^{-3}]$, can be written as,
\begin{equation}
T^{(3)}_{\alpha\beta\gamma} = (-3r^{-7})(5r_{\alpha}r_{\beta}r_{\gamma}-r^2(r_\alpha\delta_{\beta\gamma}+r_\beta\delta_{\gamma\alpha}
                              +r_\gamma\delta_{\alpha\beta})).
\end{equation}
Thus carrying out the full contraction with the traceless octupole moment tensor 
($\mathbf{\Omega} \hspace{1mm} \mathbf{\vdots} \hspace{1mm} \mathbf{T}^{(3)}=\sum_\alpha \sum_\beta \sum_\gamma \Omega_{\alpha\beta\gamma} T^{(3)}_{\gamma\beta\alpha}$) gives an 
expression for the octupole term of the multipole expansion in the axial case,
\begin{equation}
\Phi_3 = \frac{\Omega}{2r^4}(5\cos^2{\theta}-3)\cos{\theta}.
\label{Psi_oct}
\end{equation}
The octupole moment introduced in this section $\Omega = \Omega_{zzz}$ is equal to 
$Q_3$ defined by (\ref{sphere2_new}), i.e. $Q_3 = \int d\mathbf{r} \rho(\mathbf{r}) r^3 P_3(\cos{\theta})$.  The axial
{\it magnetic} octupolar field has the same form as (\ref{Psi_oct}), with $\Omega = M_3$ where $M_3$ is obtained from the 
definition of $Q_3$ by replacing $\rho(\mathbf{r})$ by $\mathbf{r}\cdot \nabla \times \mathbf{J}/(4c)$.


\section{Electrostatic expansion using spherical tensors}\label{electro_sphere}
{\bf (This appendix is included in the astro-ph version only and will not appear in the published version of the article)}.  

\noindent Derivation of the components of the non-primitive octupole and hexadecapole moments ($l=3$ and $l=4$ respectively) 
can be found in the literature (for example \cite{sto66}), however, for the dotriacontapole ($l=5$) moment and beyond, the 
Cartesian tensor approach quickly becomes cumbersome and 
time-consuming.  Multipole expansions can also be carried out using spherical tensors, which is more straightforward.
A detailed first principles derivation can be found in \cite{gra84}.

\begin{figure}[t]
   \centering
   \includegraphics[width=65mm]{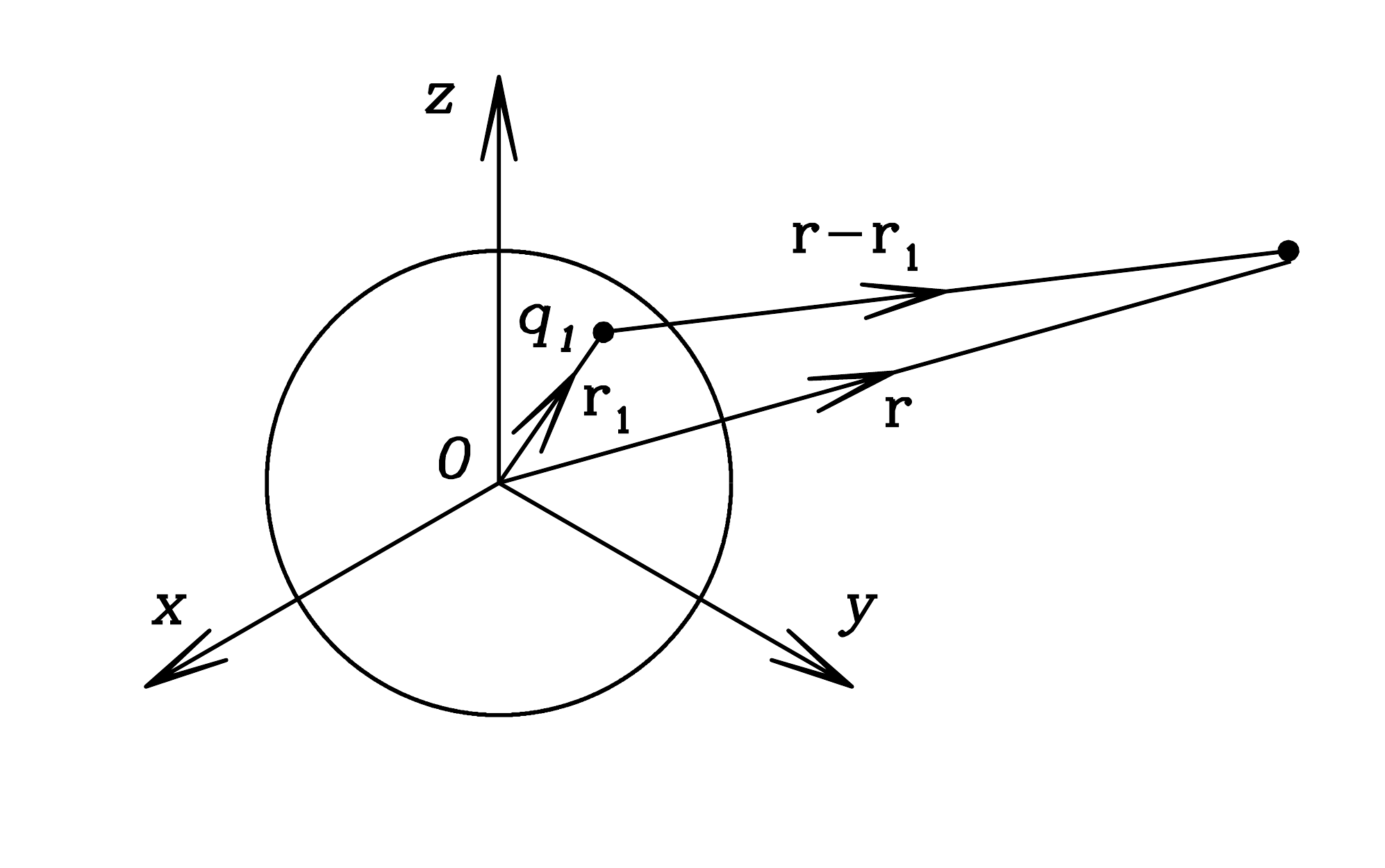} 
   \caption{Coordinates and notation for a finite charge distribution inside a volume which contains
           the origin of the coordinate system $0$.  $\mathbf{r}$ is a distant field point where the potential
           due to discrete charges $q_i$ at source points $\mathbf{r}_i$ is to be calculated.  Only the 
           first charge $q_1$ is shown.}
   \label{fieldsource_new}
\end{figure}

By applying the cosine rule to the triangle in figure \ref{fieldsource_new} the potential $\Phi(\mathbf{r})$ can be re-written as
\begin{equation}
\Phi(\mathbf{r}) = \sum_i \frac{q_i}{|\mathbf{r}-\mathbf{r}_i|}=\sum_i \frac{q_i}{r}
                \left[1+\left(\frac{r_i}{r}\right)^2-2\left(\frac{r_i}{r}\right)\cos{\theta_i}\right]^{-1/2},
\label{sphere3_new}
\end{equation}
where $\theta_i$ is the angle between $\mathbf{r}$ and $\mathbf{r}_i$. The second term in the square brackets is the 
generating function for Legendre polynomials, which allows (\ref{sphere3}) to be
be re-written as
\begin{equation}
\Phi(\mathbf{r})=\sum_i \sum_l q_i \frac{r_i^l}{r^{l+1}}P_l(\cos{\theta_i}).
\end{equation}
Using the addition theorem for spherical harmonics, which expresses the Legendre polynomials
$P_{l}(\cos{\theta})$ as the sum of the product of the spherical harmonics $Y_{lm}(\theta_i,\phi_i)$ and $Y_{lm}^{\ast}(\theta,\phi)$
over the range $m=-l,\ldots,l$, we obtain, 
\begin{equation}
\Phi(\mathbf{r})=\sum_l\sum_m \left(\frac{4\pi}{2l+1}\right)Q_{lm}Y^{\ast}_{lm}(\theta,\phi)/r^{l+1},
\label{Psi_sphere}
\end{equation}
with $Q_{lm}$ being the $m$th component of the spherical multipole moment tensor of order $l$, 
\begin{equation}
Q_{lm}=\sum_i q_i r_i^l Y_{lm}(\theta_i,\phi_i),
\label{qlm_def}
\end{equation}
where the spherical harmonics $Y_{lm}(\theta,\phi)$ are given by (\ref{sph_ham}).  The rotationally invariant form (scalar product) 
$\sum_m A_{lm}B_{lm}^{\ast}$ of two
spherical tensors occurs in (\ref{Psi_sphere}), as it must, and rotational invariance arguments can be used to give an elegant alternative 
derivation of (\ref{Psi_sphere}) \cite{gra84}.  Note also that the spherical multipole moments correspond to the non-primitive (traceless)
Cartesian moments for every $l$.  Thus for $l=2$, for example, there are five quadrupole components $Q_{2m}$ with $m=-2,-1,0,1,2$.  These can be 
written as linear combinations of the five independent Cartesian components $Q_{\alpha\beta}$ \cite{gra84}.  

For axial multipoles, as considered in this paper, we choose the space-fixed axes with the $z$-axis along the symmetry axis.  The
potential then cannot depend on the azimuthal angle $\phi$, so that only terms with $m=0$ can contribute to (\ref{Psi_sphere}), which
therefore reduces to,
\begin{equation}
\Phi(\mathbf{r})=\sum_l \left(\frac{4\pi}{2l+1}\right)^{1/2}Q_{l0}P_l(\cos{\theta})/r^{l+1}
\label{sphere1_new}
\end{equation}
where $Q_{l0}$ is given by
\begin{equation}
Q_{l0}=\left(\frac{2l+1}{4\pi}\right)^{1/2}\sum_i q_i r_i^l P_l(\cos{\theta_i})=\left(\frac{2l+1}{4\pi}\right)^{1/2}Q_l.
\label{sphere2_new}
\end{equation}  
The quantities $Q_l$ are referred to as \emph{the} multipole moments, with $Q_0 \equiv q$, $Q_1 \equiv \mu$, $Q_2 \equiv Q$ and 
$Q_3 \equiv \Omega$ with $q$ the charge, and $\mu$, $Q$ and $\Omega$ respectively the dipole, quadrupole and octupole moments.  
The quantities $q$, $\mu$ and $Q$ are defined in \ref{electro_cart} and $\Omega$ is defined in \ref{octu_appendix}.  In general $Q_{lm}$ is a complicated function 
of the orientation of the charge distribution, but for axial distributions it is a simple function of the 
orientation ($\theta$, $\phi$) of the symmetry axis \cite{gra84}, i.e.,
\begin{equation}
Q_{lm} = Q_l Y_{lm}(\theta, \phi).
\label{newQlm}
\end{equation}
An analogous expression for the traceless Cartesian multipole moments is given in \cite{gra84}.  The 
expressions (\ref{Psi_dip}), (\ref{Psi_quad}), and (\ref{Psi_oct}) derived using the Cartesian tensor 
method are obtained by using (\ref{sphere1_new}) and (\ref{sphere2_new}).  The general term 
of the electrostatic potential expansion for the axial case is then,
\begin{equation}
\Phi_l = \frac{Q_l}{r^{l+1}}P_l(\cos{\theta}).
\label{Psi_general2}
\end{equation}


\bibliographystyle{unsrt}

\section*{References}

\begin{thebibliography}{269}
	
\bibitem{ake05}Akeson R L, Walker C H, Wood K, Eisner J A, Scire E, Penprase B, Ciardi D R, van Belle G T,  Whitney B and Bjorkman J E 
2005 {\it ApJ} {\bf 622} 440--450 	

\bibitem{alt69} Altschuler M D and Newkirk G 1969 {\it Sol. Phys.} {\bf 9} 131--149

\bibitem{aly87} Aly J J 1987 {\it Sol. Phys.} {\bf 111} 287--296

\bibitem{aly90} Aly J J and Kuijpers J 1990 {\it Astron. Astrophys.} {\bf 227} 473--482

\bibitem{aly93} Aly J J and Seehafer N 1993 {\it Sol. Phys.} {\bf 144} 243--254
	
\bibitem{ant99} Antiochos S K, DeVore C R and Klimchuk J A 1999 {\it Astrophys. J.} {\bf 510} 485--493

\bibitem{ard00} Ardila D R and Basri G 2000 {\it Astrophys. J.} {\bf 539} 834--846

\bibitem{ard02} Ardila D R, Basri G, Walter F M, Valenti J A and Johns-Krull C M 2002	{\it Astrophys. J.} {\bf 567} 1013--1027 

\bibitem{arg07} Argiroffi C, Maggio A and Peres G 2007 {\it Astron. Astrophys.} {\bf 465} 5--8

\bibitem{arm96} Armitage P J and Clarke C J 1996 {\it Mon. Not. R. Astron. Soc.} {\bf 280} 458--468

\bibitem{ass02} Asseo E and Khechinashvili 2002 {\it Mon. Not. R. Astron. Soc.} {\bf 334} 743--759

\bibitem{bas92} Basri G, Marcy G W and Valenti J A 1992 {\it Astrophys. J.} {\bf 390} 622--633

\bibitem{bax08} Baxter E, Corrales L, Yamada R and Esin A A 2008 {\it Astrophys. J.} {\bf 689} 308--315

\bibitem{ber01} Beristain G, Edwards S and Kwan J 2001 {\it Astrophys. J.} {\bf 551} 1037--1064

\bibitem{bes08} Bessolaz N, Zanni C, Ferreira J, Keppens R and Bouvier J 2008 {\it Astron. Astrophys.} {\bf 478} 155--162

\bibitem{bou07aatau} Bouvier J \etal 2007 {\it Astron. Astrophys.} {\it 463} 1017--1028

\bibitem{bou93} Bouvier J, Cabrit S, Fernandez M, Martin E L, Matthews J M 1993 {\it Astron. Astrophys.} {\bf 272} 176--206

\bibitem{bou95} Bouvier J, Covino E, Kovo O, Martin E L Matthews J M, Terranegra L and Beck S C 1995 {\it Astron. Astrophys.} {\bf 299} 89--107

\bibitem{bou07pp} Bouvier J, Alencar S H P, Harries T J, Johns-Krull C M and Romanova M M 2007 {\it Protostars and Planets V (Hawai'i, October 2005)}
eds B Reipurth, D Jewitt and K Keil (University of Arizona Press, Tucson) pp~479--494

\bibitem{bou07iau} Bouvier J 2007 {\it Proc. of the IAU Symp. 243 Star-Disk Interaction in Young Stars (May 2007, Grenoble)} 
eds J Bouvier and I Appenzeller (Cambridge: Cambridge University Press) pp~231--240

\bibitem{bro71} Bronzan J B 1971 {\it American J. of Phys.} {\bf 39} 1357--1359

\bibitem{bro81} Brown D N and Landstreet J D 1981 {\it Astrophys. J.} {\bf 246} 899--904 

\bibitem{bro91} Brown S F, Donati J-F, Rees D E and Semel M 1991 {\it Astron. Astrophys.} {\bf 250} 463--474

\bibitem{bro08} Browning M K 2008 {\it Astrophys. J.} {\bf 676} 1262--1280

\bibitem{buc59} Buckingham A D 1959 {\it Quarterly Reviews} {\bf 13} 183--214

\bibitem{cai10} Cai M J 2010 {\it Astrophysics and Space Science Proc. Protostellar Jets in Context (July 2008, Rhodes)} eds
K Tsinganos, T Ray and M Stute (Berlin Heidelberg: Springer) pp~143--152

\bibitem{cam90} Camenzind M 1990 {\it Reviews in Modern Astron.} {\bf 3} 234--265

\bibitem{car07} Carr J S 2007 {\it Proc. of the IAU Symp. 243 Star-Disk Interaction in Young Stars (May 2007, Grenoble)} 
eds J Bouvier and I Appenzeller (Cambridge: Cambridge University Press) pp~135--146

\bibitem{cat07} Catala C, Alecian E, Donati J-F, Wade G A, Landstreet J D, B{\"o}hm T, Bouret J-C, Bagnulo S, Folsom C and Silvester J
2007 {\it Astron. Astrophys.} {\bf 462} 293--301

\bibitem{cha97} Chabrier G and Baraffe I 1997 {\it Astron. Astrophys.} {\bf 327} 1039--1053

\bibitem{cha06} Chabrier G and K{\"u}ker M 2006 {\it Astron. Astrophys.} {\bf 446} 1027--1037

\bibitem{chu07} Chuntonov G A, Smirnov D A and Lamzin S A 2007 {\it Astron. Letters} {\bf 33} 38--44

\bibitem{cie07} 	Cieza L and Baliber N 2007 {\it Astrophys. J.} {\bf 671} 605--615

\bibitem{cla95} Clarke C J, Armitage P J, Smith K W and Pringle J E 1995 {\it Mon. Not. R. Astron. Soc.} {\bf 273} 639--642

\bibitem{cla06} Clarke C J and Pringle J E 2006 {\it Mon. Not. R. Astron. Soc.} {\bf 370} 10--13

\bibitem{col89} Collier Cameron A and Robinson R D 1989 {\it Mon. Not. R. Astron. Soc.} {\bf 236} 57--87

\bibitem{col93} Collier Cameron A and Campbell C G 1993 {\it Astron. Astrophys.} {\bf 274} 309--318	

\bibitem{cra08} Cranmer S R 2008 {\it Astrophys. J.} {\bf 689} 316--334

\bibitem{cra09} Cranmer S R 2009 {\it Astrophys. J.} {\bf 706} 824--843

\bibitem{dev05} DeVore C R and Antiochos S K 2005 {\it Astrophys. J.} {\bf 628} 1031--1045

\bibitem{dew08} de Wijn A G, Lites B W, Berger T E, Frank Z A, Tarbell T D and Ishikawa R 2008 {\it Astrophys. J.} {\bf 684} 1469--1476

\bibitem{dao06} Daou A G, Johns-Krull C M and Valenti J A 2006 {\it Astron. J.} {\bf 131} 520--526

\bibitem{dob06} Dobler W, Stix M and Brandenburg A 2006 {\it Astrophys. J.} {\bf 638} 336--347

\bibitem{don97a} Donati J-F and Brown S F 1997 {\it Astron. Astrophys.} {\bf 326} 1135--1142 

\bibitem{don97b} Donati J-F and Collier Cameron, A 1997 {\it Mon. Not. R. Astron. Soc.} {\bf 291} 1--19

\bibitem{don92} Donati J-F, Brown S F, Semel M, Rees D E, Dempsey R C, Matthews J M, Henry G W and Hall D S 1992 {\it Astron. Astrophys.} {\bf 265} 682--700

\bibitem{don99} Donati J-F, Collier Cameron A, Hussain G A J and Semel M 1999 {\it Mon. Not. R. Astron. Soc.} {\bf 302} 437--456

\bibitem{don01} Donati J-F 2001 {\it Astrotomography, Indirect Imaging Methods in Observational Astronomy (Lecture Notes in Physics {\rm vol 573})} ed H M J Boffin \etal
(Berlin: Springer Verlag) pp~207--231 

\bibitem{don06} Donati J-F \etal 2006 {\it Mon. Not. R. Astron. Soc.} {\bf 370} 629--644

\bibitem{don07} Donati J-F \etal 2007 {\it Mon. Not. R. Astron. Soc.} {\bf 380} 1297--1312

\bibitem{don08a} Donati J-F \etal 2008a {\it Mon. Not. R. Astron. Soc.} {\bf 390} 545--560 

\bibitem{don10aatau} Donati J-F \etal 2010 {\it Mon. Not. R. Astron. Soc.} in press [astro-ph/1007.4407]

\bibitem{don08b} Donati J-F, Moutou C, Far{\`e}s R, Bohlender D, Catala C, Deleuil M, Shkolnik E, Cameron A C, Jardine M M and Walker G A H 2008b 
{\it Mon. Not. R. Astron. Soc.} {\bf 385} 1179--1185 

\bibitem{don08c} Donati J-F \etal 2008c {\it Mon. Not. R. Astron. Soc.} {\bf 386} 1234--1251 

\bibitem{don97c} Donati J-F, Semel M, Carter B D, Rees D E and Collier Cameron A 1997 {\it Mon. Not. R. Astron. Soc.} {\bf 291} 658--682

\bibitem{don09} Donati J-F and Landstreet J D 2009 {\it Annual Review of Astron. Astrophys.} {\bf 47} 333--370

\bibitem{don10} Donati J-F, Skelly M B, Bouvier J, Jardine M M, Gregory S G, Morin J, Hussain G A J, Dougados C, Menard F and Unruh Y 2010 {\it Mon. Not. R. Astron. Soc.}
{\bf 402} 1426--1436

\bibitem{dow10} Downs C, Roussev I I, van der Holst B, Lugaz N, Sokolov I V and Gombosi T I 2010 {\it Astrophys. J.} {\bf 712} 1219--1231

\bibitem{dud95} Dudorov A E 1995 {\it Astron. Reports} {\bf 39} 790--798 

\bibitem{edw94} Edwards S, Hartigan P, Ghandour L and Andrulis C 1994 {\it Astron. J.} {\bf 108} 1056--1070

\bibitem{eis10} Eisner J A, Monnier J D, Woillez J, Akeson R L, Millan-Gabet R, Graham J R, Hillenbrand L A, Pott J-U, Ragland S and Wizinowich P 2010
{\it Astrophys. J.} {\bf 718} 774--794

\bibitem{fal06} Fallscheer C and Herbst W 2006 {\it Astrophys. J.} {\bf 647} 155--158

\bibitem{fei02} Feigelson E D, Garmire G P and Pravdo S H 2002 {\it Astrophys. J.} {\bf 572} 335--349

\bibitem{fer97} Ferreira J 1997 {\it Astron. Astrophys.} {\bf 319} 340--359

\bibitem{fer00} Ferreira J, Pelletier G and Appl S 2000 {\it Mon. Not. R. Astron. Soc.} {\bf 312} 387--397

\bibitem{fer06} Ferreira J, Dougados C and Cabrit S 2006 {\it Astron. Astrophys.} {\bf 453} 785--796

\bibitem{fer08} Ferreira J 2008 {\it New Astron. Reviews} {\bf 52} 42--59

\bibitem{fis08} Fischer W, Kwan J, Edwards S and Hillenbrand L 2008 {\it Astrophys. J.} {\bf 687} 1117--1144

\bibitem{fla05} Flaccomio E, Micela G, Sciortino S, Feigelson E D, Herbst W, Favata F, Harnden F R Jr and Vrtilek S D 2005 {\it Astrophys. J. Supplement Series} {\bf 160} 450--468
	
\bibitem{fle08} Fleck R C 2008 {\it Astrophys. Space Sci.} {\bf 313} 351--356

\bibitem{gar99} Gary G A and Alexander D 1999 {\it Sol. Phys.} {\bf 186} 123--139

\bibitem{get08a} Getman K V, Feigelson E D, Broos P S, Micela G and Garmire G P 2008a {\it Astrophys. J.} {\bf 688} 418--436

\bibitem{get08b} Getman K V, Feigelson E D, Micela G, Jardine M M, Gregory S G and Garmire G P 2008b {\it Astrophys. J.} {\bf 688} 437--455

\bibitem{gho77} Ghosh P, Pethick C J and Lamb F K 1977 {\it Astrophys. J.} {\bf 217} 578--596 

\bibitem{gho79a} Ghosh P and Lamb F K 1979a {\it Astrophys. J.} {\bf 232} 259--276 

\bibitem{gho79b} Ghosh P and Lamb F K 1979b {\it Astrophys. J.} {\bf 234} 296--316 

\bibitem{gia93} Giampapa M S, Basri G S, Johns C M and Imhoff C 1993 {\it Astrophys. J. Supplement Series} {\bf 89} 321-344

\bibitem{gla85}	Glagolevskii Y V, Piskunov N E and Khokhlova V L 1985 {\it Soviet Astron. Lett.} {\bf 11} 154--156

\bibitem{gol50} Goldstein H 1950 {\it Classical mechanics (Addison-Wesley World Student Series)} (Reading, Mass.: Addison-Wesley) p~151

\bibitem{goo97} Goodson A P, Winglee R M and B{\"o}hm K-H 1997 {\it Astrophys. J.} {\bf 489} 199--209

\bibitem{gra76} Gray C G and Stiles P J 1976 {\it Can. J. Phys.} {\bf 54} 513--518

\bibitem{gra78a} Gray C G 1978a {\it Am. J. Phys.} {\bf 46} 169--179 

\bibitem{gra78b} Gray C G 1978b {\it Am. J. Phys.} {\bf 46} 582--583 

\bibitem{gra78c} Gray C G and Nickel B G 1978 {\it Am. J. Phys.} {\bf 46} 735--736

\bibitem{gra79} Gray C G 1979 {\it Am. J. Phys.} {\bf 47} 457--459

\bibitem{gra80} Gray C G 1980 {\it Am. J. Phys.} {\bf 48} 984--985

\bibitem{gra84} Gray C G and Gubbins K E 1984 {\it Theory of molecular fluids Volume 1: Fundamentals} (Oxford: Clarendon Press)

\bibitem{gra09} Gray C G, Karl G and Novikov V A 2009 {\it Am. J. Phys.} {\bf 77} 807--817

\bibitem{gra10} Gray C G, Karl G and Novikov V A 2010 {\it Am. J. Phys.} {\bf 78} in press

\bibitem{gre05} Gregory S G, Jardine M, Collier Cameron A and Donati J-F 2005 {\it Proc. of the 13th Cambridge Workshop on Cool Stars, Stellar Systems and the Sun 
(European Space Agency Special Publications)} (July 2004, Hamburg) eds F Favata, G A J Hussain and B Battrick 560 pp~191--197

\bibitem{gre06a} Gregory S G, Jardine M, Simpson I and Donati J-F 2006a {\it Mon. Not. R. Astron. Soc.} {\bf 371} 999--1013

\bibitem{gre06b} Gregory S G, Jardine M, Cameron A C and Donati J-F 2006b {\it Mon. Not. R. Astron. Soc.} {\bf 373} 827--835

\bibitem{gre07} Gregory S G, Wood K and Jardine M 2007 {\it Mon. Not. R. Astron. Soc.} {\bf 379} 35--39

\bibitem{gre08} Gregory S G, Matt S P, Donati J-F and Jardine M 2008 {\it Mon. Not. R. Astron. Soc.} {\bf 389} 1839--1850

\bibitem{gre09} Gregory S G, Flaccomio E, Argiroffi C, Bouvier J, Donati J-F, Feigelson E D, Getman K V, Hussain G A J, Jardine M and Walter F M 2009
{\it Proc. of the workshop on High Resolution X-ray Spectroscopy: Towards IXO (March 2009, Mullard Space Science Laboratory, Surrey)} eds Branduardi-Raymont G and Blustin A
 published electronically at http://www.mssl.ucl.ac.uk/$\sim$ajb/workshop3/index.html

\bibitem{gue99} Guenther E W, Lehmann H, Emerson J P and Staude J 1999 {\it Astron. Astrophys.} {\bf 341} 768--783

\bibitem{gul98} Gullbring E, Hartmann L, Brice{\~n}o C and Calvet N 1998 {\it Astrophys. J.} {\bf 492} 323--341	

\bibitem{gun07} G{\"u}nther H M, Schmitt J H M M, Robrade J and Liefke C 2007 {\it Astron. Astrophys.} {\bf 466} 1111--1121

\bibitem{har91} Hartigan P, Kenyon S J, Hartmann L, Strom S E, Edwards S, Welty A D and Stauffer J 1991 {\it Astrophys. J.} {\bf 382} 617--635

\bibitem{har02} Hartmann L 2002 {\it Astrophys. J.} {\bf 566} 29--32

\bibitem{hay96} Hayashi M R, Shibata K and Matsumoto R 1996 {\it Astrophys. J. Letters} {\bf 468} 37--40

\bibitem{hil92} Hillenbrand L A, Strom S E, Vrba F J and Keene J 1992 {\it Astrophys. J.} {\bf 397} 613--643

\bibitem{hol07} Holzwarth V and Jardine M 2007 {\it Astron. Astrophys.} {\bf 463} 11--21

\bibitem{hus01} Hussain G A J, Jardine M and Collier Cameron A 2001 {\it Mon. Not. R. Astron. Soc.} {\bf 322} 681--688

\bibitem{hus02} Hussain G A J, van Ballegooijen A A, Jardine M and Collier Cameron A 2002 {\it Astrophys. J.} {\bf 575} 1078--1086

\bibitem{hus07} Hussain G A J, Jardine M, Donati J-F, Brickhouse N S, Dunstone N J, Wood K, Dupree A K, Collier Cameron A and Favata F 
2007 {\it Mon. Not. R. Astron. Soc.} {\bf 377} 1488--1502

\bibitem{hus09} Hussain G A J \etal 2009 {\it Mon. Not. R. Astron. Soc.} {\bf 398} 189--200

\bibitem{lly09} Ilyin I, Strassmeier K G, Woche M and Hofmann A 2009 
{\it Proc. of the International Astronomical Union Symposium on Cosmic Magnetic Fields: From Planets, to 
Stars and Galaxies {\bf vol 259} (November 2008, Tenerife)}
eds K G Strassmeier, A G Kosovichev and J Beckman (Cambridge University Press) 259 pp~663--664

\bibitem{iva03} Ivanova N and Taam R E 2003 {\it Astrophys. J.} {\bf 599} 516--521

\bibitem{jar02a} Jardine M, Collier Cameron A and Donati J-F 2002a {\it Mon. Not. R. Astron. Soc.} {\bf 333} 339--346

\bibitem{jar02b} Jardine M, Wood K, Collier Cameron A, Donati J-F and Mackay D H 2002b {\it Mon. Not. R. Astron. Soc.} {\bf 336} 1364--1370

\bibitem{jar05} Jardine M and van Ballegooijen A A 2005 {\it Mon. Not. R. Astron. Soc.} {\bf 361} 1173--1179

\bibitem{jar06} Jardine M, Collier Cameron A, Donati J-F, Gregory S G and Wood K 2006 {\it Mon. Not. R. Astron. Soc.} {\bf 367} 917--927

\bibitem{jar08a} Jardine M and Cameron A C 2008 {\it Astron. Astrophys.} {\bf 490} 843--851  

\bibitem{jar08b} Jardine M M, Gregory S G and Donati J-F 2008 {\it Mon. Not. R. Astron. Soc.} {\bf 386} 688--696 

\bibitem{jia10} Jiang J, Cameron R, Schmitt D and Sch{\"u}ssler M 2010 {\it Astrophys. J.} {\bf 709} 301--307

\bibitem{joh95} Johns C M and Basri G 1995 {\it Astrophys. J.} {\bf 449} 341--364

\bibitem{joh96} Johns-Krull C M and Valenti J A 1996 {\it Astrophys. J. Lett.} {\bf 459} 95--98

\bibitem{joh97} Johns-Krull C M and Hatzes A P 1997 {\it Astrophys. J.} {\bf 487} 896--915

\bibitem{joh99a} Johns-Krull C M, Valenti J A, Hatzes A P and Kanaan A 1999a {\it Astrophys. J.} {\bf 510} 41--44 

\bibitem{joh99b} Johns-Krull C M and Valenti J A and Koresko C 1999b {\it Astrophys. J.} {\bf 516} 900--915

\bibitem{joh02} Johns-Krull C M and Gafford A D 2002 {\it Astrophys. J.} {\bf 573} 685--698

\bibitem{joh04} Johns-Krull C M, Valenti J A and Saar S H 2004 {\it Astrophys. J.} {\bf 617} 1204--1215

\bibitem{joh07} Johns-Krull C M 2007 {\it Astrophys. J.} {\bf 664} 975--985

\bibitem{joh10} Johnstone C, Jardine M and Mackay D H 2010 {\it Mon. Not. R. Astron. Soc.} {\bf 404} 101--109

\bibitem{joh86} Johnstone R M and Penston M V 1986 {\it Mon. Not. R. Astron. Soc.} {\bf 219} 927--941
	
\bibitem{joh87} Johnstone R M and Penston M V 1987 {\it Mon. Not. R. Astron. Soc.} {\bf 227} 797--800   

\bibitem{ken94} Kenyon S J, Dobrzycka D and Hartmann L 1994 {\it Astron. J.} {\bf 108} 1872--1880

\bibitem{kie72} Kielich S 1972 {\it Journal of Molecular Structure, Dielectric and related Molecular Processes} vol 1 ed M Davies (London: Specialist
Periodical Reports The Chemical Society) pp~192--387

\bibitem{koc04} Kochukhov O, Bagnulo S, Wade G A, Sangalli L, Piskunov N, Landstreet J D, Petit P and Sigut T A A 2004 {\it Astron. Astrophys.} {\bf 414} 613--632

\bibitem{koc02} Kochukhov O and Piskunov N 2002 {\it Astron. Astrophys.} {\bf 388} 868--888

\bibitem{koc09} Kochukhov O and Piskunov N 2009  {\it Proc. of the Conf. on Solar Polarization 5 (ASP Conf. Series) {\rm vol 405} (September 2007, Ascona, Switzerland)}
eds S V Berdyugina, K N Nagendra and R Ramelli (San Francisco: Astron. Soc. of the Pacific) pp~539--542

\bibitem{koc02b} Kochukhov O, Piskunov N, Ilyin I, Ilyina S and Tuominen I 2002 {\it Astron. Astrophys.} {\bf 389} 420--438

\bibitem{koc10} Kochukhov O and Wade G A 2010 {\it Astron. Astrophys.} {\bf 513} A13

\bibitem{kol02a} Koldoba A V, Lovelace R V E, Ustyugova G V and Romanova M M 2002a {\it Astron. J.} {\bf 123} 2019--2026

\bibitem{kol02b} Koldoba A V, Romanova M M, Ustyugova G V and Lovelave R V E 2002b {\it Astrophys. J.} {\bf 576} 53--56	

\bibitem{kol08} Koldoba A V, Ustyugova G V, Romanova M M and Lovelace R V E 2008 {\it Mon. Not. R. Astron. Soc.} {\bf 388} 357--366

\bibitem{kol09} Kolenberg K and Bagnulo S 2009 {\it Astron. Astrophys.} {\bf 498} 543--550

\bibitem{kom03} Kompaneyets A S and Yankovsky G 2003 {\it Theoretical Physics, 2nd rev. engl. ed} (Mineola: Dover Publications Inc.) pp~131--133

\bibitem{kon91} K{\"o}nigl A 1991 {\it Astrophys. J.} {\bf 370} 39--43

\bibitem{kra09} Krasnopolsky R, Shang H and Li Z-Y 2009 {Astrophys. J.} {\bf 703} 1863--1871 

\bibitem{kre06} Kreyszig E 2006 {\it Advanced Engineering Mathematics, 8th ed.} (Singapore: John Wiley \& Sons Inc) p~208

\bibitem{kuk99} K{\"u}ker M and R{\"u}diger G 1999 {\it Astron. Astrophys.} {\bf 346} 922--928

\bibitem{kuk03} K{\"u}ker M, Henning T and R{\"u}diger G 2003 {\it Astrophys. J.} {\bf 589} 397--409

\bibitem{kuk04} K{\"u}ker M, Henning T and R{\"u}diger G 2004 {\it Astrophys. J.} {\bf 614} 526-526

\bibitem{kul08} Kulkarni A K and Romanova M M 2008 {\it Mon. Not. R. Astron. Soc.} {\bf 386} 673--687

\bibitem{kwa07} Kwan J, Edwards S and Fischer W 2007 {\it Astrophys. J.} {\bf 657} 897--915

\bibitem{lan75} Landau L D and Lifshitz E M 1975 {\it The Classical Theory of Fields, 4th rev. engl. ed.} (Oxford: Pergamon Press) pp~105-107

\bibitem{lan98} Landolfi M, Bagnulo S and Landi degl'Innocenti M 1998 {\it Astron. Astrophys.} {\bf 338} 111--121
 
\bibitem{lan08} Lanza A F 2008 {\it Astron. Astrophys.} {\bf 487} 1163--1170

\bibitem{lan09} Lanza A F 2009 {\it Astron. Astrophys.} {\bf 505} 339--350 

\bibitem{lei67} Leibowitz E and Steinitz R 1967 {\it The Magnetic and Related Stars: Proc. of the AAS-NASA Symp. on the Magnetic and Other 
Peculiar and Metallic-Line A Stars (November 1965, Greenbelt)} ed R C Cameron (Baltimore: Mono Book Corp.) pp~89--96

\bibitem{lev82} Levine R H, Schulz M and Frazier E N 1982 {\it Sol. Phys.} {\bf 77} 363--392

\bibitem{li96a} Li J 1996 {\it Astrophys. J.} {\bf 456} 696--707

\bibitem{li96b} Li J, Wickramasinghe D T and R{\"u}diger G 1996 {\it Astrophys. J.} {\bf 469} 765--775

\bibitem{lin96} Lin D N C, Bodenheimer P and Richardson D C 1996 {\it Nature} {\bf 380} 606--607

\bibitem{lin99} Linker J A, Miki{\'c} Z, Biesecker D A, Forsyth R J, Gibson S E, Lazarus A J, Lecinski A, Riley P, Szabo A and Thompson B J	
1999 {\it J. of Geophys. Research} {\bf 104} 9809--9830

\bibitem{liu08} Liu Y and Lin H 2008 {\it Astrophys. J.} {\bf 680} 1496--1507

\bibitem{lon05} Long M, Romanova M M and Lovelace R V E 2005 {\it Astrophys. J.} {\bf 634} 1214--1222

\bibitem{lon07} Long M, Romanova M M and Lovelace R V E 2007 {\it Mon. Not. R. Astron. Soc.} {\bf 374} 436--444

\bibitem{lon08} Long M, Romanova M M and Lovelace R V E 2008 {\it Mon. Not. R. Astron. Soc.} {\bf 386} 1274--1284

\bibitem{lon10} Long M, Romanova M M and Lamb F K 2010 {\it Mon. Not. R. Astron. Soc.} submitted [astro-ph/0911.5455]

\bibitem{lov95} Lovelace R V E Romanova M M and Bisnovatyi-Kogan G S 1995 {\it Mon. Not. R. Astron. Soc.} {\bf 275} 244--254

\bibitem{lov08} Lovelace R V E, Romanova M M and Barnard A W 2008 {\it Mon. Not. R. Astron. Soc.} {\bf 389} 1233--1239

\bibitem{luf10} L{\"u}ftinger T, Kochukhov O, Ryabchikova T, Piskunov N, Weiss W W and Ilyin I 2010 {\it Astron. Astrophys.} {\bf 509} A71

\bibitem{mil07} Millan-Gabet R, Malbet F, Akeson R, Leinert C, Monnier J, Waters R 2007 {\it Protostars and Planets V (Hawai'i, October 2005)}
eds B Reipurth, D Jewitt and K Keil (University of Arizona Press, Tucson) pp~539--554

\bibitem{mat10} Matt S P, Pinz{\'o}n G, de la Reza R and Greene T P 2010 {\it Astrophys. J.} {\bf 714} 989--1000

\bibitem{mat04} Matt S and Pudritz R E 2004 {\it Astrophys. J.} {\bf 607} 43--46

\bibitem{mat05a} Matt S and Pudritz R E 2005a {\it Mon. Not. R. Astron. Soc.} {\bf 356} 167--182

\bibitem{mat05b} Matt S and Pudritz R E 2005b {\it Astrophys. J. Lett.} {\bf 632} 135--138

\bibitem{mat07} Matt S and Pudritz R E 2007 {\it Star-Disk Interaction in Young Stars, Proc. of the Int. Astron. Union, IAU Symposium {\rm vol 243}
(May 2007, Grenoble)} eds J Bouvier and I Appenzeller (Cambridge University Press) pp~299--306

\bibitem{mat08a} Matt S and Pudritz R E 2008a {\it Astrophys. J.} {\bf 678} 1109--1118

\bibitem{mat08b} Matt S and Pudritz R E 2008b {\it Astrophys. J.} {\bf 681} 391--399

\bibitem{mat09} Matt S P and Pudritz R E 2009 {\it Proc. of the 15th Cambridge Workshop on Cool Stars, Stellar
Systems and the Sun (American Institute of Physics Conf. Series) {\rm vol 1094} (July 2008, St Andrews)} 
ed E Stemples (Melville, New York: American Institue of Physics) pp~369--372 

\bibitem{mci06} McIvor T, Jardine M and Holzwarth V 2006 {\it Mon. Not. R. Astron. Soc.} {\bf 367} 1--5

\bibitem{mel91} Melrose D B and McPhedran R C 1991 {\it Electromagnetic Processes in Dispersive Media}
(Cambridge: Cambridge University Press) pp~26--31

\bibitem{men99} M{\'e}nard F and Bertout C 1999 {\it The Origin of Stars and Planetary Systems} 
eds C J Lada and N D Kylafis (Kluwer Academic Publishers) pp~341--375

\bibitem{mil97} Miller K A and Stone J M 1997 {\it Astrophys. J.} {\bf 489} 890--902

\bibitem{moh08} Mohanty S and Shu F H 2008 {\it Astrophys. J.} {\bf 687} 1323--1338

\bibitem{mor08} Morin J \etal 2008 {\it Mon. Not. R. Astron. Soc.} {\bf 390} 567--581

\bibitem{mor10} Morin J, Donati J-F, Petit P, Delfosse X, Forveille T and Jardine M M 2010 {\it Mon. Not. R. Astron. Soc.} in press [astro-ph/1005.5552]

\bibitem{mos03} Moss D 2003 {\it Astron. Astrophys.} {\bf 403} 693--697	

\bibitem{mou07} Moutou C, Donati J-F, Savalle R, Hussain G, Alecian E, Bouchy F, Catala C, Collier Cameron A, Udry S and Vidal-Madjar A 2007 
{\it Astron. Astrophys.} {\bf 473} 651--660

\bibitem{naj94} Najita J R and Shu F H 1994 {\it Astrophys. J.} {\bf 429} 808--825

\bibitem{naj03} Najita J, Carr J S and Mathieu R D 2003 {\it Astrophys. J.} {\bf 589} 931--952

\bibitem{ngu09} Nguyen D C, Scholz A, van Kerkwijk M H, Jayawardhana R and Brandeker A 2009 {\it Astrophys. J. Lett.}  {\bf 694} 153--157

\bibitem{ost95} Ostriker E C and Shu F H 1995 {\it Astrophys. J.} {\bf 447} 813--828

\bibitem{paa96} Paatz G and Camenzind M 1996 {\it Astron. Astrophys.} {\bf 308} 77--90

\bibitem{par58} Parker E N 1958 {\it Astrophys. J.} {\bf 128} 664--675

\bibitem{pas07} Pascucci I \etal 2007 {\it Astrophys. J.} {\bf 663} 383--393

\bibitem{pev03} Pevtsov A A, Fisher G H, Acton L W, Longcope D W, Johns-Krull C M, Kankelborg C C and Metcalf T R 2003 {\it Astrophys. J.} {\bf 598} 1387--1391

\bibitem{pha09} Phan-Bao N, Lim J, Donati J-F, Johns-Krull C M and Mart{\'i}n E L 2009 {\it Astrophys. J.} {\bf 704} 1721--1729

\bibitem{pin08} Pinte C, M{\'e}nard F, Berger J P, Benisty M and Malbet F 2008 {\it Astrophys. J.} {\bf 673} 63--66

\bibitem{pis85} Piskunov N E 1985 {\it Soviet Astron. Lett.} {\it 11} 18--21

\bibitem{pis02} Piskunov N and Kochukhov O 2002 {\it Astron. Astrophys.} {\bf 381} 736--756

\bibitem{pis03} Piskunov N and Kochukov O 2003 {\it Proc. of the Conf. on Solar Polarization (ASP Conf. Proc.) {\rm vol 307} (October 2002, Tenerife)} 
eds J Trujillo-Bueno and J S Almeida (San Francisco: Astron. Soc. of the Pacific) pp~539--548

\bibitem{pne71} Pneuman G W and Kopp R A 1971 {\it Sol. Phys.} {\bf 18} 258--270

\bibitem{pre05} Preibisch T \etal 2005 {\it Astrophys. J. Supplement Series} {\bf 160} 401--422

\bibitem{raa05} Raab R E and de Lange O L 2005 {\it Multipole Theory in Electromagnetism (Int. Series of Monographs on Physics {\rm vol 128})} 
(Oxford: Oxford University Press) pp~1--31

\bibitem{rei06} Reiners A and Basri G 2006 {\it Astrophys. J.} {\bf 644} 497--509

\bibitem{rei07} Reiners A and Basri G 2007 {\it Astrophys. J.} {\bf 656} 1121--1135

\bibitem{rei09} Reiners A and Basri G 2009 {\it Astron. Astrophys.} {\bf 496} 787--790

\bibitem{rob07} Robitaille T P, Whitney B A, Indebetouw R and Wood K 2007 {\it Astrophys. J. Supplement Series} {\bf 169} 328--352 

\bibitem{rob80a} Robinson R D, Worden S P and Harvey J W 1980 {\it Astrophys. J. Lett.} {\bf 236} 155--158

\bibitem{rob80b} Robinson Jr R D 1980 {\it Astrophys. J.} {\bf 239} 961--967

\bibitem{rom02} Romanova M M, Ustyugova G V, Koldoba A V and Lovelace R V E 2002 {\bf 578} 420--438

\bibitem{rom03} Romanova M M, Ustyugova G V, Koldoba A V, Wick J V and Lovelace R V E 2003 {\it Astrophys. J.} {\bf 595} 1009--1031

\bibitem{rom04} Romanova M M, Ustyugova G V, Koldoba A V and Lovelace R V E 2004 {Astrophys. J.} {\bf 610} 920--932

\bibitem{rom06} Romanova M M and Lovelace R V E 2006 {\it Astrophys. J.} {\bf 645} 73--76

\bibitem{rom08} Romanova M M, Kulkarni A K and Lovelace R V E 2008 {\it Astrophys. J. Letters} {\bf 673} 171--174

\bibitem{rom10} Romanova M M, Long M, Lamb F K, Kulkarni A K and Donati J-F 2010 {\it Mon. Not. R. Astron. Soc.} submitted [astro-ph/0912.1681]

\bibitem{ril06} Riley P, Linker J A, Miki{\'c} Z, Lionello R, Ledvina S A and Luhmann J G 2006 {\it Astrophys. J.} {\bf 653} 1510--1516

\bibitem{rua08} Ruan P, Wiegelmann T, Inhester B, Neukirch T, Solanki S K and Feng L 2008 {\it Astron. Astrophys.} {\bf 481} 827--834

\bibitem{saa88} Saar S H 1988 {\it Astrophys. J.} {\bf 324} 441--465

\bibitem{saa94} Saar S H 1994 {\it Proc. of the IAU Symp. 154 Infrared Solar Physics (March 1992, Tucson)} 
ed D M Rabin \etal (Dordrecht: Kluwer Academic) pp~493--497

\bibitem{saa85} Saar S H and Linsky J L 1985 {\it Astrophys. J. Lett.} {\bf 299} 47--50

\bibitem{saf98} Safier P N 1998 {\it Astrophys. J.} {\bf 494} 336--341

\bibitem{san09} Sandman A W, Aschwanden M J, Derosa M L, W{\"u}lser J P and Alexander D 2009 {\it Sol. Phys.} {\bf 259} 1--11

\bibitem{sch69} Schatten K H, Wilcox J M and Ness N F 1969 {\it Sol. Phys.} {\bf 6} 442--455

\bibitem{sch71} Schatten K H 1971 {\it in Solar Wind}, eds C P Sonett, P J Coleman Jr. and J M Wilcox (Washington D.C.: NASA SP-308) pp~44--54

\bibitem{sch75} Sch{\"u}$\upbeta$ler M 1975 {\it Astron. Astrophys.} {\bf 38} 263--270

\bibitem{sch78} Schulz M, Frazier E N and Boucher D J Jr 1978 {\it Sol. Phys.} {\bf 60} 83--104

\bibitem{sch96} Schulz M and McNab M C 1996 {\it J. of Geophys. Research} {\bf 101} 5095--5118

\bibitem{sch97} Schulz M 1997 {\it Annales Geophysicae} {\bf 15} 1379--1387

\bibitem{sch06} Sch{\"u}ssler M and Baumann I 2006 {\it Astron. Astrophys.} {\bf 459} 945--953

\bibitem{sem89} Semel M 1989 {\it Astron. Astrophys.} {\bf 225} 456--466

\bibitem{shk08} Shkolnik E, Bohlender D A, Walker G A H and Collier Cameron A 2008 {\it Astrophys. J.} {\bf 676} 628--638

\bibitem{shk09} Shkolnik E, Aigrain S, Cranmer S, Fares R, Fridlund M, Pont F, Schmitt J, Smith A and Suzuki T 2009 
{\it Proc. of the 15th Cambridge Workshop on Cool Stars, Stellar Systems and the Sun (American Institute of Physics Conf. Series) 
{\rm vol 1094} (July 2008, St Andrews)} ed E Stemples (Melville, New York: American Institue of Physics) pp~275--282

\bibitem{shu94a} Shu F, Najita J, Ostriker E, Wilkin F, Ruden S and Lizano S 1994a {\it Astrophys. J.} {\bf 429} 781--796

\bibitem{shu94b} Shu F H, Najita J, Ruden S P and Lizano S 1994b {\it Astrophys. J.} {\bf 429} 797--807	

\bibitem{shu95} Shu F H and Najita J, Ostriker E C and Shang H 1995 {\it Astrophys. J. Letters} {\bf 455} 155--158

\bibitem{shu96} Shu F H, Najita J, Ostriker E C, Shang H 1996 {\it Astrophys. J. Letters} {\bf 459} 43--43

\bibitem{sie00} Siess L, Dufour E and Forestini M 2000 {\it Astron. Astrophys.} {\bf 358} 593--599

\bibitem{smi03} Smirnov D A, Fabrika S N, Lamzin S A and Valyavin G G 2003 {\it Astron. Astrophys.} {\bf 401} 1057--1061

\bibitem{smi04} Smirnov D A, Lamzin S A, Fabrika S N and Chuntonov G A 2004 {\it Astron. Letters} {\bf 30} 456--460

\bibitem{smi05} Smirnov D A, Romanova M M and Lamzin S A 2005 {\it Astron. Letters} {\bf 31} 335--339

\bibitem{sni08} Snik F, Jeffers S, Keller C, Piskunov N, Kochukhov O, Valenti J and Johns-Krull C 2008
{\it Ground-based and Airborne Instrumentation for Astronomy II (Society of Photo-Optical Instrumentation Engineers (SPIE) Conf. Series) {\rm vol 701} (June 2008, Marseille)}
ed I S McLean and M M Casali (Bellingham, Washington: SPIE) pp~70140O-70140O-10

\bibitem{sta99} Stassun K G, Mathieu R D, Mazeh T and Vrba F J 1999 {\it Astron. J.} {\bf 117} 2941--2979 

\bibitem{sta01} Stassun K G, Mathieu R D, Vrba F J, Mazeh T and Henden A 2001 {\it Astron. J.} {\bf 121} 1003--1012

\bibitem{sto66} Stogryn D E and Stogryn A P 1966 {\it Molecular Physics} {\bf 11} 371--393

\bibitem{str04} Strassmeier K G, Pallavicini R, Rice J B and Andersen M I 2004 {\it Astronomische Nachrichten} {\bf 325} 278--298

\bibitem{str05} Strassmeier K G, Rice J B, Ritter A, K{\"u}ker M, Hussain G A J, Hubrig S and Shobbrook R 2005 {\it Astron. Astrophys.} {\bf 440} 1105--1118

\bibitem{sym05} Symington N H, Harries T J, Kurosawa R and Naylor T 2005 {\it Mon. Not. R. Astron. Soc.} {\bf 358} 977--984	

\bibitem{tad09} Tadesse T, Wiegelmann T and Inhester B 2009 {\it Astron. Astrophys.} {\bf 508} 421--432

\bibitem{tay87} Tayler R J 1987 {\it Mon. Not. R. Astron. Soc.} {\bf 227} 553--561

\bibitem{tor02} Torres del Castillo G F 2002 {\it Revista Mexicana de Fisica} {\bf 48} 348--354

\bibitem{uch81} Uchida Y and Low B C 1981 {\it J. of Astrophys. Astron.} {\bf 2} 405--419

\bibitem{uch84} Uchida Y and Shibata K 1984 {\it Publication Astron. Soc. Japan} {\bf 36} 105--118
 
\bibitem{udd09} Ud-Doula A, Owocki S P and Townsend R H D 2009 {\it Mon. Not. R. Astron. Soc.} {\bf 392} 1022--1033

\bibitem{usm96} Usmanov A V 1996 {\it Proc. of the 8th International Solar Wind Conf. (American Institute of Physics Conf. Proc.) {\rm vol 382}}
eds D Winterhalter, J T Gosling, S R Habbal, W S Kurth and M Neugebauer (New York: American Institute of Physics) pp~141--144

\bibitem{val03} Valenti J A, Johns--Krull C M and Hatzes A P 2003 {\it The Future of Cool-Star Astrophysics: 12th Cambridge Workshop on Cool Stars, 
Stellar Systems, and the Sun} (July 2001, Boulder) eds A Brown, G M Harper and T R Ayres pp~729--734

\bibitem{val04} Valenti J A and Johns-Krull C M 2004 {\it Astrophys. Space Sci.} {\bf 292} 619--629

 \bibitem{van94} van Ballegooijen A A 1994 {\it Space Science Reviews} {\bf 68} 299--307

\bibitem{vid09} Vidotto A A, Opher M, Jatenco-Pereira V and Gombosi T I 2009 {\it Astrophys. J.} {\bf 703} 1734--1742

\bibitem{von04} von Rekowski B and Brandenburg A 2004 {\it Astron. Astrophys.} {\bf 420} 17--32

\bibitem{von06a} von Rekowski B and Brandenburg A 2006 {\it Astronomische Nachrichten} {\bf 327} 53--71

\bibitem{von06b} von Rekowski B and Piskunov N 2006 {\it Astronomische Nachrichten} {\bf 327} 340--354

\bibitem{wal08} Walker G A H \etal 2008 {\it Astron. Astrophys.} {\bf 482} 691--697

\bibitem{wan97} Wang Y-M 1997 {\it Astrophys. J.} {\bf 487} 85--88

\bibitem{wan97_2} Wang Y-M \etal 1997 {\it Astrophys. J.} {\bf 485} 419--429

\bibitem{wil82} Willis D M and Osborne A D 1982 {\it Geophysical J.} {\bf 68} 765--776

\bibitem{win05} Winch D E, Ivers D J, Turner J P R and Stening R J 2005 {\it Geophys. J. Int.} {\bf 160} 487--504

\bibitem{wol85} Wolfson R 1985 {\it Astrophys. J.} {\bf 288} 769--778

\bibitem{yan05} Yang H, Johns-Krull C M and Valenti J A 2005 {\it Astrophys. J.} {\bf 635} 466--475

\bibitem{yan07} Yang H, Johns-Krull C M and Valenti J A 2007 {\it Astron. J.} {\bf 133} 73--80

\bibitem{yan08} Yang H, Johns-Krull C M and Valenti J A 2008 {\it Astron. J.} {\bf 136} 2286--2294

\bibitem{yan09} Yang H and Johns-Krull C M 2009 {\it Proc. of the 15th Cambridge Workshop on Cool Stars, Stellar
Systems and the Sun (American Institute of Physics Conf. Series) {\rm vol 1094} (July 2008, St Andrews)} 
ed E Stemples (Melville, New York: American Institue of Physics) pp~736--739 

\bibitem{yel06} Yelenina T G, Ustyugova G V and Koldoba A V 2006 {\it Astron. Astrophys.} {\bf 458} 679--686 

\bibitem{zan09} Zanni C and Ferreira J 2009 {\it Astron. Astrophys.} {\bf 508} 1117--1133

\bibitem{zha94} Zhao X and Hoeksema J T 1994 {\it Sol. Phys.} {\bf 151} 91--105 

\bibitem{zha95} Zhao X and Hoeksema J T 1995 {\it J. of Geophys. Research} {\bf 100} 19--33

\bibitem{zwe06} Zweibel E G, Hole K T and Mathieu R D 2006 {\it Astrophys. J.} {\bf 649} 879--887

\end{thebibliography}

\end{document}